\documentclass[reprint, amsmath,amssymb,showpacs,floatfix,longbibliography, aps, twocolumn, superscriptaddress,nofootinbib]{revtex4-2}

\usepackage{amsthm}

\newtheorem{definition}{Definition}%
\usepackage[utf8]{inputenc}
\usepackage{amsmath,amssymb,amsfonts,stmaryrd}
\usepackage{physics}
\usepackage{dsfont}
\usepackage{cancel}
\usepackage{mathrsfs}
\usepackage{relsize}
\usepackage{nameref}
\usepackage{hyperref}
\hypersetup{colorlinks=true,citecolor=blue,linkcolor=blue, urlcolor=blue, breaklinks=true}
\hypersetup{linktocpage}
\usepackage[all]{xy}
\usepackage{yhmath}
\usepackage{blkarray}
\usepackage[table]{xcolor}

\usepackage{diagbox}

\usepackage{graphics}
\usepackage{graphicx}
\usepackage{capt-of}     

\usepackage{bm} 
\usepackage{subfigure} 
\usepackage{environ}
\usepackage{url}
\usepackage{scalerel}
\usepackage{stackengine,wasysym}
\usepackage{enumitem}
\usepackage{tikz}

\allowdisplaybreaks

\usepackage{mathtools}

\newcommand{\ZZ}{{\mathbb Z}}

\newcommand{\eps}{\epsilon}

\newcommand{\bx}{{\overline{x}}}
\newcommand{\by}{{\overline{y}}}
\newcommand{\mX}{{\mathcal{X}}}
\newcommand{\mZ}{{\mathcal{Z}}}
\newcommand{\mS}{{\mathcal{S}}}
\newcommand{\mP}{{\mathcal{P}}}

\newcommand{\bigsubstack}[1]{%
  \begin{array}{@{}c@{}}%
    #1%
  \end{array}%
}

\graphicspath{{./Figures/}}

\NewEnviron{eqs}{%
\begin{equation}\begin{split}
    \BODY
\end{split}\end{equation}}

\newcommand{\change}{\color{black}}

\begin{document}

\author{Zijian Liang}
\affiliation{International Center for Quantum Materials, School of Physics, Peking University, Beijing 100871, China}

\author{Jens Niklas Eberhardt}
\email[E-mail: ]{mail@jenseberhardt.com}
\affiliation{Institute of Mathematics, Johannes Gutenberg-Universität Mainz, Germany}

\author{Yu-An Chen}
\email[E-mail: ]{yuanchen@pku.edu.cn}
\affiliation{International Center for Quantum Materials, School of Physics, Peking University, Beijing 100871, China}

\date{\today}
\title{Planar quantum low-density parity-check codes with open boundaries}

\begin{abstract}

Although high-threshold and low-overhead quantum low-density parity-check (qLDPC) codes, such as bivariate bicycle (BB) codes, can reduce the physical-qubit cost by an order of magnitude compared to the Kitaev toric code, their torus layout remains difficult for physical implementation. In this work, we introduce the first systematic procedure to convert BB codes into fully planar, open-boundary qLDPC codes, preserving their performance.
We present planar code families with logical dimensions $6 \leq k\leq13$, e.g., $[[78, 6, 6]]$, $[[107, 7, 7]]$, $[[268, 8, 12]]$, $[[405, 9, 15]]$, $[[348, 10, 13]]$, $[[450, 11, 15]]$, $[[386, 12, 12]]$, $[[362, 13, 11]]$, all with geometrically local weight-6 stabilizers. Allowing weight‑8 stabilizers produces a $[[282,12,14]]$ code, exhibiting an efficiency metric ($kd^2/n$) an order of magnitude higher than the surface code.
The construction combines boundary anyon condensation with the ``lattice grafting'' optimization, yielding high-performance qLDPC codes natively compatible with planar hardware architectures.
It also uncovers Sierpinski‑type fractal logical operators whose distance scales with the fractal area on finite lattices.
These planar qLDPC codes provide an implementable route to resource-efficient, high-threshold fault tolerance and a flexible framework for future code design on realistic two-dimensional hardware.




\end{abstract}

\maketitle

\tableofcontents

\section{Introduction}
The realization of large-scale fault-tolerant quantum computers relies on quantum error correction (QEC) to protect fragile quantum information from noise~\cite{Shor1995Scheme, Steane1996Error, Knill1997Theory, gottesman1997stabilizer, kitaev2003fault}.
While the surface code has emerged as a main candidate due to its planar geometry and local stabilizer checks~\cite{bravyi1998quantum, dennis2002topological, terhal2015quantum, semeghini2021probing, Verresen2021PredictionTC, bluvstein2022quantum, google2023suppressing, Google2023NonAbelian, Google2024surface, iqbal2023topological, iqbal2024NonAbelian, Cong2024EnhancingTO}, its resource overhead remains a significant challenge, requiring potentially millions of physical qubits for practical algorithms~\cite{Fowler2012Surfacecodes, Litinski2019gameofsurfacecodes}.
For the rotated surface code~\cite{delfosse2016generalized}, the scaling is limited by the ratio ${k d^2}/{n} \approx 1$,
where $n$, $k$, and $d$ denote the numbers of physical qubits, logical qubits, and the code distance, respectively.
This motivates the search for alternative quantum codes that are also implementable with local interactions on a 2D planar architecture but offer better resource efficiency.

While general quantum low-density parity-check (qLDPC) codes can achieve linear scaling of both the number of logical qubits $k$ and the code distance $d$ with the number of physical qubits $n$~\cite{breuckmann2021quantum, panteleev2021degenerate, breuckmann2021balanced, Panteleev2022goodqldpc, Lin2022c3Locally, Dinur2023Good}, two-dimensional local qLDPC codes are restricted by the Bravyi–Poulin–Terhal bound~\cite{bravyi2009no, Bravyi2010Tradeoffs}:
\begin{equation}
    \frac{kd^2}{n} \in \mathcal{O}(1).
\end{equation}
Within this constraint, we aim to significantly improve the \emph{constant factor} over the surface code.
In this context, bivariate bicycle (BB) codes have recently attracted considerable attention. With a fixed parity-check weight of $6$, they can achieve a ratio of $kd^2/n \approx 10$ on a torus, thereby representing an order-of-magnitude improvement over the toric code~\cite{Kovalev2013QuantumKronecker, Pryadko2022DistanceGB, Bravyi2024HighThreshold, wang2024coprime, Wang2024Bivariate, tiew2024low, wolanski2024ambiguity, gong2024toward, maan2024machine, cowtan2024ssip, shaw2024lowering, cross2024linear, voss2024multivariate, berthusen2025toward, Eberhardt2025Logical, lin2025single, liang2025generalized, chen2025anyon}.
However, implementing torus geometry on planar hardware (e.g., superconducting architectures) is challenging, as it requires non-local stabilizer checks that undermine the geometric locality advantage of the surface code.

To overcome the geometric constraints of BB codes, we will use insights from condensed matter physics, specifically the theory of topological order.
In this framework, 2D translation-invariant stabilizer codes, including BB codes, can be understood as equivalent (via finite-depth local quantum circuits) to multiple copies of the Kitaev toric code---the standard $\mathbb{Z}_2$ lattice gauge theory~\cite{bombin_Stabilizer_14, haah_module_13, haah2016algebraic, haah_classification_21, ruba2024homological}. This allows us to employ topological quantum field theory (TQFT) concepts—such as partition functions (capturing logical dimensions), anyon Wilson lines (representing logical operators), and, most importantly, anyon condensation of a Lagrangian subgroup at gapped boundaries (defining the boundary stabilizer Hamiltonian)~\cite{dijkgraaf1990topological, Wen1993Topologicalorder, kitaev2006anyons, bombin2006topological, Levin2006Detecting, Chen2011Complete,  levin2012braiding, chen2012symmetry, Jiang2012Identifying, Cincio2013Characterizing, gu2014effect, gu2014lattice, jian2014layer, bombin2015gauge, wang2015topological, Ye2015Vortex, yoshida2016topological, ye2016topological, Kapustin2017Higher, Chen2018Exactbosonization, Lan2018Classification, wang2018towards, cheng2018loop, Chan2018Borromean, Chen2019Free, Han2019GeneralizedWenZee, Wang2019Anomalous, Chen2019Bosonization, Lan2019Classification, Chen2020Exactbosonization, wang2020construction, Chen2021Disentangling, Barkeshli2022Classification, Johnson2022Classification, ellison2022pauli, Chen2023HigherCup, chen2023exactly, maissam2023codimension, Kobayashi2024CrossCap, Maissam2024Higher, Maissam2024Highergroup, kobayashi2024universal, hsin2024classifying}.
We make use of the TQFT perspective to devise a systematic approach for adapting BB codes defined on a torus to planar geometries with open boundaries. 
Our approach ensures that the resulting codes achieve the maximal logical dimensions predicted by TQFT, preserve the locality and weight of stabilizers, and exhibit code distances that scale with the length of the system.
Moreover, we introduce a "lattice grafting" technique that reduces the number of physical qubits by removing boundary edges while preserving all other code properties.

We explicitly demonstrate our approach by constructing families of planar qLDPC codes derived from BB codes.
Notably, all stabilizer checks in these planar codes remain local, with check weights not exceeding 6, and the ratio $kd^2/n$ outperforms the surface code by an order of magnitude. Moreover, the resulting codes offer more flexible lattice sizes than the BB codes designed on a torus. This represents the first construction, to our knowledge, of 2D local qLDPC codes exhibiting such a substantial advantage over the surface code~\cite{eberhardt2024pruning}. We remark that similar codes have been constructed simultaneously by Steffan et al.~\cite{steffan2025tile}.

Furthermore, our analysis reveals intriguing properties of the logical operators in these planar qLDPC codes. In finite-sized systems relevant for near-term implementations (e.g., $100\sim 500$ qubits), we observe that the minimal-weight logical operators exhibit fractal structures, reminiscent of Sierpinski triangles embedded within the finite lattice. This fractal regime, where the code distance is proportional to the fractal area, precedes the expected emergence of string-like logical operators in the thermodynamic limit~\cite{Dua2019Compactifyingfracton, Dua2020Bifurcating, Sullivan2021Fractonic, ruba2024homological, liang2023extracting, liang2024operator, liang2025generalized, chen2025anyon}.

In the future, the open boundaries inherent to our construction facilitate not only physical implementation but potentially enable advanced QEC techniques such as lattice surgery~\cite{Litinski2019gameofsurfacecodes, Lawrence2022surgery, cowtan2025parallel} and magic state distillation~\cite{Bravyi2005magic, wang2025magic}. Moreover,  investigating the transition between the fractal and string-like regimes could provide helpful intuition for understanding the code distances of these 2D qLDPC codes~\cite{aitchison2023no}.
Overall, our work provides a strategy for designing high-performance, hardware-compatible quantum error-correcting codes, paving the way towards more resource-efficient fault-tolerant quantum computation on realistic boundary-constrained architectures.

\subsection*{Summary of results}
\label{sec: Summary of results}

{\change
We present explicit families of high-performance planar quantum low-density parity-check (qLDPC) codes with open boundaries that outperform the surface code in resource efficiency (measured by $kd^2/n$) while preserving geometric locality suitable for 2D architectures. Our approach systematically adapts bivariate bicycle (BB) codes from the torus to planar geometries via boundary anyon condensation motivated by topological order, and further refines them through a ``lattice grafting'' optimization.

As a primary example, Sec.~\ref{sec: [[288, 8, 12]] planar code} constructs the $[[288,8,12]]$ code on a $12\times12$ open square lattice (Fig.~\ref{fig: [[288, 8, 12]] code}), and explicitly identifies its lowest-weight logical operators, which follow a Sierpinski-triangle pattern.

Sec.~\ref{sec: Topological order perspectives} outlines the construction, derives logical operators from the anyon viewpoint, compares our codes with generalized toric codes, describes the code search procedure, and introduces the ``lattice grafting'' optimization. For small fixed logical dimension $k$ and code distance $d$, Table~\ref{tab: GTC large weight and stabilizer} reports the minimal $n$ and the corresponding stabilizers realizing $[[n,k,d]]$ planar codes produced by our method (prior to grafting).

Sec.~\ref{sec: examples of planar qLDPC codes} analyzes planar code families with $6 \leq k \leq 9$, and Appendix~\ref{app: Families of planar qLDPC codes} covers families with $10 \leq k \leq 13$. Focusing on stabilizers with weight $\leq 6$, representative planar codes include:
\begin{itemize}[leftmargin=2em]
    \item $k=6$ (Fig.~\ref{fig: [[88, 6, 6]] code}): $[[54, 6, 4]]$, $[[88, 6, 6]]$,~$[[247, 6, 12]]$.
    \item $k=7$ (Fig.~\ref{fig: [[131, 7, 7]] code}): $[[131, 7, 7]]$,~$[[205, 7, 10]]$,~$[[337, 7, 14]]$.
    \item $k=8$ (Fig.~\ref{fig: [[188, 8, 9]] code}): $[[188, 8, 9]]$,~$[[326, 8, 13]]$,~$[[368, 8, 15]]$.
    \item $k=9$ (Figs.~\ref{fig: [[441, 9, 15]]} and~\ref{fig: [[432, 9, 14]]}): $[[225, 9, 9]]$,~$[[265, 9, 10]]$,
    $[[432, 9, 14]]$,~~$[[441, 9, 15]]$.
    \item $k=10$ (Figs.~\ref{fig: [[324, 10, 11]]} and~\ref{fig: [[381, 10, 13]]}): $[[226, 10, 9]]$,~$[[324, 10, 11]]$, 
    $[[381, 10, 13]]$, $[[490, 10, 15]]$.
    \item $k=11$ (Fig.~\ref{fig: [[494, 11, 15]]}): $[[248, 11, 9]]$,~$[[494, 11, 15]]$.
    \item $k=12$ (Figs.~\ref{fig: [[432, 12, 12]]} and~\ref{fig: [[279, 12, 9]]}): $[[279, 12, 9]]$,~$[[432, 12, 12]]$, 
    $[[486, 12, 13]]$, $[[588, 12, 15]]$.
    \item $k=13$ (Figs.~\ref{fig: [[392, 13, 11]]} and~\ref{fig: [[495, 13, 13]]}): $[[286, 13, 9]]$,~$[[392, 13, 11]]$, 
    $[[495, 13, 13]]$, $[[648, 13, 16]]$,
\end{itemize}
Sec.~\ref{sec: [[292, 12, 14]] planar code} additionally introduces a $k=12$ code family with weight-8 stabilizers (Fig.~\ref{fig: [[292, 12, 14]] code}), including $[[240,12,12]]$, $[[292,12,14]]$, and $[[399,12,18]]$.

Applying lattice grafting (Sec.~\ref{sec: Lattice grafting}) preserves locality and stabilizer weight while reducing $n$. For the examples above, the grafted parameters are summarized below (see Appendix~\ref{app: Grafted planar qLDPC codes} for detailed constructions):
\begin{eqs}
    [[54, 6, 4]] \longrightarrow ~ [[44,6,4]]:& \quad \frac{kd^2}{n} = 2.18, \\
    [[88, 6, 6]] \longrightarrow ~ [[78,6,6]]:& \quad \frac{kd^2}{n} = 2.77, \\
    [[131, 7, 7]] \longrightarrow ~ [[107,7,7]]:& \quad \frac{kd^2}{n} = 3.21, \\
    [[188, 8, 9]] \longrightarrow ~ [[173,8,9]]:& \quad \frac{kd^2}{n} = 3.75, \\
    [[288, 8, 12]] \longrightarrow ~ [[268,8,12]]:& \quad \frac{kd^2}{n} = 4.30, \\
    [[441, 9, 15]] \longrightarrow ~ [[405,9,15]]:& \quad \frac{kd^2}{n} = 5.00, \\
    [[381,10,13]] \longrightarrow ~ [[348,10,13]]:& \quad \frac{kd^2}{n} = 4.86, \\
    [[494,11,15]] \longrightarrow ~ [[450,11,15]]:& \quad \frac{kd^2}{n} = 5.50, \\
    [[432,12,12]] \longrightarrow ~ [[386,12,12]]:& \quad \frac{kd^2}{n} = 4.48, \\
    [[392,13,11]] \longrightarrow ~ [[362,13,11]]:& \quad \frac{kd^2}{n} = 4.35, \\
    [[292,12,14]] \longrightarrow ~ [[282,12,14]]:& \quad \frac{kd^2}{n} = 8.34.
\nonumber
\end{eqs}
We conclude with a discussion in Sec.~\ref{sec: discussion} and provide tables of code parameters in the appendices.
}

\begin{figure*}[thb]
    \centering
    \hspace{-0.98cm}
    \subfigure[~100 bulk $X$-stabilizers]{\raisebox{-0.07cm}{\includegraphics[scale=0.05]{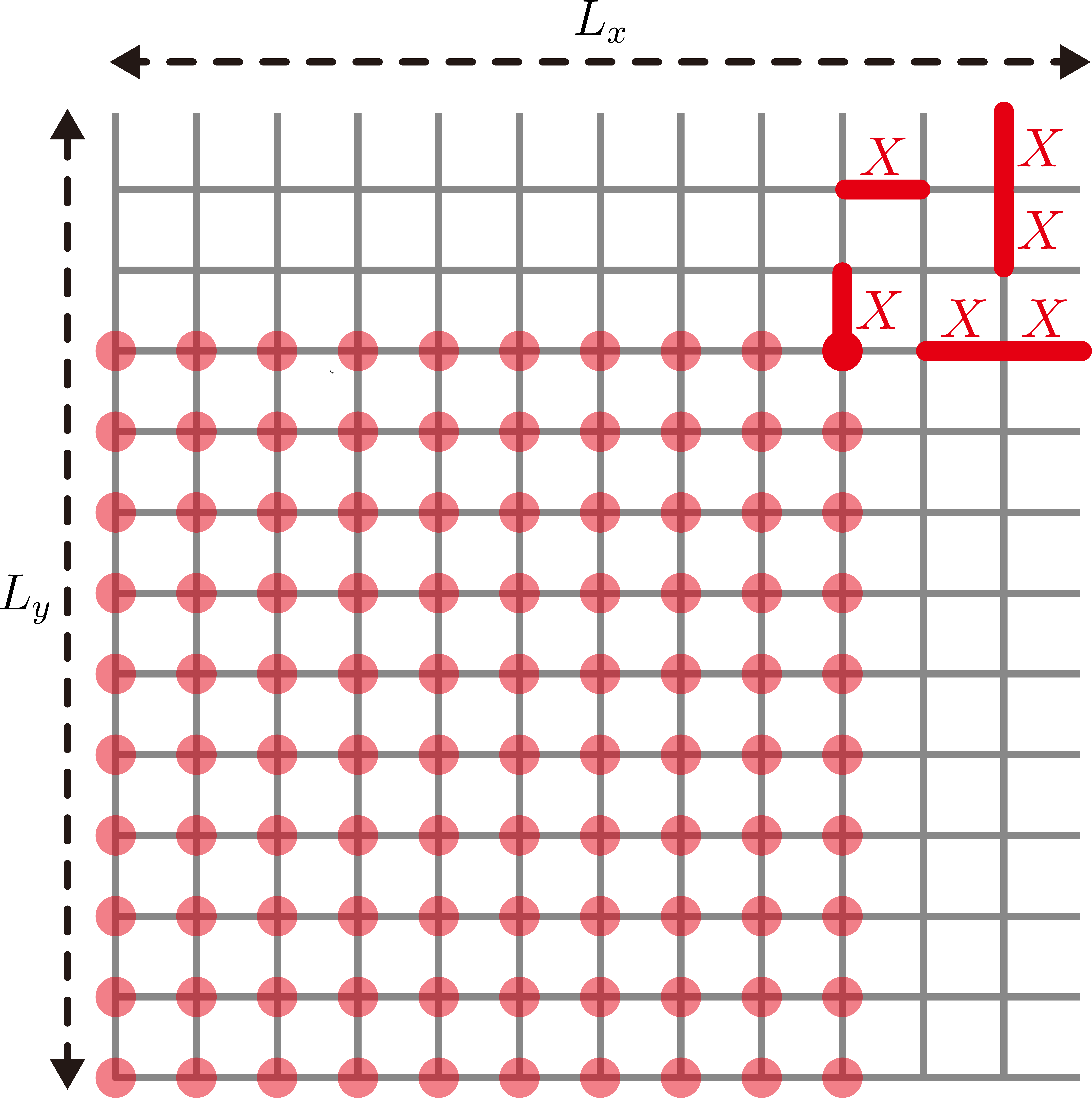}}}
    \hspace{1.25cm}
    \subfigure[~100 bulk $Z$-stabilizers]{\includegraphics[scale=0.05]{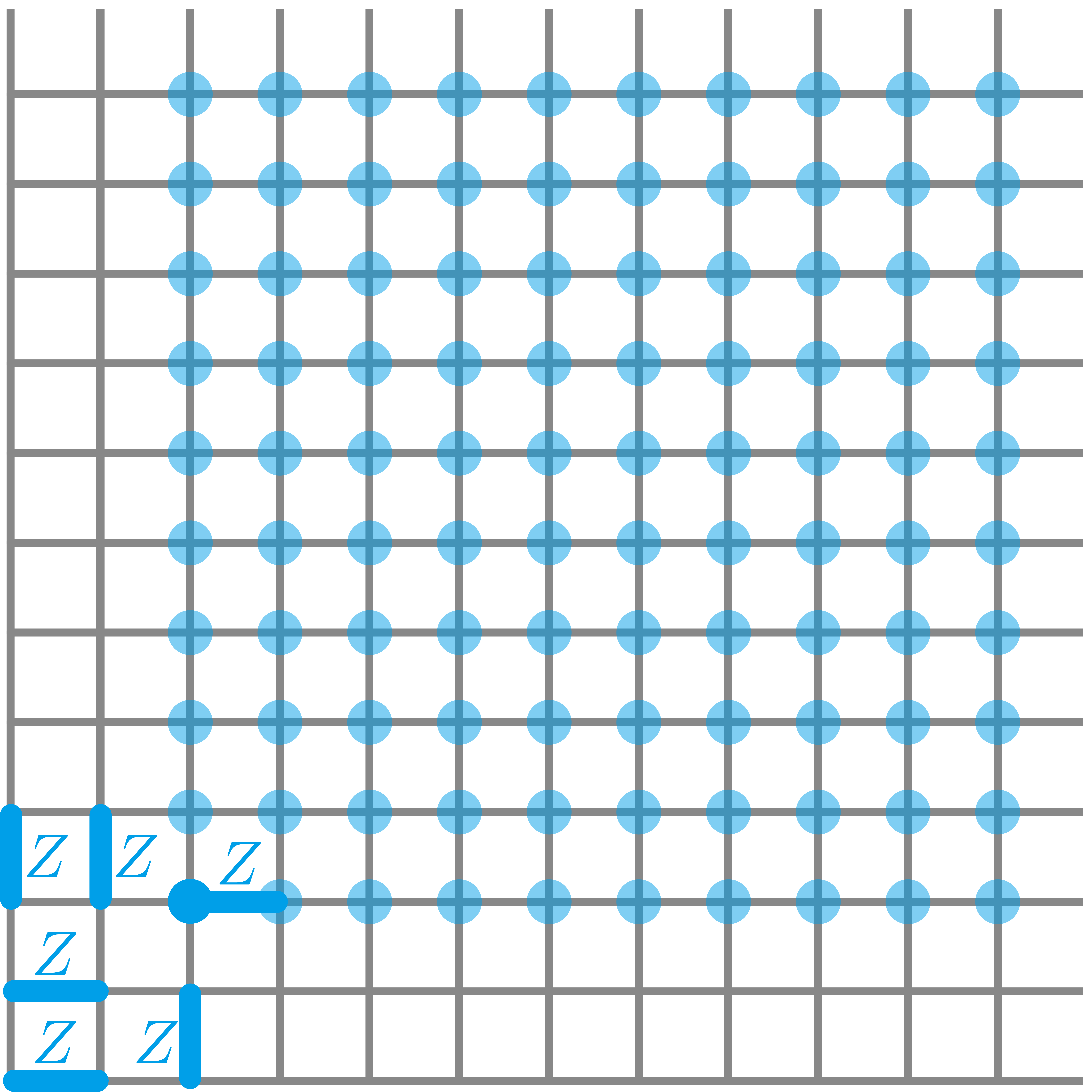}}\\
    \vspace{-0.1cm}
    \subfigure[~40 boundary $X$-stabilizers]{\includegraphics[scale=0.05]{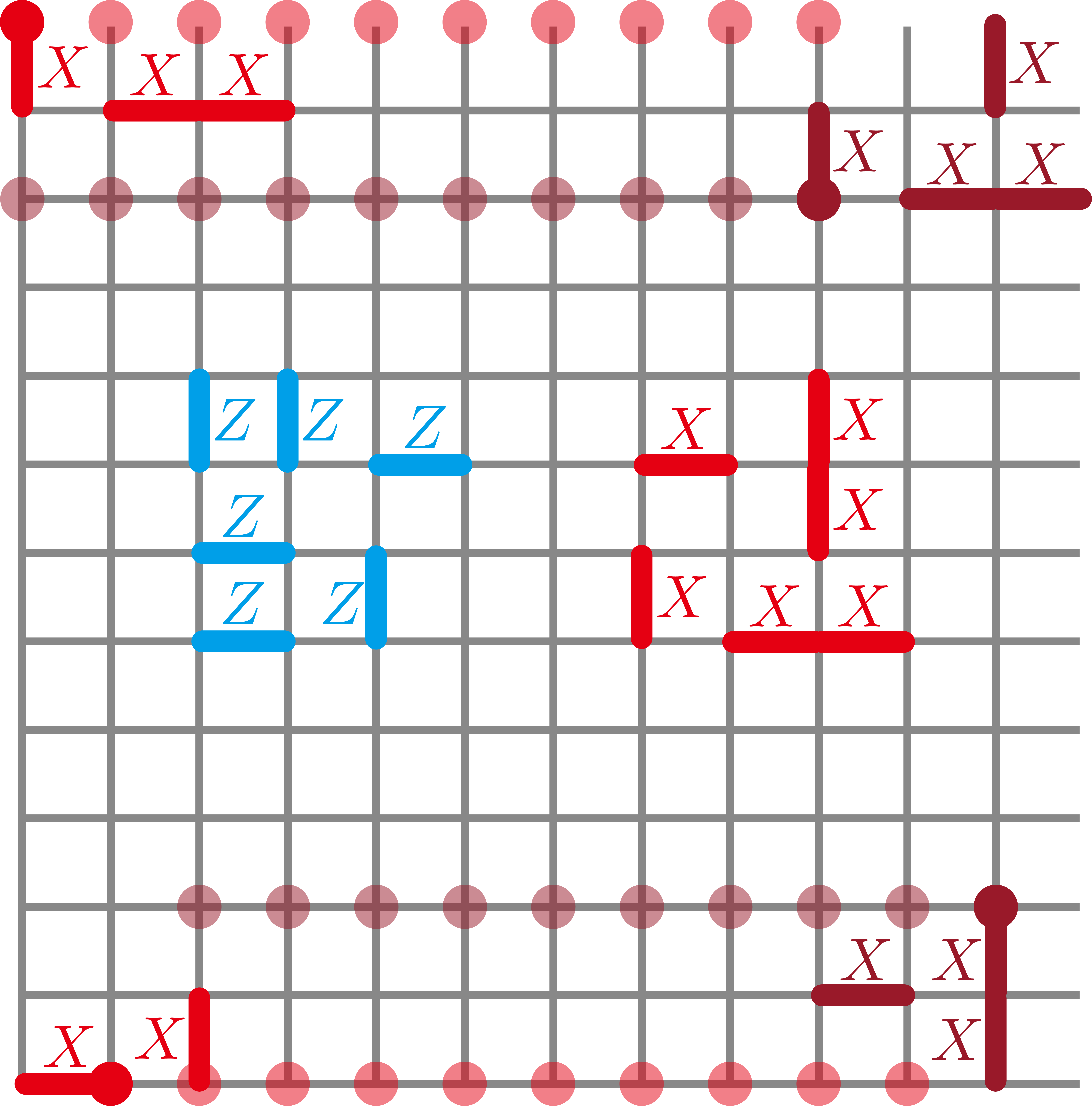}}
    \hspace{1.25cm}
    \subfigure[~40 boundary $Z$-stabilizers]{\includegraphics[scale=0.05]{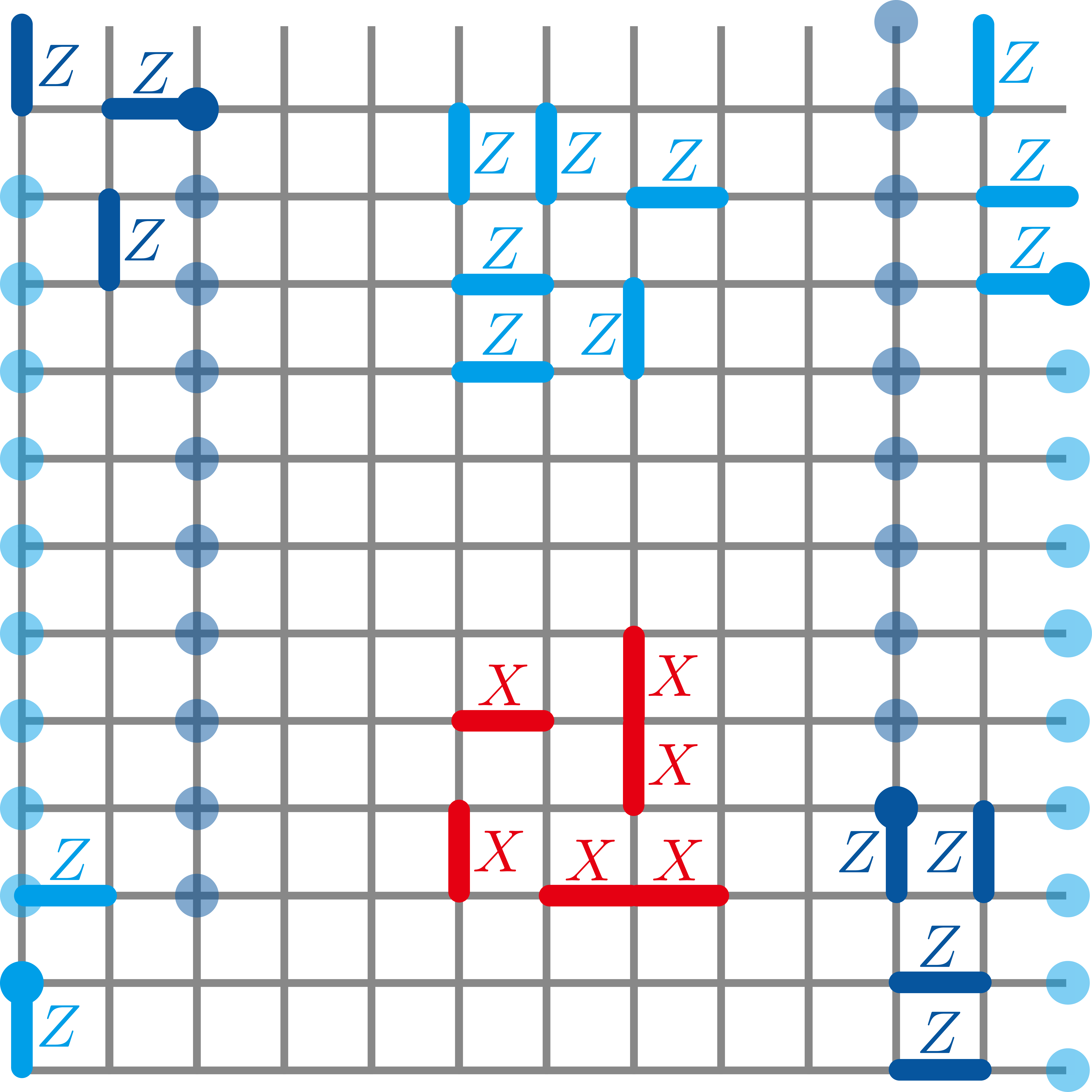}}
    \vspace{-0.2cm}
    \caption{{\change The $[[288,8,12]]$ code on an open square lattice. Bulk stabilizers are specified by the polynomials $f(x,y) = x(1+x+x^{-1}y^2)$ and $g(x,y) = x^2y(1+y+x^{-2}y^{-1})$ (see Appendix~\ref{Appendix: Review of the Laurent polynomial formalism} for a review of the polynomial formalism).}
    (a) Bulk $X$-stabilizers. Each red dot represents a weight-6 Pauli operator. With translational symmetry, there are 100 distinct bulk $X$-stabilizers. (b) 100 bulk $Z$-stabilizers arranged similarly. (c) Boundary $X$-stabilizers near the top and bottom boundaries. Two distinct types are present at each boundary, exhibiting translational symmetry (10 stabilizers each), resulting in 40 boundary $X$-stabilizers. (d) 40 boundary $Z$-stabilizers near the left and right boundaries. The independence of these bulk and boundary stabilizers yields a total logical dimension of $ k = n-\text{\# of stabilizers} =288-2\times100-2\times40=8$. The lattice sizes $L_x$ and $L_y$ are flexible, generating a code family with fixed logical dimension $k=8$: $[[2L_x L_y, ~8, ~d(L_x, L_y)]]$. The code distances $d(L_x, L_y)$ are listed in Table~\ref{tab: d(Lx, Ly) for [[288,8,12]]} in Appendix~\ref{app: Families of planar qLDPC codes}, and are computed exactly using the integer programming approach~\cite{landahl2011fault, Bravyi2024HighThreshold}.}
    \label{fig: [[288, 8, 12]] code}
\end{figure*}

\section{$[[288,8,12]]$ planar code}\label{sec: [[288, 8, 12]] planar code}

\begin{figure*}[thb]
    \centering
    \vspace{-0.25cm}
    \subfigure[~Lowest-weight logical $X$ operator]{\includegraphics[scale=0.05]{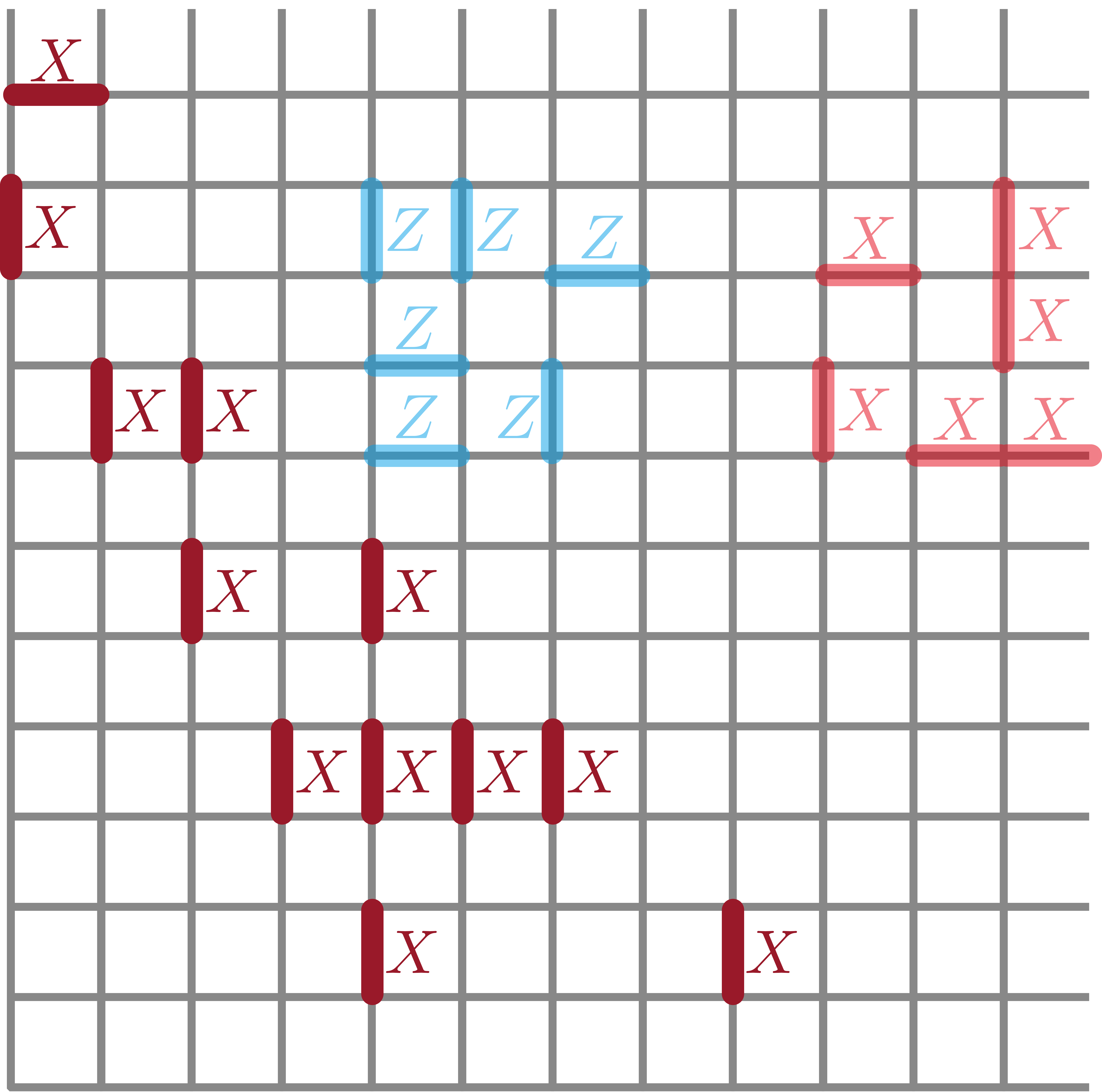}}
    \hspace{1cm}
    \subfigure[~Lowest-weight logical $Z$ operator]{\includegraphics[scale=0.05]{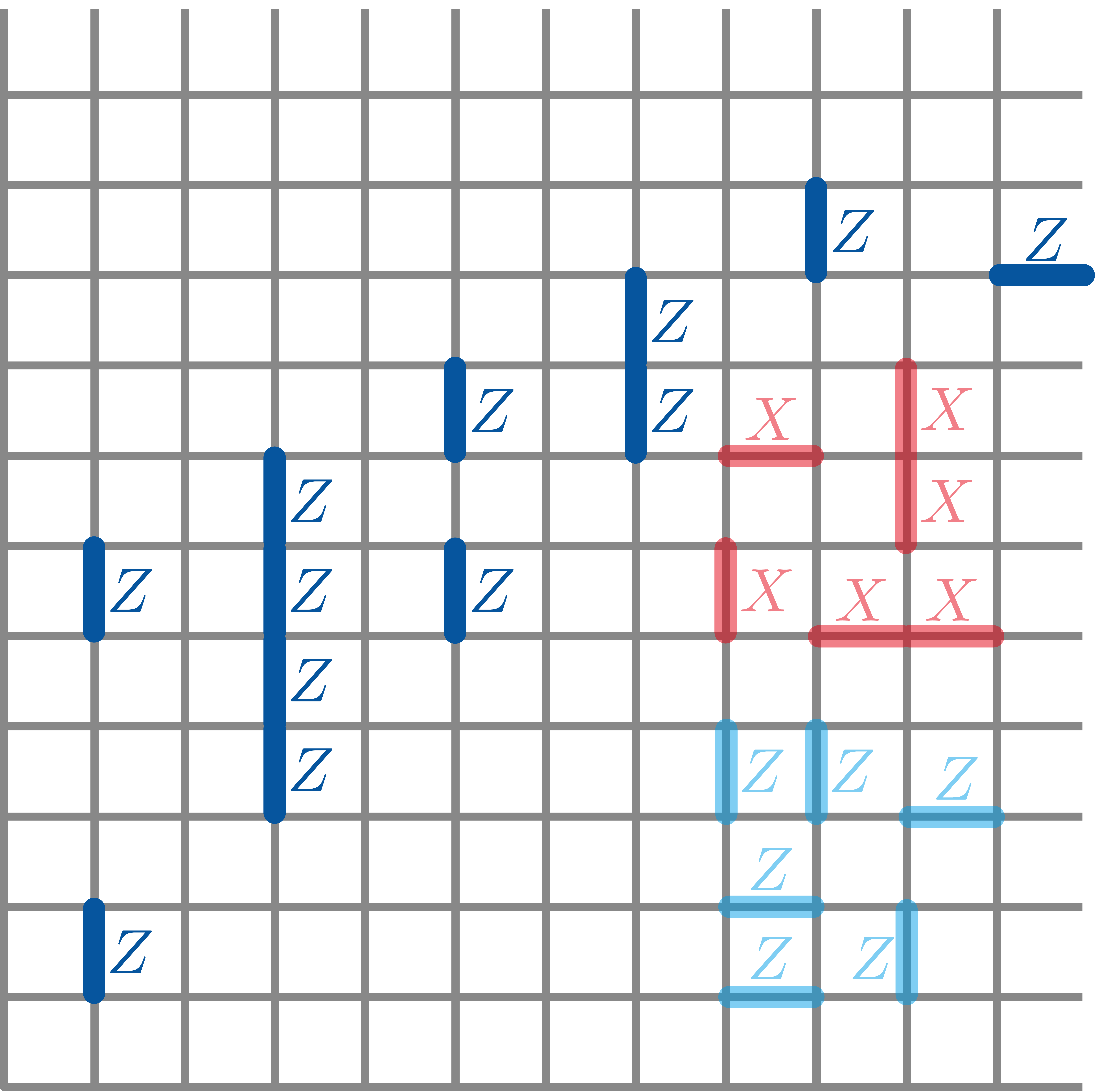}}\\
    \hspace{-0.25cm}
    \subfigure[~Sierpinski triangle induced by the stabilizers]{\includegraphics[width=0.6\textwidth]{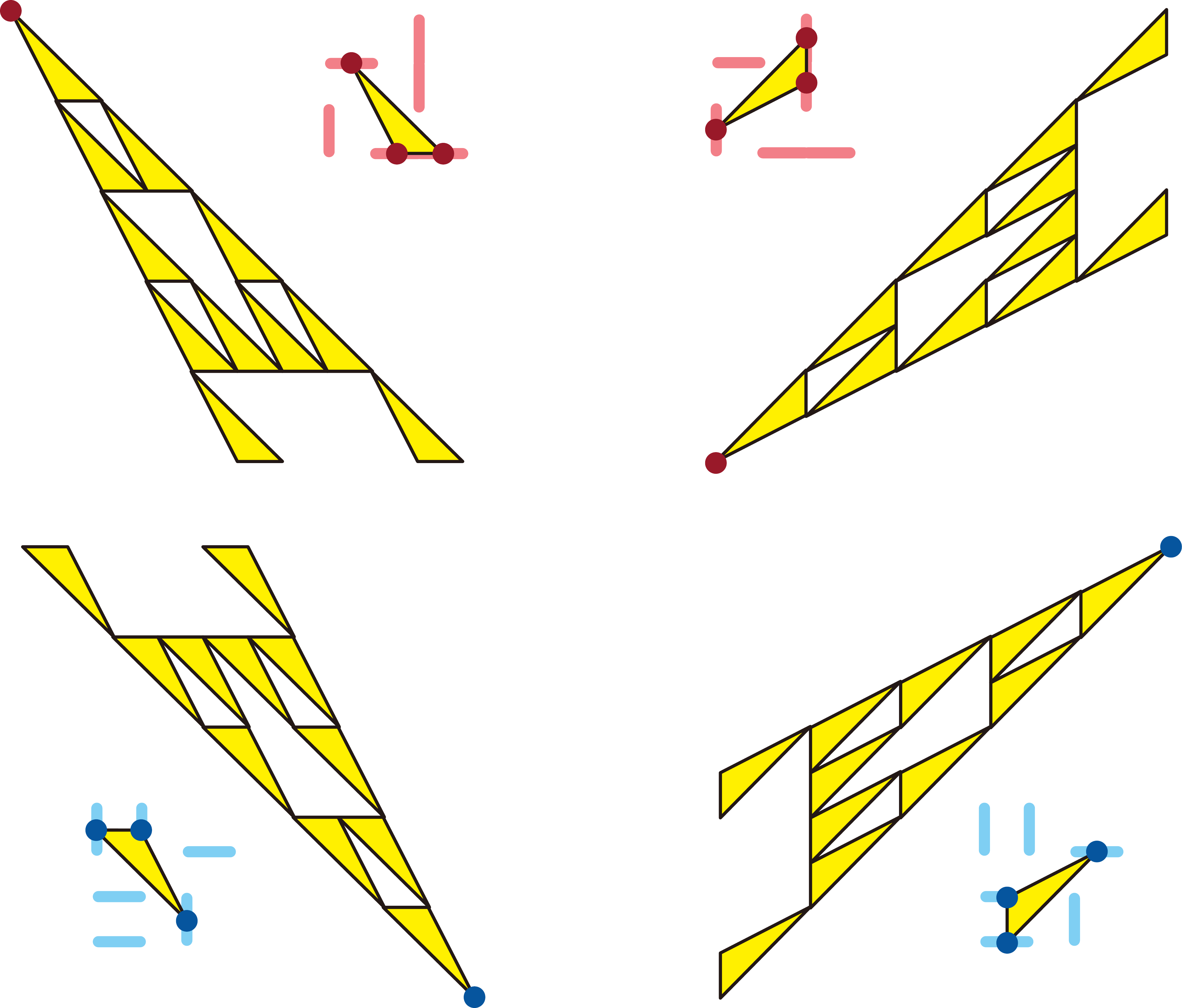}\label{fig: fractal}}
    \vspace{-0.3cm}
    \caption{Illustrations of the lowest-weight logical operators and the emergence of the fractal structure in the $[[288, 8, 12]]$ code. (a) The lowest-weight non-trivial logical $X$ operator (deep red). (b) The lowest-weight non-trivial logical $Z$ operator (deep blue). (c) The truncated Sierpinski triangle corresponding to the bulk stabilizers.}
    \label{fig: logical operator of [[288, 8, 12]]}
\end{figure*}

In this section, we begin with the construction of the $[[288,8,12]]$ code, illustrated in Fig.~\ref{fig: [[288, 8, 12]] code}.
We consider a truncated square lattice with a single qubit at each edge. Both bulk and boundary stabilizers are explicitly depicted. These stabilizers exhibit translational symmetry: bulk stabilizers are uniformly distributed across the two-dimensional plane, while boundary stabilizers can shift along directions parallel to the boundary, provided they remain fully supported within the finite lattice. Specifically, the top and bottom boundaries contain Pauli $X$ operators, and the left and right boundaries host Pauli $Z$ operators. The boundary stabilizers are carefully designed to ensure that they commute with each other around the four corners.
These stabilizers are independent; hence, the logical dimension $k$ is equal to the number of physical qubits $n$ minus the number of stabilizers: $k = 288 - 2 \times 100 - 2 \times 40 = 8$, as illustrated in Fig.~\ref{fig: [[288, 8, 12]] code}.

The $12 \times 12$ square lattice can be readily generalized to any $L_x \times L_y$ rectangle by inserting additional rows or columns, yielding a family of qLDPC codes with $n=2L_xL_y$ physical qubits and a fixed logical dimension of $k=8$.
Representative examples of this family on small lattices include:
\begin{enumerate}
    \item $L_x = L_y = 6$: $[[72, 8, 4]]$, ~${kd^2}/{n} = 1.78$.
    \item $L_x = L_y = 8$: $[[128, 8, 6]]$, ~${kd^2}/{n} = 2.25$.
    \item $L_x = L_y = 10$: $[[200, 8, 9]]$, ~${kd^2}/{n} = 3.24$.
    \item $L_x = L_y = 12$: $[[288, 8, 12]]$, ~${kd^2}/{n} = 4$.
    \item $L_x = L_y = 13$: $[[338, 8, 13]]$, ~${kd^2}/{n} = 4$.
    \item $L_x = L_y = 14$: $[[392, 8, 15]]$, ~${kd^2}/{n} = 4.59$.
\end{enumerate}
A comprehensive table detailing the code distances $d$ for each combination of $L_x$ and $L_y$ is provided in Appendix~\ref{app: Families of planar qLDPC codes}. {\change The exact code distances were computed using the integer programming approach~\cite{landahl2011fault}.}

\begin{figure}[tb]
\centering
    \includegraphics[width= \linewidth]{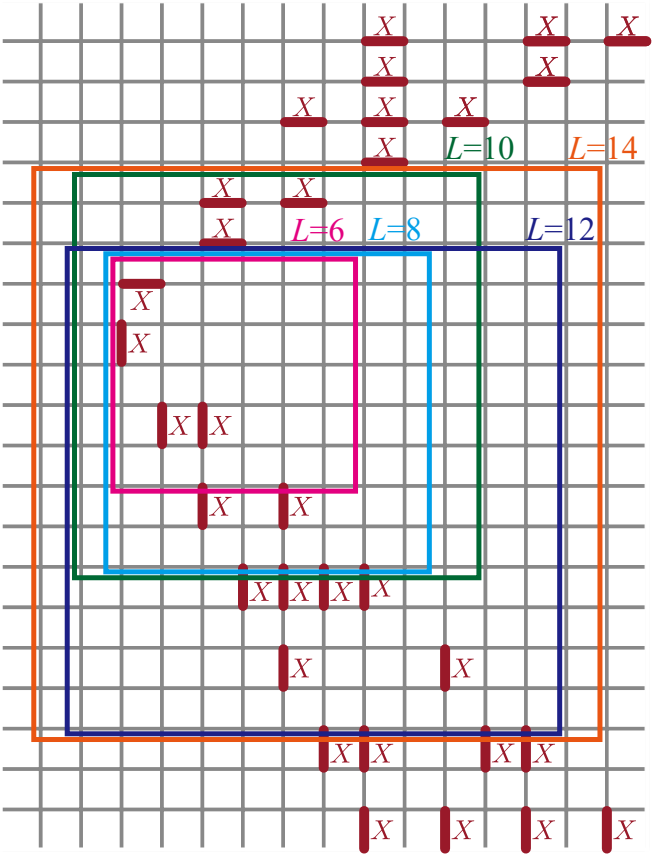}
     \caption{The red edges depict a fractal operator that transports an anyon across the infinite plane. On finite lattices, logical operators arise from truncating this infinite fractal operator. The minimal number of $X$-edges (with positions defined at their midpoints) enclosed within $L \times L$ squares for $L = 6, 8, 10, 12, 14$ are $4, 6, 9, 12, 15$, respectively, as shown in the figure (with the constraint that the fractal operator may exit the finite square only through its top and bottom boundaries).  These values exactly match the code distances $d(L_x, L_y)$ of the $[[288,8,12]]$ code family listed in Table~\ref{tab: d(Lx, Ly) for [[288,8,12]]}.}
     \label{fig: fractal_distance}
\end{figure}

For the $[[288, 8, 12]]$ code, Fig.~\ref{fig: logical operator of [[288, 8, 12]]} depicts the corresponding non-trivial logical operators with the lowest weight. Notably, rather than exhibiting the string {\change operator} anticipated from the anyon theory (unitary modular tensor category~\cite{rowell2009classification, wang2010topological, Wang2022in, Plavnik2023Modular}), the logical operators manifest as fractal patterns. More precisely, they follow the Sierpinski triangle that emerged from the stabilizers. Although for a large system size $L$, the weight of a string operator, $\Omega(L)$, is asymptotically more favorable than that of a Sierpinski triangle operator with the weight $\Omega(L^{1.59})$\footnote{The Hausdorff dimension of the Sierpinski triangle is $d=\frac{\log 3}{\log 2}$, obtained by solving $2^d = 3$.}, in our finite system the Sierpinski triangle operators have lower weight compared to the string operator.

We verify that the numerically obtained code distances match the number of small triangles of the Sierpinski fractal embedded in the open square lattice.
{\change More precisely, each single Pauli operator violates three stabilizers, corresponding to the vertices of a small yellow triangle in Fig.~\ref{fig: fractal}. A natural way to cancel these violations is to construct a semi-infinite Sierpinski triangle, which leaves only a single violation at its apex. By combining two such Sierpinski triangles, the remaining violations at their apexes cancel, as illustrated in Fig.~\ref{fig: fractal_distance}.}
When the combined operator is embedded into the finite lattice such that it intersects the boundary only along the top and bottom edges, the resulting truncated operator commutes with all stabilizers and thus forms a logical operator.
Because this operator transports an anyon (i.e., the violation), it is nontrivial, and its weight provides an upper bound for the code distance. Remarkably, for the small lattices listed in Table~\ref{tab: d(Lx, Ly) for [[288,8,12]]}, this bound is tight; the code distance exactly equals the number of small triangles from the combined Sierpinski triangles contained within the finite lattice.

As shown in Fig.~\ref{fig: fractal_distance}, the minimal number of small triangles enclosed by an $L\times L$ square for $L=6, 8, 10, 12, 14$ is $4, 6, 9, 12,$ and $15$, respectively. These values precisely match the code distances of the representative examples discussed earlier. This correspondence demonstrates that these planar codes, constructed on small open square lattices, remain in the “fractal regime” and have not yet reached the thermodynamic limit characterized by string operators. {\change Such Sierpinski-type fractal logical operators commonly arise whenever the stabilizers are weight-$6$, with three Pauli operators acting on horizontal edges and three on vertical edges.}
For the bulk stabilizers of the $[[288, 8, 12]]$ code, string operators are stable at $L_x = L_y = 217$ (the $(-2,2)$-BB code in Ref.~\cite{liang2024operator}).\footnote{ {\change For $L>16$, the minimal-weight operator begins to transition from fractal-like to string-like (though the resulting string remains relatively wide and irregular). As $L$ increases, the Pauli weight of fractal operators grows superlinearly (scaling approximately as $L^{1.59}$), making such operators quickly unfavorable.}}

{\change Finally, we emphasize that these codes can be further optimized using ``lattice grafting,'' a technique that removes qubits near the boundary, which will be discussed in Sec.~\ref{sec: Lattice grafting}.} Our numerical experiments indicate that this approach can reduce the total number of qubits $n$ by $O(L_x + L_y)$. For example, by applying lattice grafting to the $[[288,8,12]]$ code to remove boundary qubits, we obtain the $[[268,8,12]]$ code, as illustrated in Fig.~\ref{fig: [[268, 8, 12]] code} in Appendix~\ref{app: Grafted planar qLDPC codes}.

\section{Planar codes from topological order}\label{sec: Topological order perspectives}

In this section, we introduce our general approach to constructing planar qLDPC codes applicable to arbitrary bulk stabilizers from a topological order viewpoint.
We begin in Sec.~\ref{sec: Boundary anyons} by reviewing the concepts of boundary gauge operators and boundary anyons, which allow us to achieve anyon condensation on the boundary and ensure the topological order (TO) condition is satisfied.
In Sec.~\ref{sec: Planar code construction}, we detail the layout of interleaving $\{m_i\}$-condensed and $\{e_i\}$-condensed boundaries, and examine the geometric structure of logical operators associated with these boundaries.
Sec.~\ref{sec: Comparison between open and periodic} provides a comparative analysis between open and periodic boundary conditions, and discusses their implications for designing high-performance qLDPC codes.
Lastly, in Sec.~\ref{sec: Lattice grafting}, we introduce lattice grafting, which enables us to further reduce the number of physical qubits in the examples presented in the previous section without altering any other parameters.

\subsection{Boundary gauge operators and anyons}
\label{sec: Boundary anyons}

This section reviews the essential terminology from Ref.~\cite{liang2024operator} that underpins our construction of planar qLDPC codes.
{\change We begin by defining the notion of the topological order condition:
\begin{definition}
    The \textbf{topological order (TO) condition}\footnote{\change This condition arises from the TQO-1 condition as defined in Refs.~\cite{bravyi2010topological, bravyi2011short} for a broader range of commuting projector Hamiltonians.} for a stabilizer code requires that any local operator $\mathcal{O}$ that commutes with all stabilizers must itself be a local product of stabilizers. That is, if $[\mathcal{O}, S_i]=0$ for all stabilizers $S_i$, then there exists a finite subset $A$ of stabilizers such that
    \begin{equation}
        \mathcal{O} = \prod_{i \in A} S_i~.
    \end{equation}
\end{definition}
Starting with a qLDPC code on a torus, one can truncate it to an open lattice by retaining only those bulk stabilizers that are fully supported within the finite lattice. Near the boundaries, however, the TO condition is violated, meaning that there exist local operators commuting with all bulk stabilizers that are not products of these stabilizers.}
These operators are referred to as \textbf{boundary gauge operators}.
More precisely, the set of local operators that commute with the bulk stabilizers, modulo the bulk stabilizers, forms the group $\mathcal{G}$ of boundary gauge operators.
Note that these operators may not necessarily commute with each other.\footnote{This terminology is inspired by subsystem codes, where the gauge operators do not commute, and their commutants form the stabilizer group.}
{\change Therefore, to ensure that the boundary is gapped, we need to incorporate a \emph{subset} of boundary gauge operators into the Hamiltonian such that they mutually commute and restore the TO condition near the boundary. In the language of subsystem codes, this step is called ``gauge fixing.'' \cite{Bombín2015Gauge, Poulin2023Stabilizer, Bacon2006Operator} What we seek is an effective way to perform this gauge fixing so that the resulting stabilizer codes have favorable code parameters.}

In this framework, bulk and boundary anyons can be defined as follows~\cite{Ellison2023paulitopological, ruba2024homological, liang2023extracting, liang2024operator, liang2025generalized}:
\begin{definition}
    A \textbf{bulk anyon} is defined as a finite violation of stabilizers on the lattice. It is a homomorphism $\phi$ from the stabilizer group $ \mathcal{S} $ to $\ZZ_2$:
    \begin{eqs}
        \phi: \mathcal{S} \rightarrow \ZZ_2= \{1,-1\},
    \label{eq: bulk homomorphism}
    \end{eqs}
    such that $\phi(S_i) = 1$ for all but a finite number of $S_i \in \mathcal{S}$.
\end{definition}
\begin{definition}
A \textbf{boundary anyon} is defined as the local ``syndrome pattern'' of boundary gauge operators that indicates how boundary gauge operators are violated:
\begin{eqs}
    \varphi: \mathcal{G} \rightarrow \ZZ_2= \{1,-1\},
\end{eqs}
such that the homomorphism $\varphi(G_i) = 1$ for all but a finite number of $G_i \in \mathcal{G}$.
\end{definition}
By quotienting out trivial anyons—those created by local operators—the resulting equivalence classes form a finite Abelian fusion group.
{\change Nontrivial anyons can be realized at the endpoints of string operators, corresponding to twisted sectors under the 1-form symmetries generated by closed anyon worldlines.}
In particular, boundary anyons arise at the endpoints of a \textbf{boundary string operator}—a product of boundary gauge operators along a boundary segment.
There exists a one-to-one correspondence between bulk and boundary anyons, allowing boundary anyons to move into the bulk and vice versa~\cite{liang2024operator}.

Each gapped boundary in an Abelian topological order (i.e., anyon theory $\mathcal{A}$) is associated with a \textbf{Lagrangian subgroup} $\mathcal{L} \subset \mathcal{A}$, defined as the maximal set of bosonic anyons that braid trivially with one another \cite{etingof2010fusion, Kapustin2011Topological, Kong2014Anyon, kaidi2022higher}.
The condensation of the bosons in $\mathcal{L}$ on the boundary allows their string operators to terminate at the boundary without any energy cost.
Constructing the boundary stabilizer Hamiltonian involves two steps:
\begin{enumerate}
    \item \textbf{Boundary anyon condensation:} Incorporate mutually commuting short boundary string operators corresponding to the bosons in the Lagrangian subgroup into the Hamiltonian.
    \item \textbf{Topological order (TO) completion:} Introduce extra local boundary gauge operators into the Hamiltonian to ensure that the topological order condition near the boundary is fully satisfied.
\end{enumerate}
Although the second step is overlooked in the literature, it is critical for constructing a lattice Hamiltonian with gapped boundaries, particularly because not all boundary gauge operators arise from truncating bulk stabilizers. For a detailed discussion of these two steps and a comprehensive algorithm outlining the procedure, see Ref.~\cite{liang2024operator}.

\subsection{Planar code construction and logical operators}
\label{sec: Planar code construction}

\begin{figure}
\centering
    \includegraphics[width=.48\textwidth]{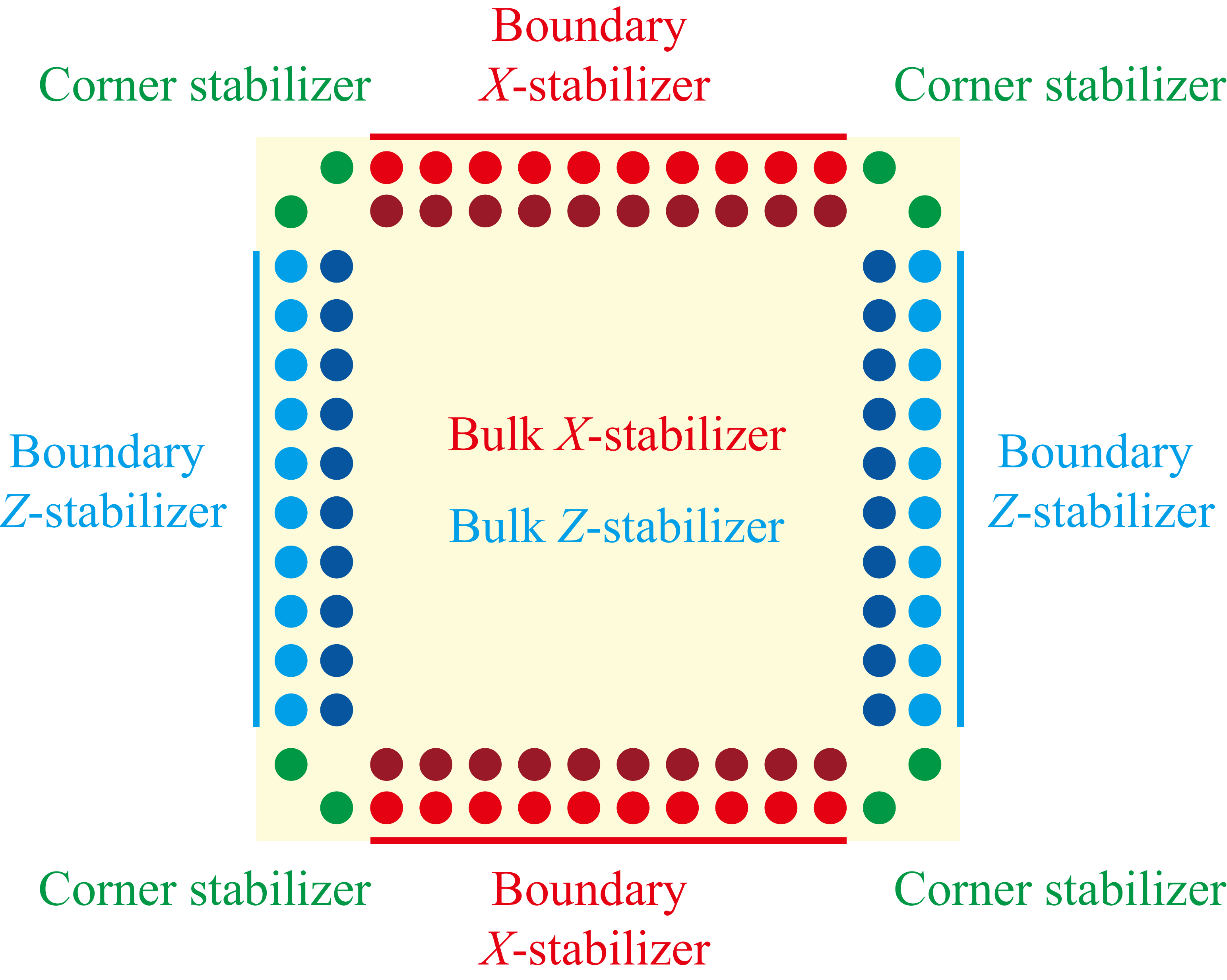}
     \caption{Planar CSS code construction. The $\{e_i\}$-anyons are condensed along the left and right boundaries, while the $\{m_i\}$-anyons are condensed along the top and bottom boundaries. In addition to incorporating the short boundary string operators of these bosons into the stabilizer group, extra local boundary gauge operators are introduced near the boundaries and corners to remove any local logical operators. This crucial ``topological order completion'' step ensures the code distance scales with the system size.
     The $\{m_i\}$-anyon strings can terminate on the top and bottom boundaries without any energy cost, and similarly, the $\{e_i\}$-anyon bulk string operators can terminate on the left and right boundaries. Vertical $\{m_i\}$-strings and horizontal $\{e_i\}$-strings serve as the logical $X$ and $Z$ operators, respectively, so that the logical dimension $k$ equals the number of $\{e_i\}$-anyons.}
     \label{fig: Planar code construction}
\end{figure}

Any two-dimensional translation-invariant stabilizer code satisfying the topological order conditio is equivalent to a direct sum of Kitaev toric codes via a finite-depth Clifford circuit~\cite{bombin_Stabilizer_14, haah_module_13, haah2016algebraic, haah_classification_21, Chen2023Equivalence, ruba2024homological}. Consequently, the corresponding topological order (anyon theory) is characterized by the fusion group $A = \mathbb{Z}_2^q \times \mathbb{Z}_2^q$, generated by the anyons $\{e_i\}$ and $\{m_i\}$ for $1 \leq i \leq q$. These anyons are bosons with the braiding statistics
\begin{equation}
    B(e_i, m_i) = -1, \quad \forall 1 \leq i \leq q,
\end{equation}
while $B(a, b) = 1$ for all other pairs.

Although the general procedure applies to arbitrary stabilizer codes, we focus on CSS codes for simplicity. 
The planar code layout in Fig.~\ref{fig: Planar code construction} depicts an open square with two distinct Lagrangian subgroups arranged on opposite edges: $\mathcal{L}_1 = \{e_i\}$ (violations of $X$-stabilziers) and $\mathcal{L}_2 = \{m_i\}$ (violations of $Z$-stabilziers).
More intricate geometries can be considered, such as puncturing additional holes into the open square or considering polygons with interleaving Lagrangian subgroups (e.g., $\mathcal{L}_1$, $\mathcal{L}_2$, $\mathcal{L}_3$, etc.)~\cite{delfosse2016generalized}. These approaches are standard techniques for increasing the logical dimension in surface codes, but the code performance ratio $kd^2/n$ usually decreases. Therefore, we focus on the simplest layout in Fig.~\ref{fig: Planar code construction}.

According to the procedure outlined in the previous section, one should first identify the boundary anyons $\{e_i\}$ or $\{m_i\}$, incorporate their corresponding short boundary operators as stabilizers, and then perform topological order completion to eliminate any local logical operators.
However, for an arbitrary CSS code as depicted in Fig.~\ref{fig: Planar code construction}, the boundary construction algorithm can be simplified. Specifically, along the left and right boundaries, all $\{e_i\}$ anyons are generated exclusively by $Z$-type operators, allowing the direct imposition of all $Z$-type boundary gauge operators as stabilizers. Similarly, along the top and bottom boundaries, the $\{m_i\}$ anyons are created solely by $X$-type operators, enabling the direct inclusion of all $X$-type boundary gauge operators as stabilizers. Consequently, topological order completion is required only in the vicinity of the four corners to resolve the anti-commutation between the $X$ and $Z$ boundary gauge operators.

In this setting, the logical $X$-operators correspond to $\{m_i\}$-string operators, with one endpoint on the top boundary and the other on the bottom boundary. In practice, we first solve the anyon equation~\cite{liang2023extracting} to obtain the vertical $X$-type anyon string operators (which transport $\{m_i\}$-anyons) on an infinite plane. These operators are then embedded into the finite lattice by truncating the portions that lie outside it. This embedding yields logical operators because the truncation does not affect the bulk stabilizers, ensuring that the truncated string operators continue to commute with all bulk stabilizers on the finite lattice.
Moreover, the top and bottom boundaries consist solely of boundary $X$-stabilizers, which naturally commute with the truncated $X$-string operators. A similar argument applies to the horizontal $\{e_i\}$-string operators truncated by the left and right boundaries; therefore, logical dimension $k$ equals the number of independent $\{e_i\}$-anyons.
Algebraically, the number of $\{e_i\}$-anyons can be computed as~\cite{liang2025generalized}:
\begin{equation}
    k_{\mathrm{open}} = \dim\left(
    \frac{\mathbb{Z}_2[x,y,x^{-1},y^{-1}]}{\langle f(x,y),~g(x,y) \rangle}
    \right),
\label{eq: logical dimension k for open boundary}
\end{equation}
where $f(x,y)$ and $g(x,y)$ are the defining polynomials of the bulk stabilizers, labeling the positions of the horizontal and vertical edges of each stabilizer~\cite{liang2023extracting}. Intuitively, a single Pauli operator can violate the stabilizers at the locations specified by $f(x,y)$ or $g(x,y)$, creating a trivial anyon. Consequently, an equivalence class of anyons corresponds to the violation pattern Eq.~\eqref{eq: bulk homomorphism}, represented as a polynomial in $\mathbb{Z}_2[x,y,x^{-1},y^{-1}]$, modulo the trivial anyon configurations defined by $f(x,y)$ and $g(x,y)$.
Notably, the logical dimension $k$ is independent of the system size as long as the top and bottom boundaries condense $\{m_i\}$ anyons and the left and right boundaries condense $\{e_i\}$ anyons.

Finally, we emphasize that the term ``string operators'' above refers to operators that move anyons. It has been shown that, in two-dimensional systems, every particle excitation can be moved by a string operator~\cite{ruba2024homological}.
Thus, our construction of planar qLDPC codes can be scaled to arbitrarily large systems, with the code distance growing proportionally to the system size in the thermodynamic limit. However, in small lattices, the ``string operators'' may not exhibit a conventional string-like form.
For instance, Fig.~\ref{fig: fractal} illustrates a fractal pattern for moving an anyon, demonstrating the emergence of fractal operators.

\subsection{Comparison between open and periodic boundary conditions}\label{sec: Comparison between open and periodic}

The same stabilizer code can exhibit markedly different behavior depending on whether open or periodic boundary conditions are imposed. The code parameters $[[n,k,d]]$ can vary significantly:
\begin{enumerate}
    \item {\bf Logical dimension $k$}:
    For open boundary conditions, all anyon string operators are explicitly realized on a finite lattice, regardless of the lattice's shape or size (provided the lattice is larger than the support of the stabilizers). This is characterized by Eq.~\eqref{eq: logical dimension k for open boundary}.  In contrast, under periodic boundary conditions, the logical dimension is given by~\cite{haah_module_13, liang2025generalized}:
    \begin{eqs}
        \qquad& k_{\mathrm{torus}}=\\
        \qquad& 2 \dim\left(
        \frac{\mathbb{Z}_2[x,y,x^{-1},y^{-1}]}{\langle f(x,y),~g(x,y), ~x^{L_x}-1, ~y^{L_y}-1 \rangle}
        \right),
    \label{eq: logical dimension k for torus}
    \end{eqs}
    where $f(x,y)$ and $g(x,y)$ are the polynomials of the bulk stabilizers implemented on an $L_x \times L_y$ torus.
    Two key distinctions emerge:
    \begin{itemize}
        \item The factor of $2$ reflects the fact that on a torus, vertical logical operators can be realized as both $e$-string and $m$-string operators, whereas open boundaries permit only one type of logical operator.
        \item On a torus, sites are identified modulo $(L_x, 0)$ and $(0, L_y)$, requiring the quotient of the Laurent polynomial ring by $x^{L_x}-1$ and $y^{L_y}-1$. Consequently, achieving the maximal logical dimension requires a torus of an appropriate size.
    \end{itemize}
    For example, our planar $[[288, 8, 12]]$ code shares the same bulk stabilizers as the $(-2, 2)$-BB code in Ref.~\cite{liang2024operator}. On a torus, achieving the maximal $k=16$ would require a lattice of size $L_x = L_y = 217$.\footnote{If the lattice dimensions are not multiples of 217, the maximal logical dimension is not achieved. For example, on a $7\times7$ torus the $(-2, 2)$-BB code yields a $[[98,6,12]]$ code~\cite{liang2025generalized}.}
    Thus, while toric codes can support a larger logical dimension, they impose stringent requirements on lattice size.
    In contrast, open lattices, though limited to half the maximal logical dimension, do not require compatibility between the bulk stabilizers and the system size, offering a more practical alternative. This trade-off illustrates that, despite their reduced logical dimension, planar qLDPC codes on open lattices offer greater flexibility in lattice size.

    \item {\bf Code distance $d$}:
    On an open lattice, anyons can be pumped from one side to the other, and the corresponding operator embedded in the finite lattice serves as a logical operator of the planar qLDPC code. As shown in Fig.~\ref{fig: logical operator of [[288, 8, 12]]}, the lowest-weight logical operators in small lattices take the form of a Sierpinski triangle, with the code distance corresponding to the number of such triangles embedded in the lattice. If the range of the stabilizers is reduced (resulting in smaller triangles), more triangles can be accommodated within the finite lattice, leading to larger code distances for a fixed lattice size.

    In contrast, the code distances of toric codes are determined by the string operators that transport anyons vertically and horizontally~\cite{chen2025anyon}. Consequently, the sparsity or density of Pauli operators in these string operators is the key factor in determining the code distance. This criterion differs from planar qLDPC codes with open boundaries, implying that the optimal families for planar and toric codes are distinct. For example, the well-known toric code, the $[[144,12,12]]$ code from Ref.~\cite{Bravyi2024HighThreshold}, does not yield a favorable family of planar codes.

    \item {\bf Physical qubits $n$}:
    For open boundaries, it is necessary to complete the stabilizer Hamiltonian at the corners to ensure the topological order condition. 
    Multiple options exist for these corner terms, some of which may involve only a single Pauli operator. 
    Although selecting such single-qubit corner operators can reduce the number of physical qubits, it may also decrease the code distance. To balance these trade-offs, we prefer higher-weight corner terms to preserve the code distance. 
    Next, we apply lattice grafting, see Sec.~\ref{sec: Lattice grafting}, to further reduce the number of physical qubits while maintaining the code distance. 
    This approach is feasible because most boundary edges participate in only a few stabilizers, so their removal and the subsequent rearrangement of stabilizers do not degrade the code distance or the locality of the stabilizers. Consequently, the number of physical qubits, $n$, is flexible and can often be significantly reduced. 
    For example, our $[[288,8,12]]$ planar code can be reduced to a $[[268,8,12]]$ code without affecting the locality or the maximal weight of the stabilizers.
    Under periodic boundary conditions, however, the number of physical qubits is fixed by the lattice dimensions as $n=2L_xL_y$, leaving no room for further optimization.
\end{enumerate}

\begin{table*}[h]
\setlength{\tabcolsep}{0pt} 
\centering
\definecolor{mycolor1}{RGB}{231,  76,  60}  
\definecolor{mycolor2}{RGB}{241, 196,  15}   
\definecolor{mycolor3}{RGB}{ 46, 204, 113}   
\definecolor{mycolor4}{RGB}{ 52, 152, 219}   
\definecolor{mycolor5}{RGB}{50, 250, 250}   
\definecolor{mycolor6}{RGB}{230, 126,  34}   
\definecolor{mycolor7}{RGB}{ 52,  73,  94}  
\definecolor{mycolor8}{RGB}{174, 217, 69}  
\definecolor{mycolor9}{RGB}{241, 111, 130}  
\definecolor{mycolor10}{RGB}{ 93, 173, 226}  
\definecolor{mycolor11}{RGB}{190, 144, 212}  
\definecolor{mycolor12}{RGB}{247, 220, 111}  
\definecolor{mycolor13}{RGB}{112, 161, 115}  
\definecolor{mycolor14}{RGB}{165, 105,  79}  
\renewcommand{\arraystretch}{1.2}
\begin{tabular}{|c|c|c|c|c|c|c|c|c|c|c|c|c|c|}
\hline
\diagbox[height=1cm, innerwidth=1cm]{\large ~$k$}{\raisebox{-0.5 em}{\large $d$~}}
   & 4 & 5 & 6 & 7 & 8 & 9 & 10 & 11 & 12 & 13 & 14 & 15 & 16 \\
\hline
6
&$\bigsubstack{54\\ x^{-1}y^{-2}\\ xy^{-1}}$\cellcolor{cyan!16}
&$\bigsubstack{77\\ x^{-1}y^{-2}\\ xy^{-1}}$\cellcolor{cyan!16}
&$\bigsubstack{88\\ ~x^{-1}y^{-2}~{} \\ xy^{-1}}$\cellcolor{cyan!16}
&$\bigsubstack{117\\ ~x^{-1}y^{-2}~{} \\ xy^{-1}}$\cellcolor{cyan!16}
& $\bigsubstack{150\\ ~x^{-1}y^{-2}~{} \\ xy^{-1}}$\cellcolor{cyan!16}
& $\bigsubstack{165\\ ~x^{-1}y^{-2}~{} \\ xy^{-1}}$\cellcolor{cyan!16}
& $\bigsubstack{204\\ ~x^{-1}y^{-2}~{} \\ xy^{-1}}$\cellcolor{cyan!16}
& $\bigsubstack{228\\ ~x^{-1}y^{-2}~{} \\ xy^{-1}}$\cellcolor{cyan!16}
& $\bigsubstack{247\\ ~x^{-1}y^{-2}~{} \\ xy^{-1}}$\cellcolor{cyan!16}
& $\bigsubstack{294\\ ~x^{-1}y^{-2}~{} \\ xy^{-1}}$\cellcolor{cyan!16}
& $\bigsubstack{345\\ ~x^{-1}y^{-2}~{} \\ xy^{-1}}$\cellcolor{cyan!16}
& $\bigsubstack{400\\ ~x^{-1}y^{-2}~{} \\ xy^{-1}}$\cellcolor{cyan!16}
& $\bigsubstack{425\\ ~x^{-1}y^{-2}~{} \\ xy^{-1}}$\cellcolor{cyan!16}
\\ \hline
7
&$\bigsubstack{55\\ x^{-1}y\\ xy^{3}}$\cellcolor{mycolor2!75}
&$\bigsubstack{89\\ x^{-1}y\\ xy^{3}}$\cellcolor{mycolor2!75}
&$\bigsubstack{118\\ x^{-1}y\\ xy^{3}}$\cellcolor{mycolor2!75}
&$\bigsubstack{131\\ x^{-1}y\\ xy^{3}}$\cellcolor{mycolor2!75}
&$\bigsubstack{166\\ x^{-1}y\\ xy^{3}}$\cellcolor{mycolor2!75}
& $\bigsubstack{202\\ x^{-2}y^{-1}\\ x^{-1}y^{2}}$
&$\bigsubstack{205\\ x^{-1}y\\ xy^{3}}$\cellcolor{mycolor2!75}
&$\bigsubstack{248\\ x^{-1}y\\ xy^{3}}$\cellcolor{mycolor2!75}
&$\bigsubstack{267\\ x^{-1}y\\ xy^{3}}$\cellcolor{mycolor2!75}
&$\bigsubstack{316\\ x^{-1}y\\ xy^{3}}$\cellcolor{mycolor2!75}
&$\bigsubstack{337\\ x^{-1}y\\ xy^{3}}$\cellcolor{mycolor2!75}
&$\bigsubstack{392\\ x^{-1}y\\ xy^{3}}$\cellcolor{mycolor2!75}
&$\bigsubstack{451\\ x^{-1}y\\ xy^{3}}$\cellcolor{mycolor2!75}
\\ \hline
8
&$\bigsubstack{59\\ x^{-1}y^{2}\\ xy^{3}}$\cellcolor{mycolor3!75}
&$\bigsubstack{94\\ x^{-1}y^{2}\\ xy^{3}}$\cellcolor{mycolor3!75}
&$\bigsubstack{105\\ x^{-1}y^{2}\\ xy^{3}}$\cellcolor{mycolor3!75}
&$\bigsubstack{137\\ x^{-1}y^{2}\\ xy^{3}}$\cellcolor{mycolor3!75}
&$\bigsubstack{173\\ x^{-1}y^{2}\\ xy^{3}}$\cellcolor{mycolor3!75}
&$\bigsubstack{188\\ x^{-1}y^{2}\\ xy^{3}}$\cellcolor{mycolor3!75}
&$\bigsubstack{230\\ x^{-1}y^{2}\\ xy^{3}}$\cellcolor{mycolor3!75}
&$\bigsubstack{276\\ x^{-1}y^{2}\\ xy^{3}}$\cellcolor{mycolor3!75}
&$\bigsubstack{288\\ x^{-1}y^{2}\\ ~x^{-2}y^{-1}}~ $\cellcolor{mycolor4!75}
&$\bigsubstack{326\\ x^{-1}y^{2}\\ xy^{3}}$\cellcolor{mycolor3!75}
&$\bigsubstack{347\\ x^{-1}y^{2}\\ xy^{3}}$\cellcolor{mycolor3!75}
&$\bigsubstack{368\\ x^{-1}y^{2}\\ xy^{3}}$\cellcolor{mycolor3!75}
&$\bigsubstack{426\\ x^{-1}y^{2}\\ xy^{3}}$\cellcolor{mycolor3!75}
\\ \hline
9
&$\bigsubstack{78\\ x^{-3}y^{-1}\\ xy^{-1}}$\cellcolor{mycolor6!50}
&$\bigsubstack{110\\ x^{-1}y^{3}\\ xy^{3}}$\cellcolor{mycolor5!75}
&$\bigsubstack{132\\ x^{-1}y^{3}\\ xy^{3}}$\cellcolor{mycolor5!75}
&$\bigsubstack{164\\ x^{-3}y^{-1}\\ xy^{-1}}$\cellcolor{mycolor6!50}
&$\bigsubstack{204\\ x^{-3}y^{-1}\\ xy^{-1}}$\cellcolor{mycolor6!50}
&$\bigsubstack{225\\ x^{-1}y^{3}\\ xy^{3}}$\cellcolor{mycolor5!75}
&$\bigsubstack{265\\ x^{-3}y^{-1}\\ xy^{-1}}$\cellcolor{mycolor6!50}
&$\bigsubstack{334\\ x^{-3}y^{-1}\\ xy^{-1}}$\cellcolor{mycolor6!50}
&$\bigsubstack{353\\ x^{-3}y^{-1}\\ xy^{-1}}$\cellcolor{mycolor6!50}
&$\bigsubstack{411\\ x^{-3}y^{-1}\\ xy^{-1}}$\cellcolor{mycolor6!50}
& $\bigsubstack{432\\ x^{-3}y^{-1}\\ xy^{-1}}$\cellcolor{mycolor6!50}
&$\bigsubstack{441\\ x^{-1}y^{3}\\ xy^{3}}$\cellcolor{mycolor5!75}
&$\bigsubstack{506\\ x^{-1}y^{3}\\ xy^{3}}$\cellcolor{mycolor5!75}
\\ \hline
10
&$\bigsubstack{78\\ ~x^{-2}y^{-2}~{} \\ x^{-2}y^{2}}$\cellcolor{mycolor8!50}
&$\bigsubstack{119\\ ~x^{-2}y^{-2}~{} \\ x^{-2}y^{2}}$\cellcolor{mycolor8!50}
&$\bigsubstack{133\\ x^{-1}y^{2}~{} \\ xy^{4}}$\cellcolor{mycolor7!40}
&$\bigsubstack{170\\ x^{-1}y^{2}~{} \\ xy^{4}}$\cellcolor{mycolor7!40}
&$\bigsubstack{211\\ x^{-1}y^{2}~{} \\ xy^{4}}$\cellcolor{mycolor7!40}
&$\bigsubstack{226\\ x^{-1}y^{2}~{} \\ xy^{4}}$\cellcolor{mycolor7!40}
&$\bigsubstack{273\\ x^{-1}y^{2}~{} \\ xy^{4}}$\cellcolor{mycolor7!40}
&$\bigsubstack{324\\ ~x^{-2}y^{-2}~{} \\ x^{-2}y^{2}}$\cellcolor{mycolor8!50}
&$\bigsubstack{362\\ x^{-1}y^{2}~{} \\ xy^{4}}$\cellcolor{mycolor7!40}
&$\bigsubstack{381\\ x^{-1}y^{2}~{} \\ xy^{4}}$\cellcolor{mycolor7!40}
&$\bigsubstack{442\\ x^{-1}y^{2}~{} \\ xy^{4}}$\cellcolor{mycolor7!40}
&$\bigsubstack{490\\ ~x^{-2}y^{-2}~{} \\ x^{-2}y^{2}}$\cellcolor{mycolor8!50}
&$\bigsubstack{530\\ x^{-1}y^{2}~{} \\ xy^{4}}$\cellcolor{mycolor7!40}
\\ \hline
11
&$\bigsubstack{79\\ x^{-2}y^{2}\\ xy^{3}}$
&$\bigsubstack{127\\ x^{-1}y^{-3}\\ xy^{-3}}$\cellcolor{mycolor9!60}
&$\bigsubstack{138\\ x^{-1}y^{-3}\\ xy^{-3}}$\cellcolor{mycolor9!60}
&$\bigsubstack{176\\ x^{-1}y^{-3}\\ xy^{-3}}$\cellcolor{mycolor9!60}
&$\bigsubstack{233\\ x^{-1}y^{-3}\\ xy^{-3}}$\cellcolor{mycolor9!60}
&$\bigsubstack{248\\ x^{-1}y^{-3}\\ xy^{-3}}$\cellcolor{mycolor9!60}
&$\bigsubstack{298\\ x^{-1}y^{-3}\\ xy^{-3}}$\cellcolor{mycolor9!60}
&$\bigsubstack{352\\ x^{-1}y^{-3}\\ xy^{-3}}$\cellcolor{mycolor9!60}
&$\bigsubstack{371\\ x^{-1}y^{-3}\\ xy^{-3}}$\cellcolor{mycolor9!60}
&$\bigsubstack{431\\ x^{-1}y^{-3}\\ xy^{-3}}$\cellcolor{mycolor9!60}
&$\bigsubstack{473\\ x^{-1}y^{-3}\\ xy^{-3}}$\cellcolor{mycolor9!60}
&$\bigsubstack{494\\ x^{-1}y^{-3}\\ xy^{-3}}$\cellcolor{mycolor9!60}
&$\bigsubstack{564\\ x^{-1}y^{-3}\\ xy^{-3}}$\cellcolor{mycolor9!60}
\\ \hline
12
&$\bigsubstack{105\\ x^{-1}y^{2}\\ xy^{5}}$\cellcolor{mycolor12!30}
&$\bigsubstack{140\\ x^{-1}y^{3}\\ x^{2}y^{3}}$\cellcolor{mycolor11!75}
&$\bigsubstack{172\\ x^{-1}y^{2}\\ xy^{5}}$\cellcolor{mycolor12!30}
&$\bigsubstack{216\\ x^{-1}y^{2}\\ xy^{5}}$\cellcolor{mycolor12!30}
&$\bigsubstack{264\\ x^{-1}y^{2}\\ xy^{5}}$\cellcolor{mycolor12!30}
&$\bigsubstack{279\\ x^{-1}y^{2}\\ xy^{5}}$\cellcolor{mycolor12!30}
&$\bigsubstack{330\\ x^{-1}y^{3}\\ x^{2}y^{3}}$\cellcolor{mycolor11!75}
&$\bigsubstack{384\\ x^{-1}y^{3}\\ x^{2}y^{3}}$\cellcolor{mycolor11!75}
&$\bigsubstack{432\\ x^{-1}y^{3}\\ x^{2}y^{3}}$\cellcolor{mycolor11!75}
&$\bigsubstack{486\\ x^{-1}y^{2}\\ xy^{5}}$\cellcolor{mycolor12!30}
&$\bigsubstack{560\\ x^{-1}y^{3}\\ x^{2}y^{3}}$\cellcolor{mycolor11!75}
&$\bigsubstack{588\\ x^{-1}y^{3}\\ x^{2}y^{3}}$\cellcolor{mycolor11!75}
&$\bigsubstack{660\\ x^{-1}y^{3}\\ x^{2}y^{3}}$\cellcolor{mycolor11!75}
\\ \hline
13
&$\bigsubstack{98\\ x^{-2}y^{2}\\ x^{2}y^{3}}$\cellcolor{mycolor14!60}
&$\bigsubstack{145\\ x^{-2}y^{2}\\ x^{2}y^{3}}$\cellcolor{mycolor14!60}
&$\bigsubstack{177\\ x^{5}y\\ x^{3}y^{-1}}$\cellcolor{green!40}
&$\bigsubstack{222\\ x^{5}y\\ x^{3}y^{-1}}$\cellcolor{green!40}
& $\bigsubstack{271\\ x^{5}y\\ x^{3}y^{-1}}$\cellcolor{green!40}
&$\bigsubstack{286\\ x^{5}y\\ x^{3}y^{-1}}$\cellcolor{green!40}
&$\bigsubstack{341\\ x^{5}y\\ x^{3}y^{-1}}$\cellcolor{green!40}
&$\bigsubstack{392\\ x^{-2}y^{2}\\ x^{2}y^{3}}$\cellcolor{mycolor14!60}
&$\bigsubstack{476\\ x^{5}y\\ x^{3}y^{-1}}$\cellcolor{green!40}
&$\bigsubstack{495\\ x^{5}y\\ x^{3}y^{-1}}$\cellcolor{green!40}
&$\bigsubstack{578\\ x^{-2}y^{2}\\ x^{2}y^{3}}$\cellcolor{mycolor14!60}
& $\bigsubstack{610\\ x^{5}y\\ x^{3}y^{-1}}$\cellcolor{green!40}
&$\bigsubstack{648\\ x^{-2}y^{2}\\ x^{2}y^{3}}$\cellcolor{mycolor14!60}
\\ \hline
\end{tabular}
\caption{\change
For each fixed logical dimension $k$ and code distance $d$, we enumerate all families of bulk stabilizers of the form $f(x,y) \propto 1 + x + x^a y^b$ and $g(x,y) \propto 1 + y + x^c y^d$, with $-5 \leq a,b,c,d \leq 6$ (see Appendix~\ref{Appendix: Review of the Laurent polynomial formalism} for stabilizer conventions), to identify quantum $[[n,k,d]]$ codes with the minimal number of physical qubits $n$. Details of the boundary and corner stabilizer construction are provided in Sec.~\ref{sec: searching optimal qLDPC}.
Each entry lists the minimal $n$, followed by the corresponding bulk stabilizers; the second and third lines indicate the terms $x^a y^b$ and $x^c y^d$, respectively. Entries sharing the same color correspond to the same bulk stabilizers. In each family, the logical dimension $k$ is fixed, while the code distance $d(L_x,L_y)$ depends on the system size. The exact code distances are computed using the integer programming approach~\cite{landahl2011fault}.
Our results show that, for each $k$, one or two dominant families provide the best performance. These optimal families, spanning $k=6$ to $k=13$, are presented in Sec.~\ref{sec: examples of planar qLDPC codes} and Appendix~\ref{app: Families of planar qLDPC codes}.
}
\label{tab: GTC large weight and stabilizer}
\end{table*}

\subsection{Enumerating planar qLDPC codes on open square lattices}\label{sec: searching optimal qLDPC}

In this section, we describe our strategy for identifying promising families of planar qLDPC codes. In the toric code case, previous works~\cite{liang2025generalized, chen2025anyon} have focused on a specific form—referred to as the generalized toric code—defined by
\begin{eqs}
    f(x,y) &= 1+x+x^a y^b, \\
    g(x,y) &= 1+y+x^c y^d.
\label{eq:weight-6_f_and_g}
\end{eqs}
This form of stabilizers can be viewed as the Kitaev toric code augmented with two extra edges, and it has been shown to yield optimal toric codes on twisted tori for $n \leq 400$. For simplicity, we focus on this particular family of bulk stabilizers, from which we construct the corresponding boundary Hamiltonian and evaluate their performance.

The procedure is summarized as follows:
\begin{enumerate}
    \item {\bf Bulk stabilizer:} Choose a bulk stabilizer with fixed parameters $-5 \leq a, b, c, d \leq 6$. We multiply by appropriate monomials in $f(x,y)$ and $g(x,y)$ to eliminate any negative powers of $x$ or $y$. For instance, the $[[288,8,12]]$ code defined in Sec.~\ref{sec: [[288, 8, 12]] planar code} uses the following normalization:
    \begin{eqs}
    1+x+x^{-1}y^2 &\rightarrow x\bigl(1+x+x^{-1}y^2\bigr), \\
    1+y+x^{-2} y^{-1} &\rightarrow x^2y\bigl(1+y+x^{-2} y^{-1}\bigr).
    \end{eqs}
    This renders the bulk stabilizers more geometrically local, which in turn facilitates the subsequent boundary construction. Note that the logical dimension $k$ is determined from Eq.~\eqref{eq: logical dimension k for open boundary}.
    
    \item {\bf Boundary anyon condensation:} Consider a finite $L_x \times L_y$ square lattice. Away from the corners, at the left and right boundaries, we determine all $Z$ boundary gauge operators and incorporate them into the Hamiltonian in a vertically translation-invariant fashion. Similarly, at the top and bottom boundaries away from the corners, we include all $X$ boundary gauge operators with horizontal translation symmetry as stabilizers.\footnote{At the corners, translation symmetry is not well-defined, and $X$ and $Z$ operators may overlap and anti-commute. To resolve this, we adopt the convention that if an $X$ term and a $Z$ term anti-commute, neither is included in the stabilizer Hamiltonian.}
    
    \item {\bf Corner topological order completion:} The exclusion of boundary terms near the corners causes the presence of local nontrivial logical operators. We remedy this issue by promoting these local operators to corner stabilizers, using the topological order completion algorithm described in Ref.~\cite{liang2024operator}. There is some flexibility in this step: given local logical $X$ and $Z$ operators, one may choose either to serve as the stabilizer. Based on our experiments, we prefer to select the operator with larger weight as the stabilizer, as it is more likely to enhance the code distance (choosing the lower-weight operator might allow its combination with other logical operators, thereby reducing their effective weights and the code distance). This choice typically affects the code distance $d$ only by $O(1)$.
    
    \item {\bf Removing single-qubit stabilizers:} In the process described above, some boundary and corner stabilizers may act on a single qubit. Since these can be disentangled from the system, we remove them from the lattice, thereby reducing the total number of physical qubits $n(L_x, L_y)$.
    
    \item {\bf Code distance table $d(L_x, L_y)$:} For a fixed bulk stabilizer (i.e., fixed $a, b, c, d$), the previous steps yield a family of planar qLDPC codes parameterized by the lattice dimensions, due to translation symmetry (one may insert an arbitrary number of rows or columns). We compute the code distances for various values of $L_x$ and $L_y$, collecting the data in the form $[[n(L_x, L_y), k, d(L_x, L_y)]]$.
    
    \item {\bf Optimal $n$ for fixed $k$ and $d$:} After enumerating the parameters $a, b, c, d$ over the finite range, we obtain a large set of $[[n, k, d]]$ data. We then address the following question: for fixed $k$ and $d$, which family of bulk stabilizers minimizes the number of physical qubits $n$? The optimal values of $n$ for each small $k$ and $d$ and the corresponding bulk stabilizers are listed in Table~\ref{tab: GTC large weight and stabilizer}.
\end{enumerate}

Caveat: The boundary and corner stabilizers obtained from our computation differ slightly from the examples presented in previous sections. The lattice derived from the computation is often more irregular, as the topological order completion algorithm at the corners does not preserve translational symmetry along the boundary, and some single-qubit stabilizers may be removed.
For clarity and convenience, we manually reconstruct this family to ensure that the lattice corners have regular shapes and that all corner stabilizers arise from the translation of boundary stabilizers.
In particular, all boundary and corner stabilizers are obtained by truncating the bulk stabilizers, simplifying the presentation of our planar codes.
Although the details around the corners differ, the overall performance remains essentially the same, particularly after the ``lattice grafting'' step described in the next section.

\subsection{Lattice grafting}
\label{sec: Lattice grafting}

We now explain how to further optimize the planar qLDPC with open boundary as considered in Sec.~\ref{sec: examples of planar qLDPC codes} and~\ref{sec: Topological order perspectives}. For this we present an algorithm that successively removes qubits from the boundary of the codes while recombining the stabilizers supported on these qubits to ensure that they still commute. The algorithm of recombining the stabilizers is reminiscent of grafting a plant---hence the name lattice grafting.

\begin{figure*}[thb]
    \centering
    \includegraphics[width=0.65\textwidth]{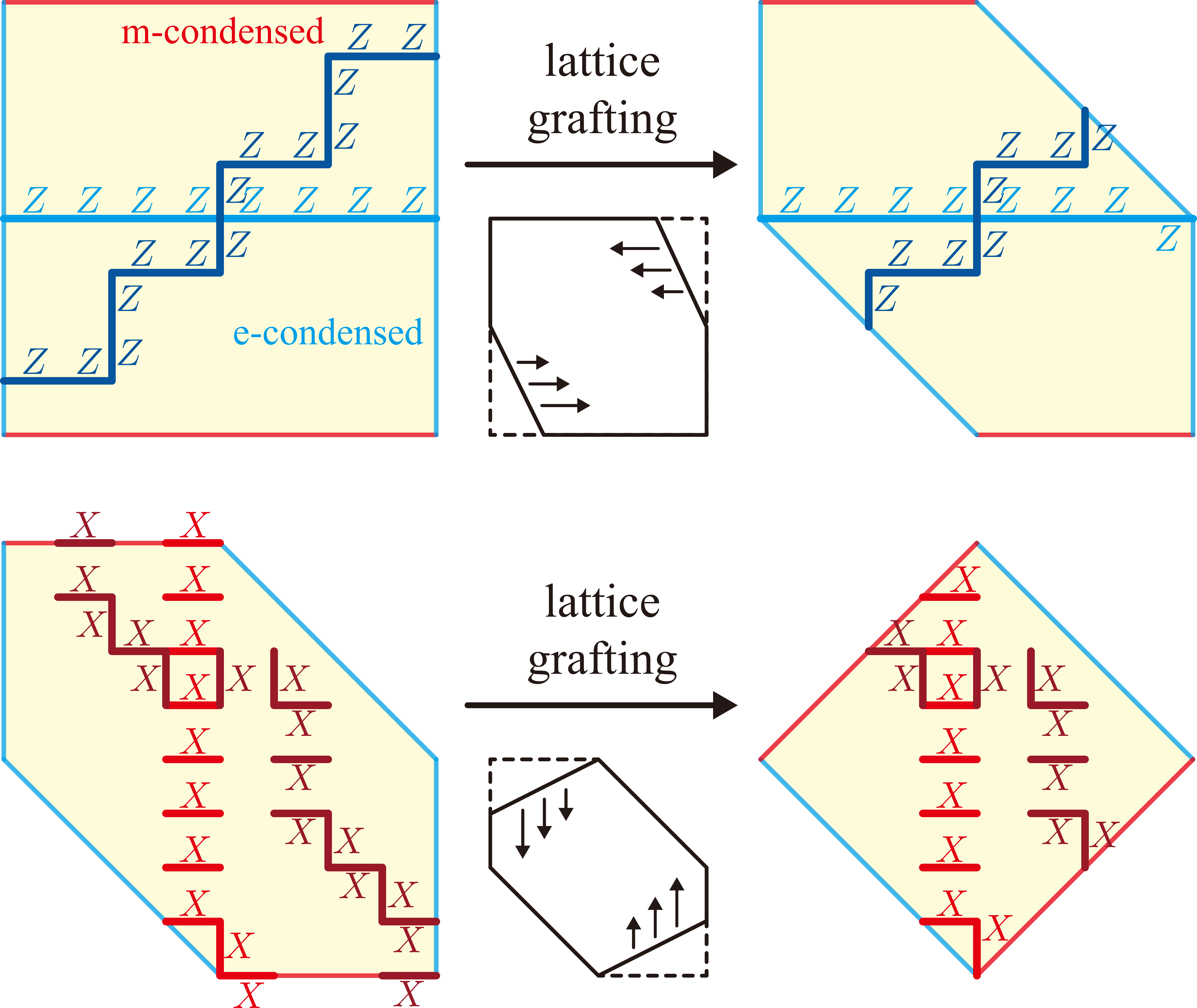}
    \caption{Lattice grafting construction of the rotated surface code. Starting from the standard surface code—with $e$‑condensed left/right boundaries and $m$‑condensed top/bottom boundaries—the logical $Z$ operator can be realized either as a straight $e$‑string (blue) or as a zig‑zag operator (deep blue) with a higher weight.
    By iteratively removing qubits from the top‑right and bottom‑left corners, we shrink the zig‑zag logical operator until its weight matches that of the straight string, without altering the code distance.
    Applying the same removal process to the top left and bottom right corners for the logical $X$ operator, which could be a (dual) straight string (red) or a zig-zag path (deep red), produces the grafted lattice that is exactly the rotated surface code.
    }
    \label{fig: lattice_grafting_surface}
\end{figure*}

In detail, the algorithm takes as an input planar code with open boundaries and performs the following steps.
\begin{enumerate}
    \item \textbf{Removing a physical qubit}. Choose a physical qubit $A$ in the boundary of the lattice which which is contained in a minimal number of stabilizers. Then remove the qubit from the code.
    \item \textbf{Grafting}. Choose a Pauli label $O\in \{X,Z\}$. Choose an enumeration $[S_1, S
_2,S_3,\dots, S_r]$ of all $O$-stabilizers  which were supported on the qubit $A$. Replace these $O$-stabilizers in the code by the stabilizers  $[S_1S_2, S_2S_3,\dots,S_{r-1}S_r]$.
     \item \textbf{Checking parameters}. Compute the weight of the new stabilizers and the distance of the code. If the weight increased or the distance decreased, undo the last two steps.
    \item{\textbf{Iterating.}} Repeat with step $1$ until all qubits have been checked.
\end{enumerate}
Pruning ensures that the stabilizers in the code constructed in each iteration still commute and that the number of stabilizers is decreased by one. The number of logicals stays constant since in each iteration the number of qubits and stabilizers is each decreased by one.

Lattice pruning is an iterative and non-deterministic algorithm. It involves, in every step, the choice of a qubit, a Pauli label, and the enumeration of stabilizers. Each choice potentially affects the successive steps. Iterating over all possibilities is not feasible.
Hence, in practice, we perform the algorithm many times with randomized choices and then pick the best result.

The general lattice grafting procedure can affect the locality and weight of stabilizers because it generates new stabilizers as products of the original ones (e.g., $S_1 S_2$). Consequently, for bivariate bicycle codes defined on a torus, lattice grafting reduces the number of physical qubits but sacrifices the locality and LDPC properties of the stabilizers.
This approach is thus not suitable for (generalized) toric codes. In contrast, for our planar qLDPC codes with open boundaries, many qubits near the boundary participate in only one $X$ or $Z$ stabilizer ($r=1$ in step~2 above).
In such cases, we can simply remove the corresponding stabilizer without introducing any new ones, thereby preserving both locality and the LDPC structure (although we must verify that the code distance $d$ remains unaffected).
We restrict ourselves to this scenario when optimizing the previously presented examples of planar qLDPC codes.
This strategy is feasible only under open boundary conditions, where the boundary qubits are sufficiently ``loose'' to allow their removal without impacting the parameters $k$, $d$, locality, or the overall LDPC properties. Explicit examples of grafted qLDPC codes are provided in Appendix~\ref{app: Grafted planar qLDPC codes}.
{\change Each code was subjected to one thousand runs of the lattice grafting optimization, and we report the grafted code with the minimal $n$ obtained.}

Our lattice grafting procedure already covers the “rotation” trick for the surface code: the rotated surface code follows directly from the standard surface code via the grafting steps illustrated in Fig.~\ref{fig: lattice_grafting_surface}. In that setting, one can manually "cut" the lattice due to the simple geometry of the logical operators, while preserving the code distance. By contrast, the logical operators in our planar qLDPC constructions, derived from BB codes, has far more intricate support. As a result, we rely on an automated lattice grafting algorithm to remove redundant qubits and verify that the code distance remains unchanged in the pruned lattice.

\section{Planar qLDPC codes}
\label{sec: examples of planar qLDPC codes}

{\change
In this section, we explicitly construct three families of planar qLDPC codes, with examples including the $[[188,8,9]]$ code in Sec.~\ref{sec: [[188, 8, 9]] planar code}, the $[[131,7,7]]$ code in Sec.~\ref{sec: [[131, 7, 7]] planar code}, and the $[[88,6,6]]$ code in Sec.~\ref{sec: [[88, 6, 6]] planar code}.
Additional planar qLDPC code families, such as the $[[381,10,13]]$, $[[494,11,15]]$, $[[432,12,12]]$, and $[[392,13,11]]$ codes, are presented in Appendix~\ref{app: Families of planar qLDPC codes}. All of these code families feature weight-6 stabilizers.
Finally, we show that by allowing weight-8 stabilizers, the code parameters can be further improved, yielding a $[[292,12,14]]$ code, as discussed in Sec.~\ref{sec: [[292, 12, 14]] planar code}.
}



%
We consider a translation-invariant $\mathbb{Z}_2$ CSS code defined on the square lattice, with bulk stabilizer generators:
\begin{equation}
    A_v = 
    \begin{bmatrix}
        f(x,y) \\
        \rule{0pt}{1.1em}g(x,y) \\
        \hline
        0 \\
        0
    \end{bmatrix}, 
    \quad
    B_p = 
    \begin{bmatrix}
        0 \\
        0 \\
        \hline
        \rule{0pt}{1.1em}\overline{g(x,y)} \\
        \overline{f(x,y)}
    \end{bmatrix},
\end{equation}
using the notation from Ref.~\cite{liang2023extracting}.
In these examples, the boundary stabilizers are obtained by truncating the bulk stabilizer generators at the lattice boundary.

\subsection{$[[188,8,9]]$ planar code}\label{sec: [[188, 8, 9]] planar code}

\begin{figure}[tbh]
    \centering
    \vspace{-0.4cm}
    \hspace{-0.77cm}
    \subfigure[~54 bulk $X$-stabilizers]{\raisebox{0.05cm}{\includegraphics[scale=0.05]{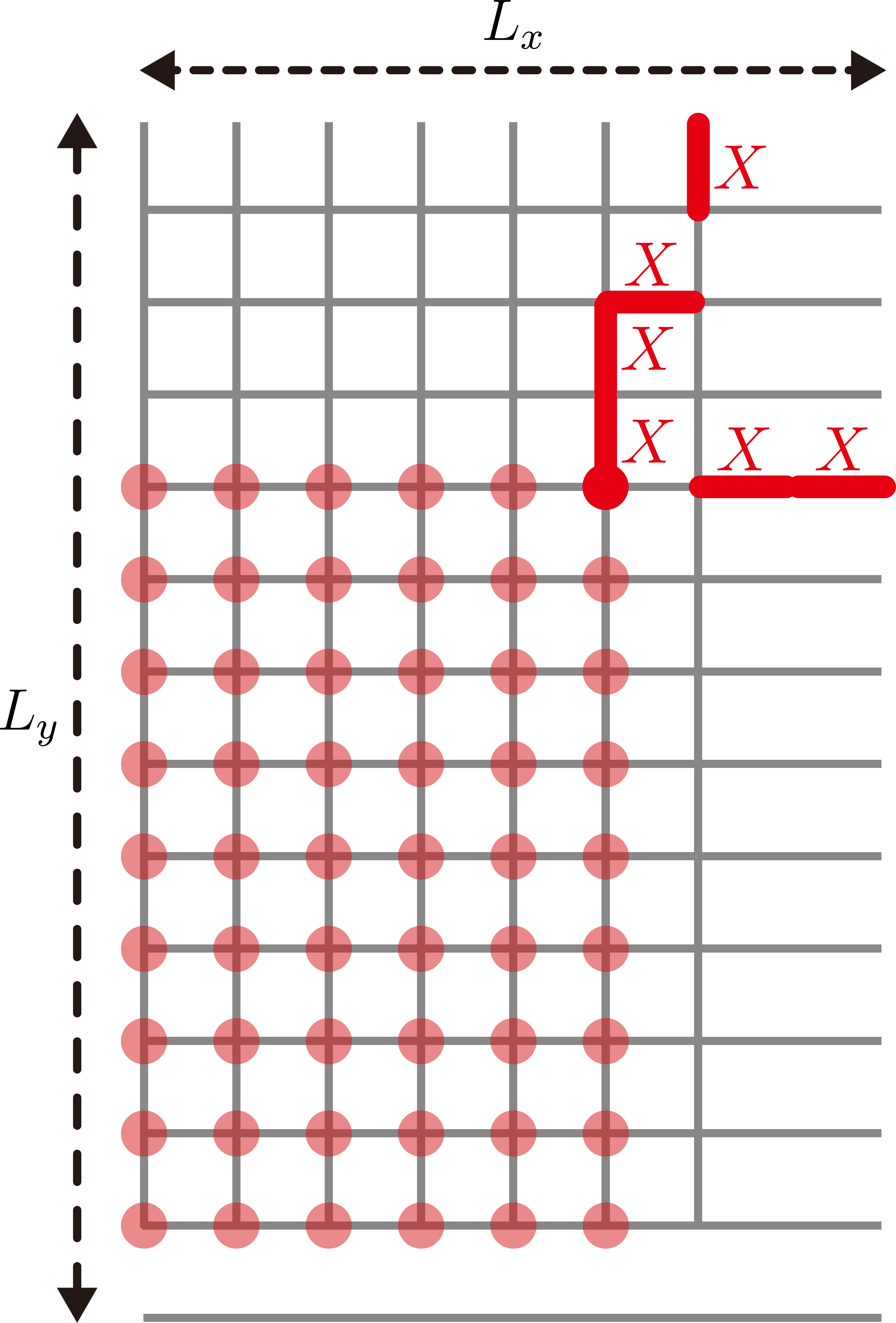}}}
    \hspace{0.0cm}
    \subfigure[~50 bulk $Z$-stabilizers]{\includegraphics[scale=0.05]{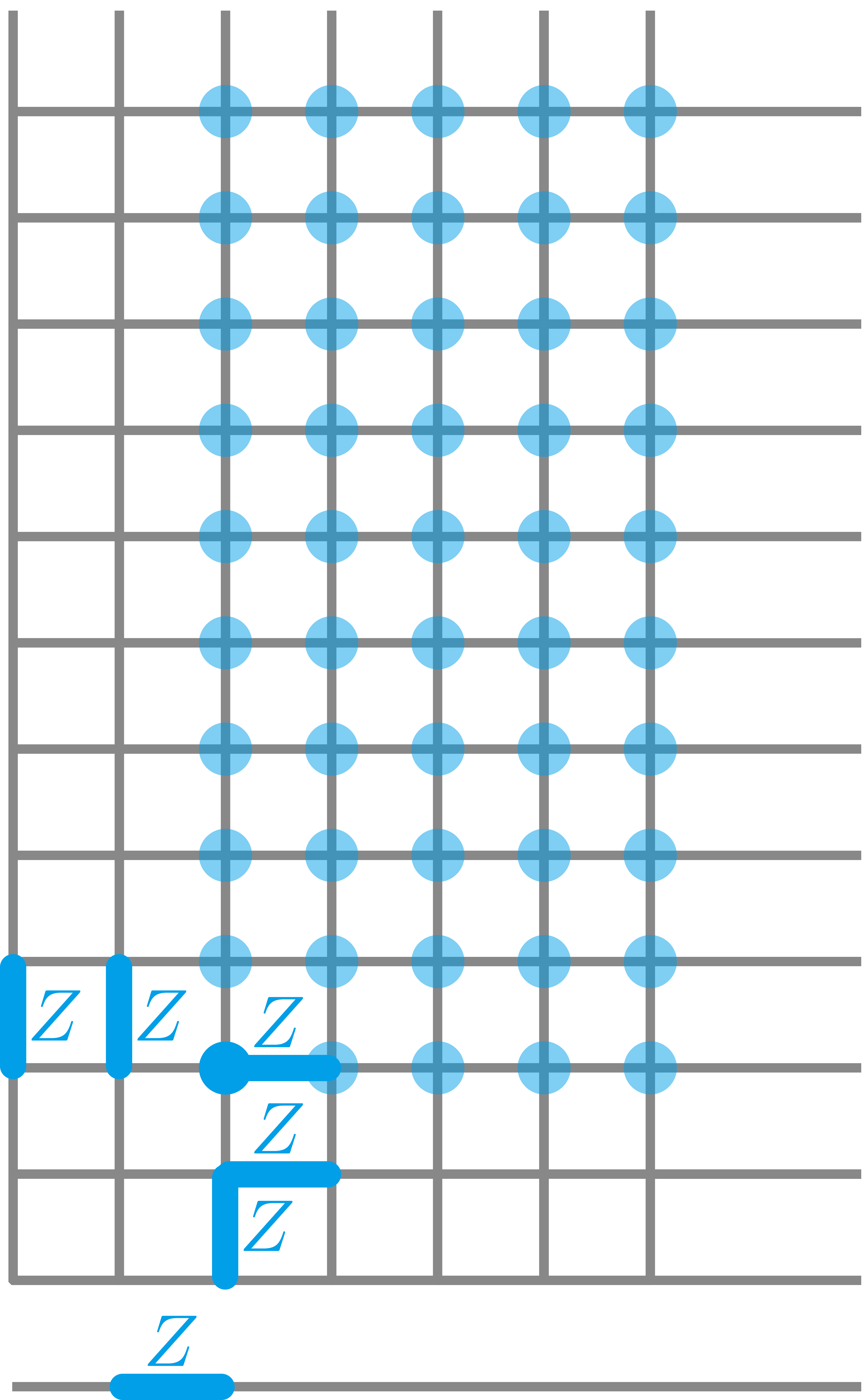}}\\
    \vspace{-0.2cm}
    \subfigure[~36 boundary $X$-stabilizers]{\includegraphics[scale=0.05]{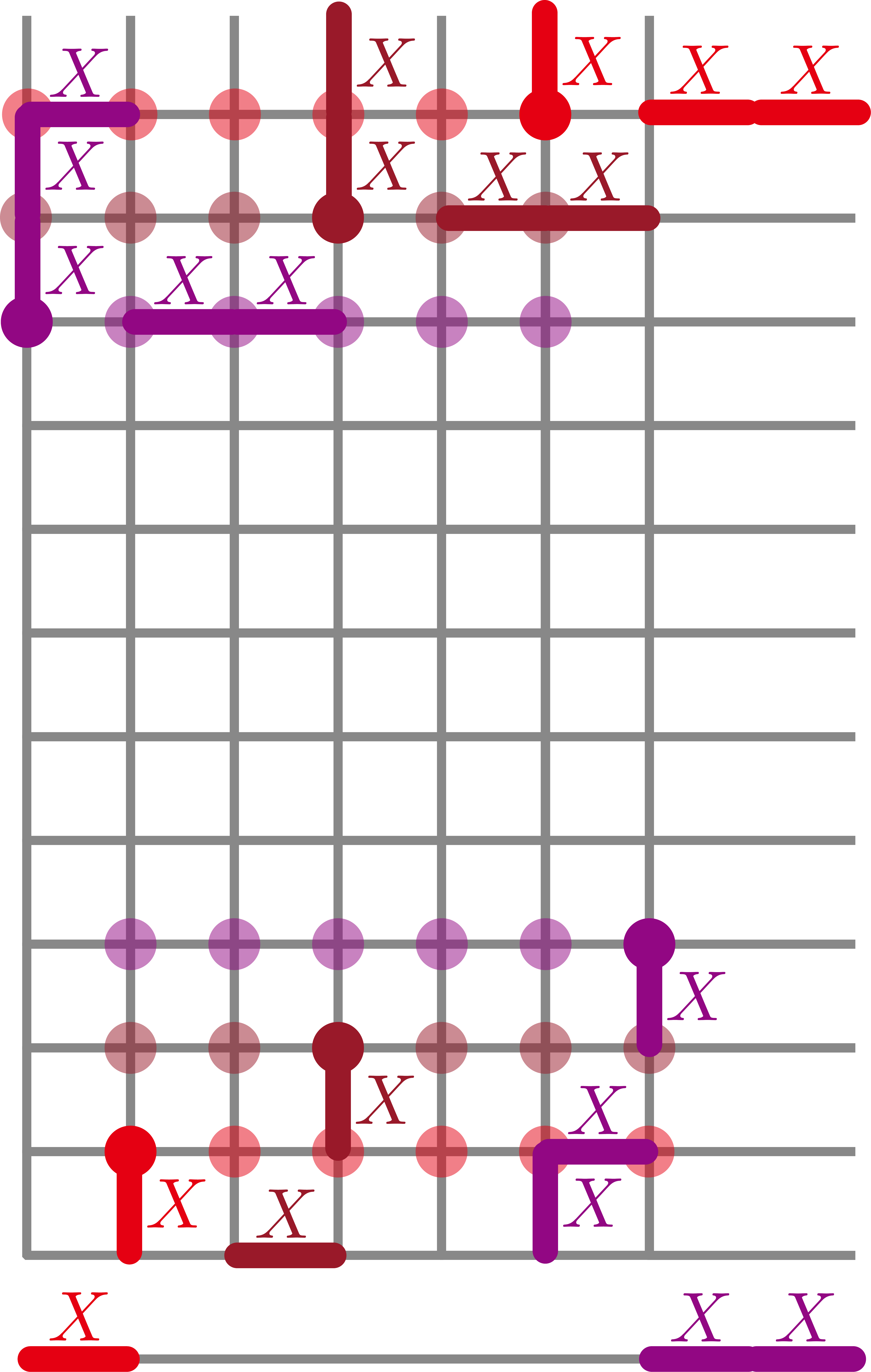}}
    \hspace{0.0cm}
    \subfigure[~40 boundary $Z$-stabilizers]{\includegraphics[scale=0.05]{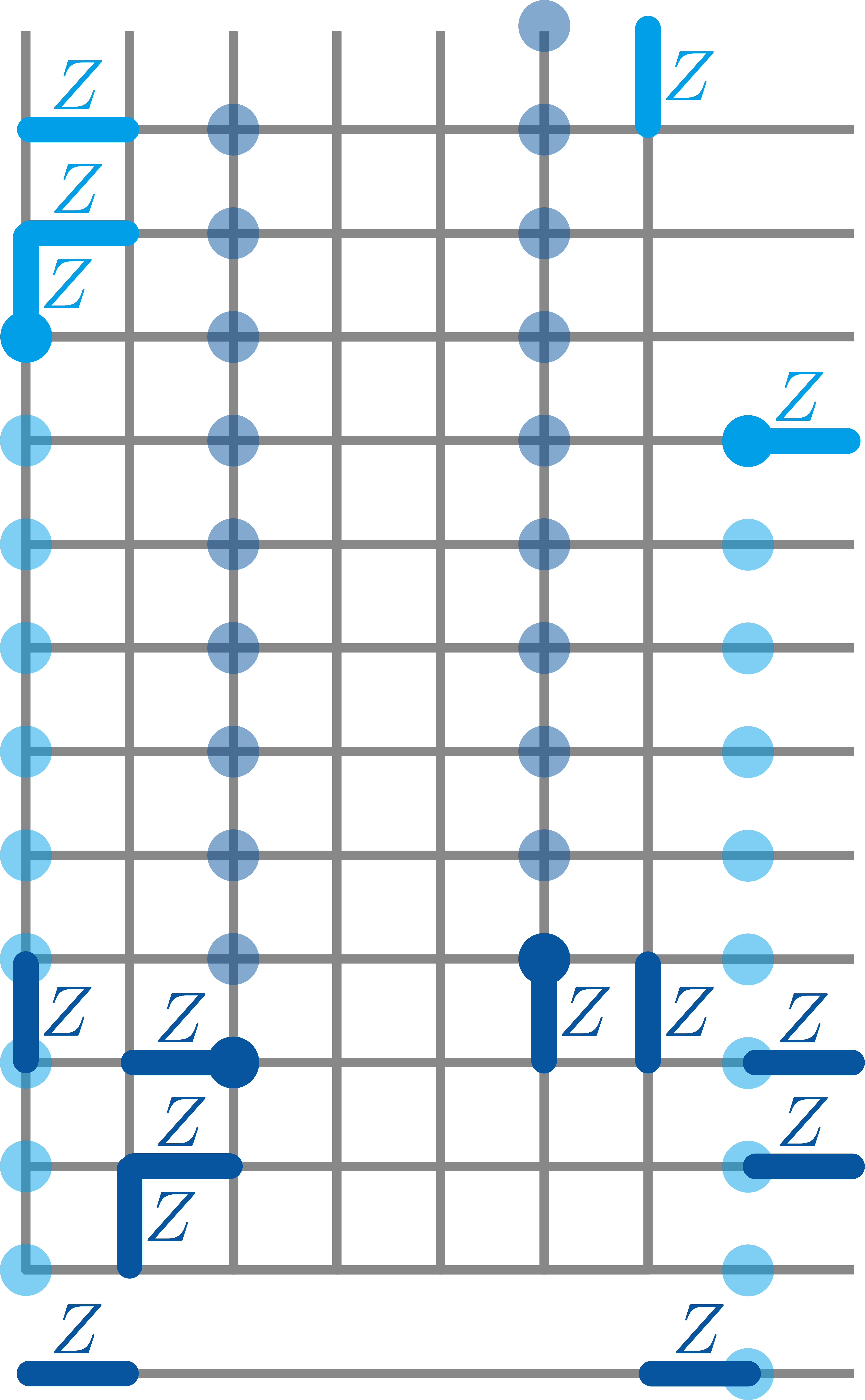}}
    \vspace{-0.35cm}
    \caption{The $[[188, 8, 9]]$ code on an open square lattice. Vertical edges of the rightmost column and the bottommost row have been removed. The bulk stabilizers are specified by $f(x,y) =x(1+x+ x^{-1}y^2)$ and $g(x,y) =1+y+xy^3$.
    (a) Bulk $X$-stabilizers consist of 54 Pauli operators that are translations of one another. (b) 50 bulk $Z$-stabilizers.
    (c) 36 boundary $X$-stabilizers near the top and bottom boundaries.
    (d) 40 boundary $Z$-stabilizers appear near the left and right boundaries. The logical dimension is $ k = 188-54-50-36-40=8$.
    This generates another code family of $k=8$: $[[2L_x L_y - L_x - L_y + 1, ~8, ~d(L_x, L_y)]]$, with code distances $d(L_x, L_y)$ listed in Table~\ref{tab: d(Lx, Ly) for [[188,8,9]]} in Appendix~\ref{app: Families of planar qLDPC codes}.
    }
    \label{fig: [[188, 8, 9]] code}
\end{figure}

The $[[188, 8, 9]]$ code is implemented on an $8 \times 13$ square lattice, as illustrated in Fig.~\ref{fig: [[188, 8, 9]] code}. In our construction, the lattice sizes $L_x$ and $L_y$ are adjustable. Note that the lattice deviates from a perfect square lattice, as certain qubits along the right and bottom boundaries have been omitted. Consequently, the number of physical qubits is given by
\begin{equation}
    n = 2L_x L_y - L_x - L_y + 1.
\end{equation}
The logical dimension remains fixed at $k=8$, determined by subtracting the number of independent stabilizers from the total number of physical qubits.
A table of the code distances $d(L_x, L_y)$ is provided in Table~\ref{tab: d(Lx, Ly) for [[188,8,9]]} in Appendix~\ref{app: Families of planar qLDPC codes}, and a few representative cases are listed below:
\begin{enumerate}
    \item $L_x = 6$, $L_y = 10$: $[[105, 8, 6]]$, ~${kd^2}/{n} = 2.74$.
    \item $L_x = 7$, $L_y = 11$: $[[137, 8, 7]]$, ~${kd^2}/{n} = 2.86$.
    \item $L_x = 8$, $L_y = 12$: $[[173, 8, 8]]$, ~${kd^2}/{n} = 2.96$.
    \item $L_x = 8$, $L_y = 13$: $[[188, 8, 9]]$, ~${kd^2}/{n} = 3.45$.
    \item $L_x = 9$, $L_y = 14$: $[[230, 8, 10]]$, ~${kd^2}/{n} = 3.48$.
    \item $L_x = 10$, $L_y = 15$: $[[276, 8, 11]]$, ~${kd^2}/{n} = 3.51$.
    \item $L_x = 10$, $L_y = 16$: $[[295, 8, 12]]$, ~${kd^2}/{n} = 3.91$.
    \item $L_x = 11$, $L_y = 16$: $[[326, 8, 13]]$, ~${kd^2}/{n} = 4.15$.
    \item $L_x = 11$, $L_y = 17$: $[[347, 8, 14]]$, ~${kd^2}/{n} = 4.52$.
    \item $L_x = 11$, $L_y = 18$: $[[368, 8, 15]]$, ~${kd^2}/{n} = 4.89$.
\end{enumerate}
{\change
This family of planar qLDPC codes has the same logical dimension $k=8$ as the previously discussed $[[288,8,12]]$ code in Sec.~\ref{sec: [[288, 8, 12]] planar code}, and generally exhibits improved performance, except in the $d=12$ case, as summarized in Table~\ref{tab: GTC large weight and stabilizer}.}

\subsection{$[[131, 7, 7]]$ planar code}\label{sec: [[131, 7, 7]] planar code}

\begin{figure}[htb]
    \centering
    \hspace{-1.3cm}
    \subfigure[~24 bulk $X$-stabilizers]{\raisebox{-0.05cm}{\includegraphics[scale=0.05]{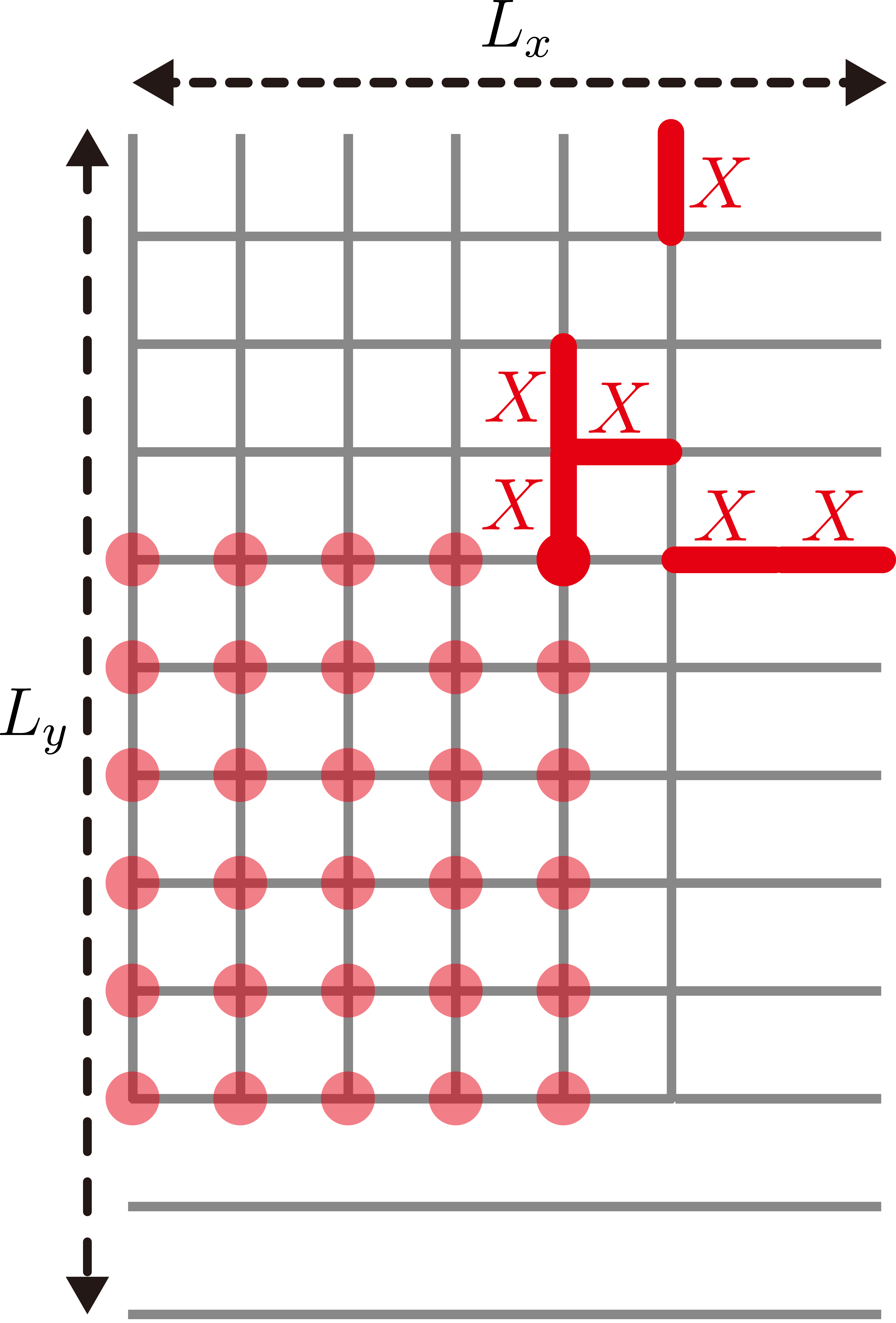}}}
    \hspace{0.5cm}
    \subfigure[~18 bulk $Z$-stabilizers]{\includegraphics[scale=0.05]{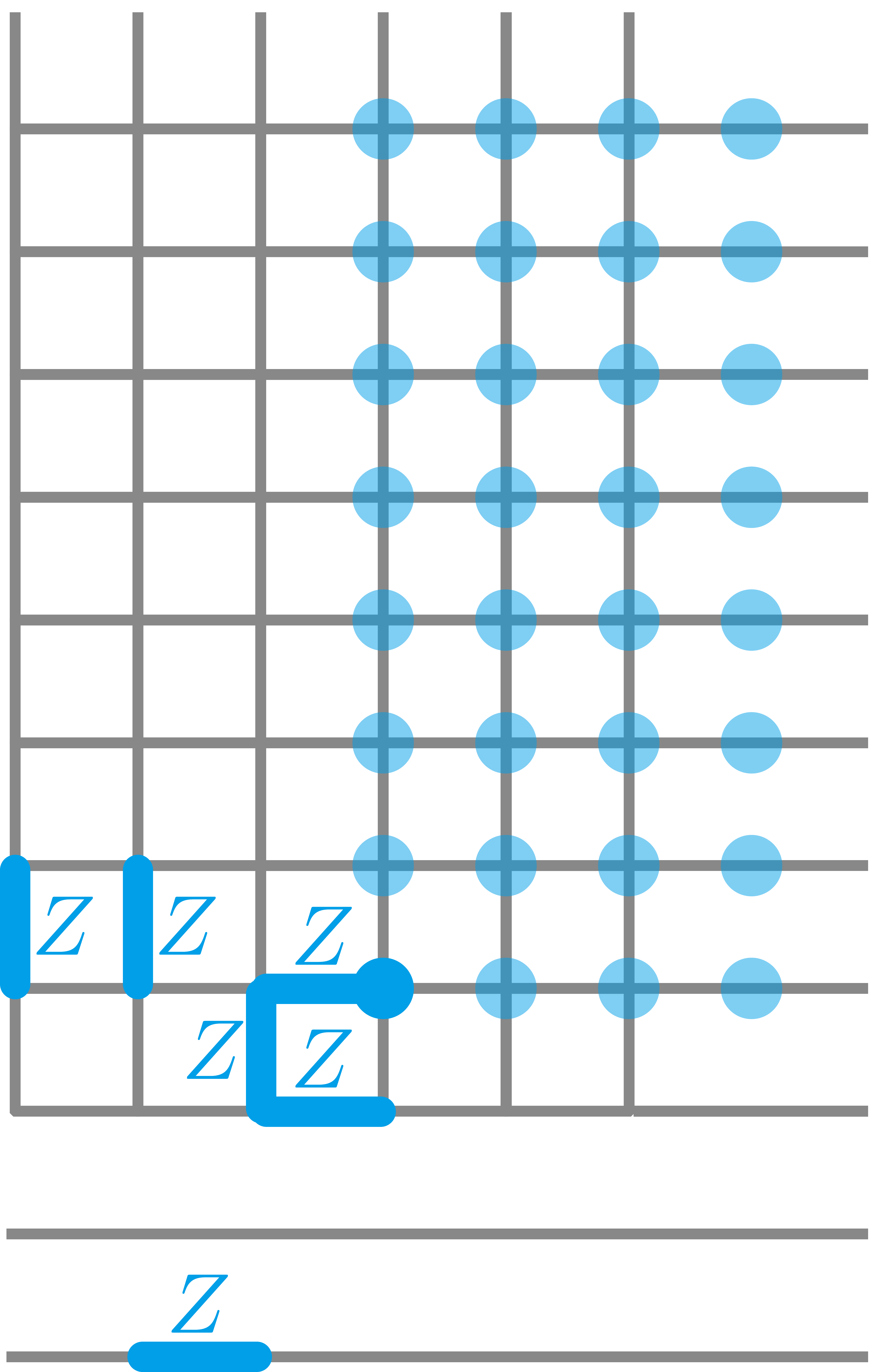}}\\
    \vspace{-0.1cm}
    \hspace{-0.3cm}
    \subfigure[~16 boundary $X$-stabilizers]{\includegraphics[scale=0.05]{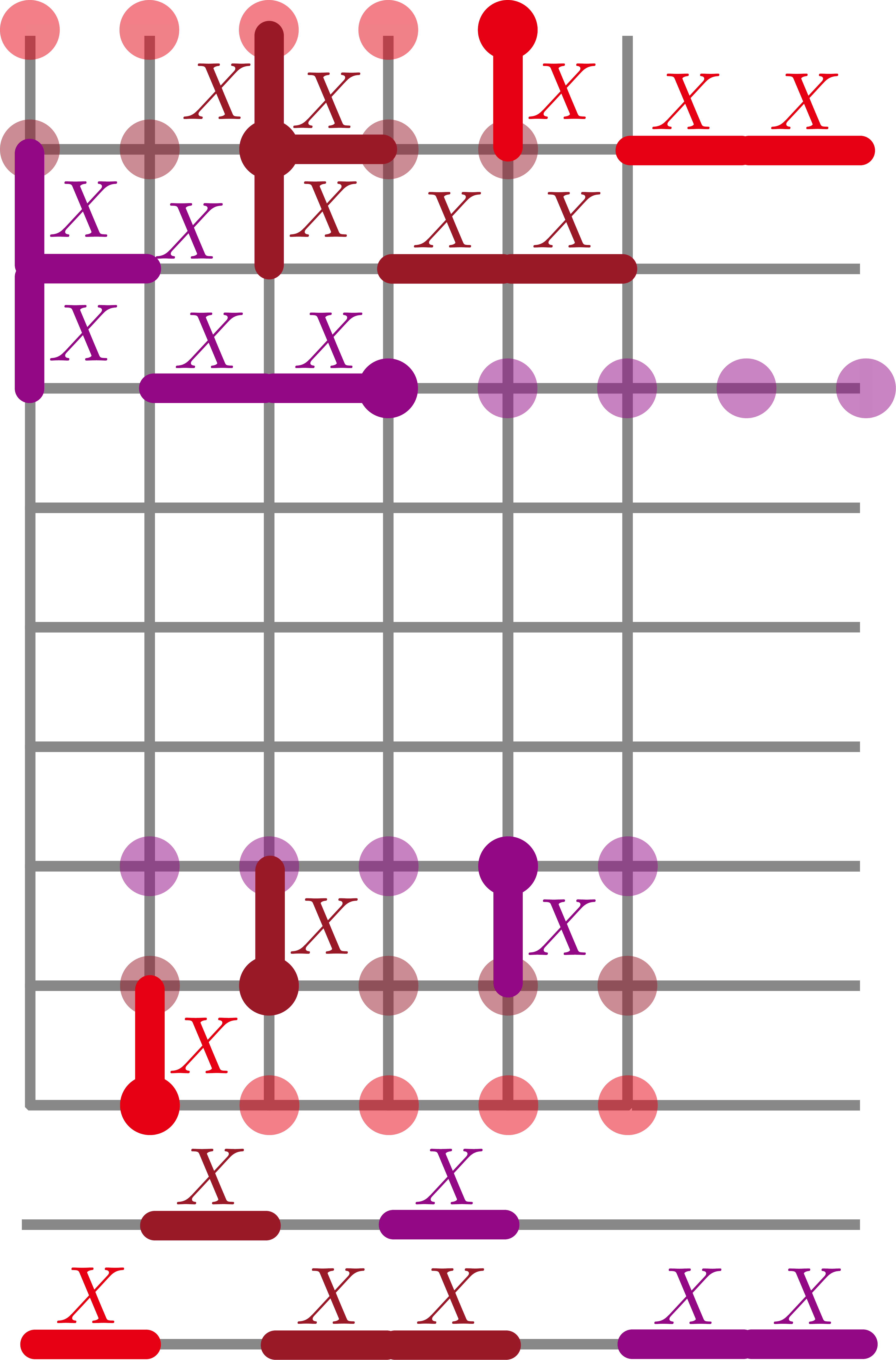}}
    \hspace{0.5cm}
    \subfigure[~24 boundary $Z$-stabilizers]{\includegraphics[scale=0.05]{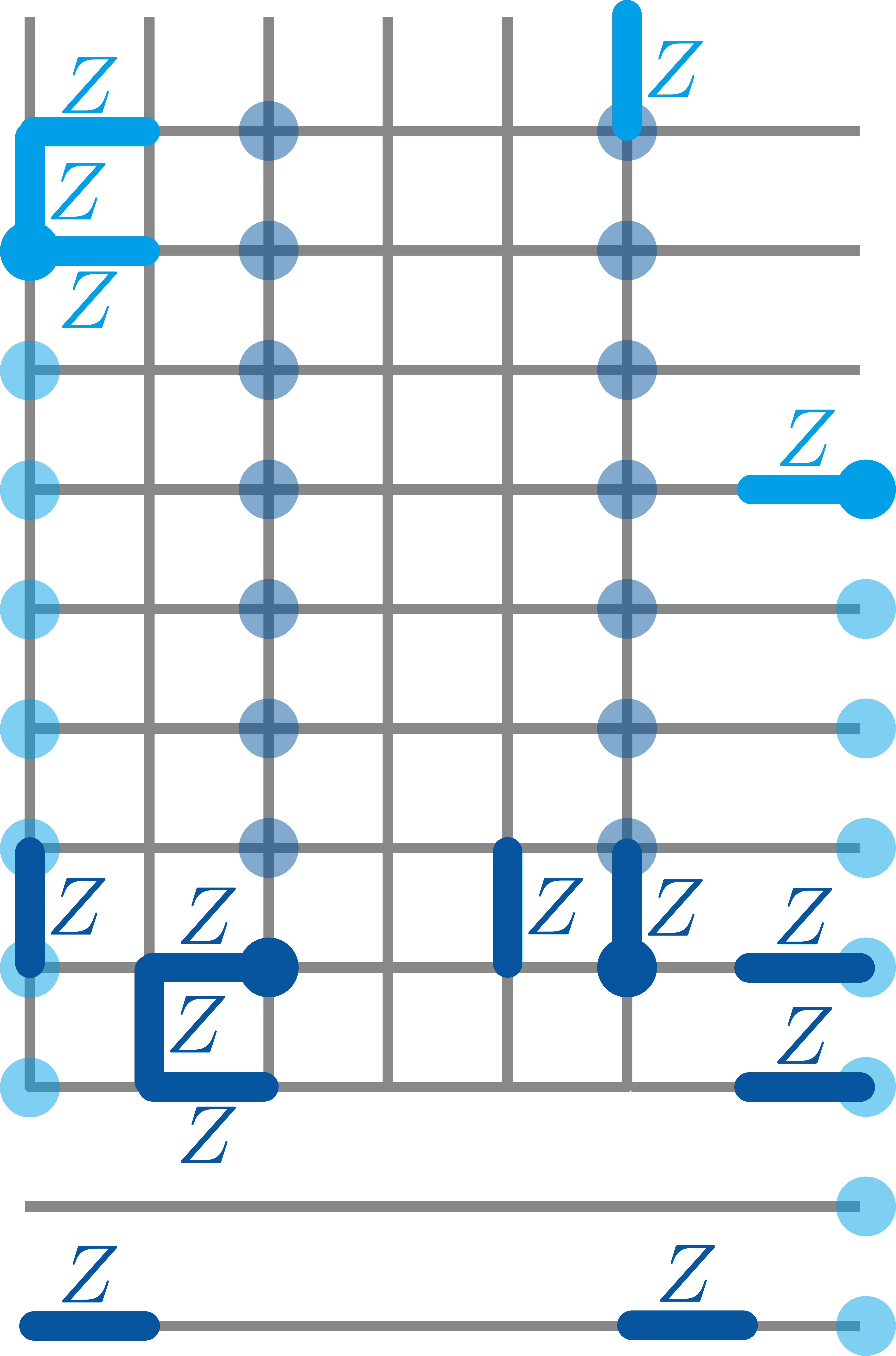}}
    \vspace{-0.3cm}
    \caption{The $[[131, 7, 7]]$ code on an open square lattice. The bulk stabilizers correspond to $f(x,y) =x(1+x+x^{-1}y)$ and $g(x,y) =1+y+xy^3$.
    (a) 30 bulk X-stabilizers. (b) 32 bulk $Z$-stabilizers. (c) 30 boundary $X$-stabilizers near the top and bottom boundaries.
    (d) 32 boundary $Z$-stabilizers appear near the left and right boundaries. The logical dimension is $ k = 131 -30-32-30-32=7$.
    This generates a code family of $k=7$: $[[2 L_x L_y -2L_x - L_y+2, ~6, ~d(L_x, L_y)]]$, with code distances $d(L_x, L_y)$ listed in Table~\ref{tab: d(Lx, Ly) for [[131,7,7]]} in Appendix~\ref{app: Families of planar qLDPC codes}.
    }
    \label{fig: [[131, 7, 7]] code}
\end{figure}

The $[[131, 7, 7]]$ code is implemented on a $7 \times 11$ square lattice, as shown in Fig.~\ref{fig: [[131, 7, 7]] code}. In this construction, the lattice dimensions $L_x$ and $L_y$ are adaptable. Note that some qubits along the rightmost column and the two bottom rows are removed, resulting in a total of
\begin{equation}
    n = 2L_x L_y - 2L_x - L_y + 2,
\end{equation}
physical qubits, while the logical dimension remains fixed at $k=7$.
The table of code distances $d(L_x, L_y)$ is provided in Table~\ref{tab: d(Lx, Ly) for [[131,7,7]]} in Appendix~\ref{app: Families of planar qLDPC codes}, with a few representative cases listed below:
\begin{enumerate}
    \item $L_x = 7$, $L_y = 11$: $[[131, 7, 7]]$, ~${kd^2}/{n} = 2.62$.
    \item $L_x = 8$, $L_y = 12$: $[[166, 7, 8]]$, ~${kd^2}/{n} = 2.70$.
    \item $L_x = 9$, $L_y = 13$: $[[205, 7, 10]]$, ~${kd^2}/{n} = 3.41$.
    \item $L_x = 10$, $L_y = 14$: $[[248, 7, 11]]$, ~${kd^2}/{n} = 3.42$.
    \item $L_x = 10$, $L_y = 15$: $[[267, 7, 12]]$, ~${kd^2}/{n} = 3.78$.
    \item $L_x = 11$, $L_y = 16$: $[[316, 7, 13]]$, ~${kd^2}/{n} = 3.74$.
    \item $L_x = 11$, $L_y = 17$: $[[337, 7, 14]]$, ~${kd^2}/{n} = 4.07$.
\end{enumerate}
For toric codes, the logical dimension $k$ must be an even integer~\cite{liang2025generalized}; therefore, constructing a planar code family with odd $k$ is noteworthy.

\begin{figure}[htb]
    \centering
    \hspace{-1.3cm}
    \subfigure[~24 bulk $X$-stabilizers]{\raisebox{-0.05cm}{\includegraphics[scale=0.055]{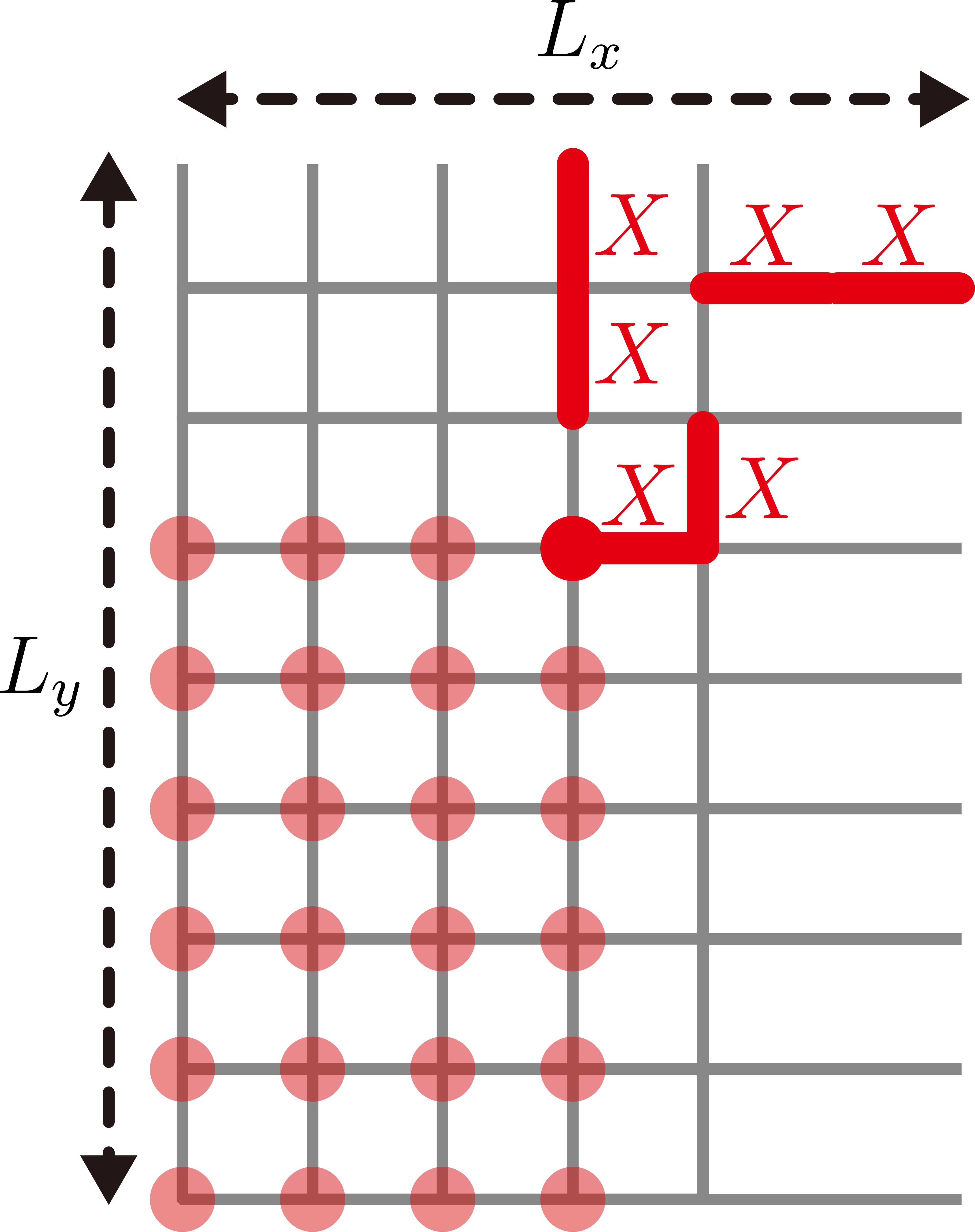}}}
    \hspace{0.5cm}
    \subfigure[~18 bulk $Z$-stabilizers]{\includegraphics[scale=0.055]{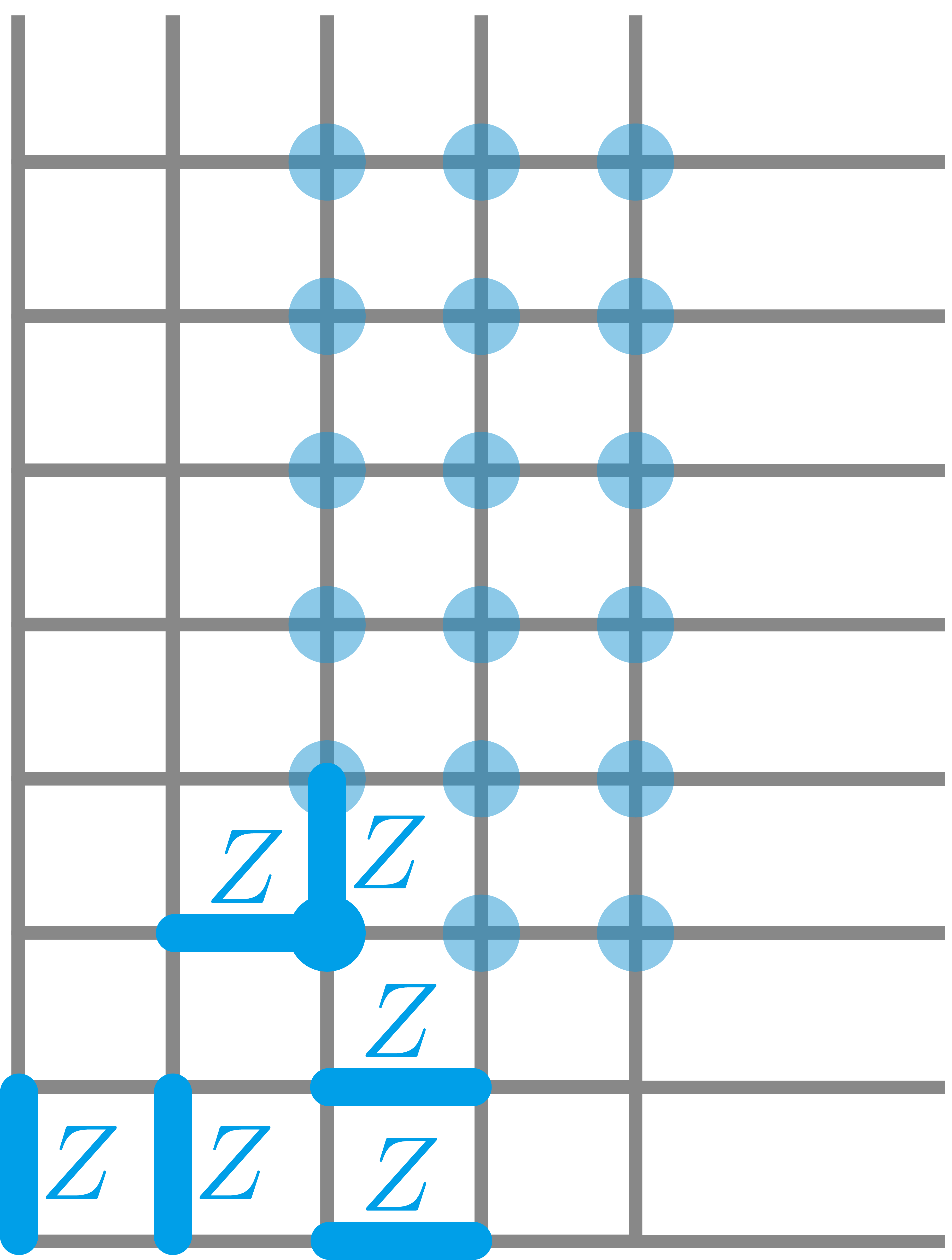}}\\
    \vspace{-0.1cm}
    \hspace{-0.3cm}
    \subfigure[~16 boundary $X$-stabilizers]{\includegraphics[scale=0.055]{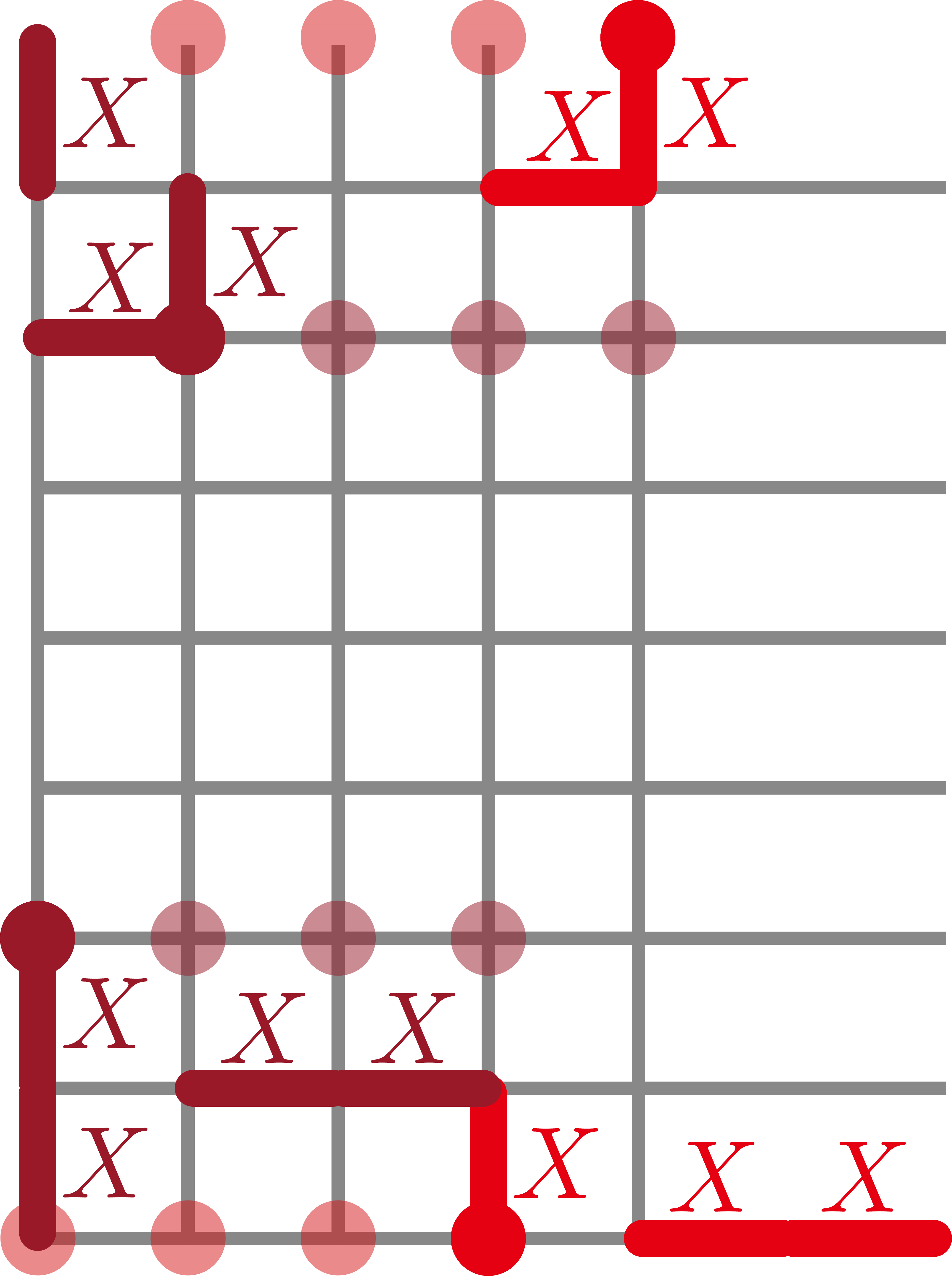}}
    \hspace{0.5cm}
    \subfigure[~24 boundary $Z$-stabilizers]{\includegraphics[scale=0.055]{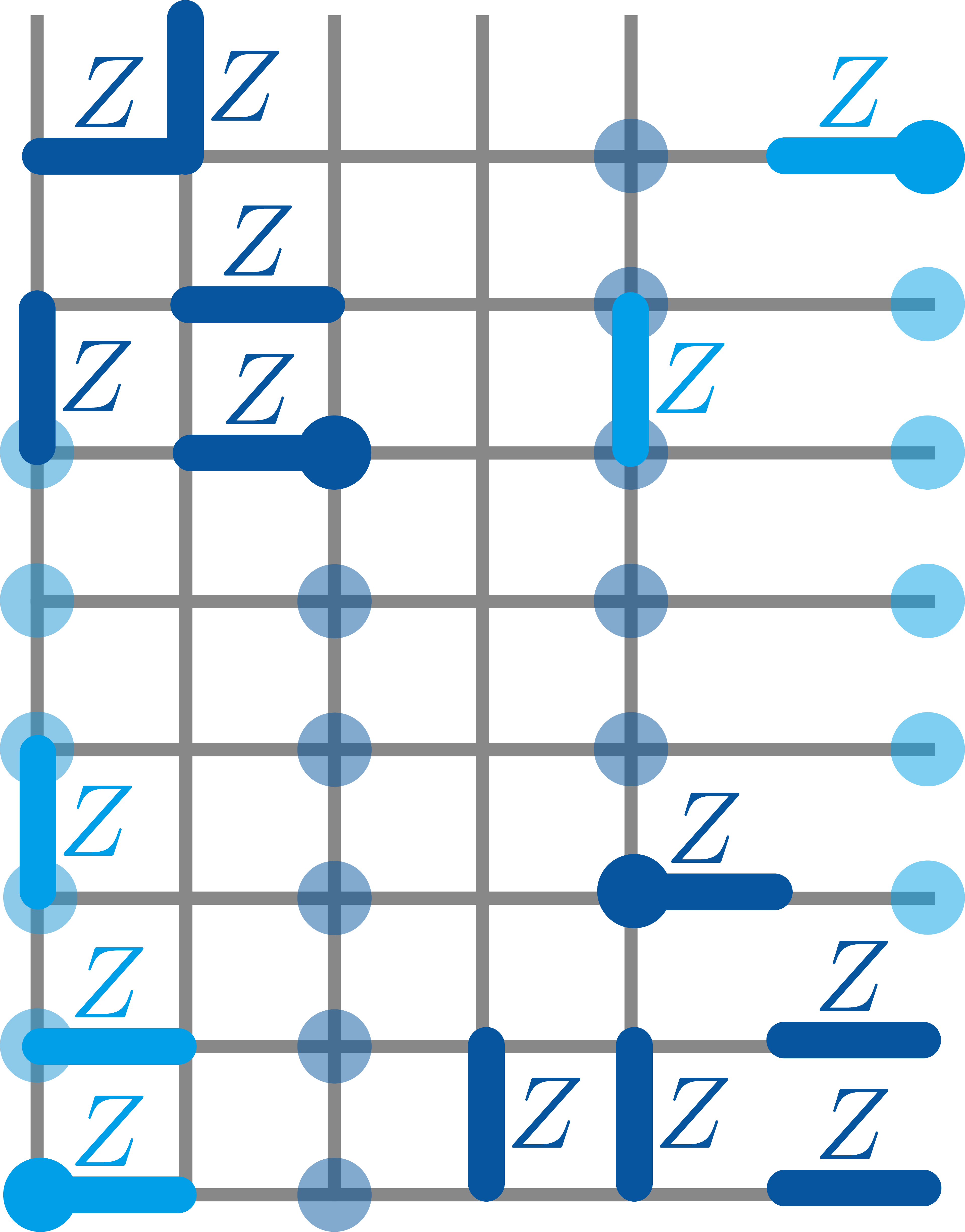}}
    \vspace{-0.3cm}
    \caption{The $[[88, 6, 6]]$ code on an open square lattice.
    The bulk stabilizers correspond to $f(x,y) = xy^2(1+x+x^{-1}y^{-2})$ and $g(x,y) = y(1+y+xy^{-1})$.
    (a) 24 bulk X-stabilizers. (b) 18 bulk $Z$-stabilizers. (c) 16 boundary $X$-stabilizers near the top and bottom boundaries.
    (d) 24 boundary $Z$-stabilizers appear near the left and right boundaries. The logical dimension is $ k = 88-24-18-16-24=6$.
    This generates a code family of $k=6$: $[[2 L_x L_y - L_y, ~6, ~d(L_x, L_y)]]$, with code distances $d(L_x, L_y)$ listed in Table~\ref{tab: d(Lx, Ly) for [[88,6,6]]} in Appendix~\ref{app: Families of planar qLDPC codes}.
    }
    \label{fig: [[88, 6, 6]] code}
\end{figure}

\begin{figure*}[h]
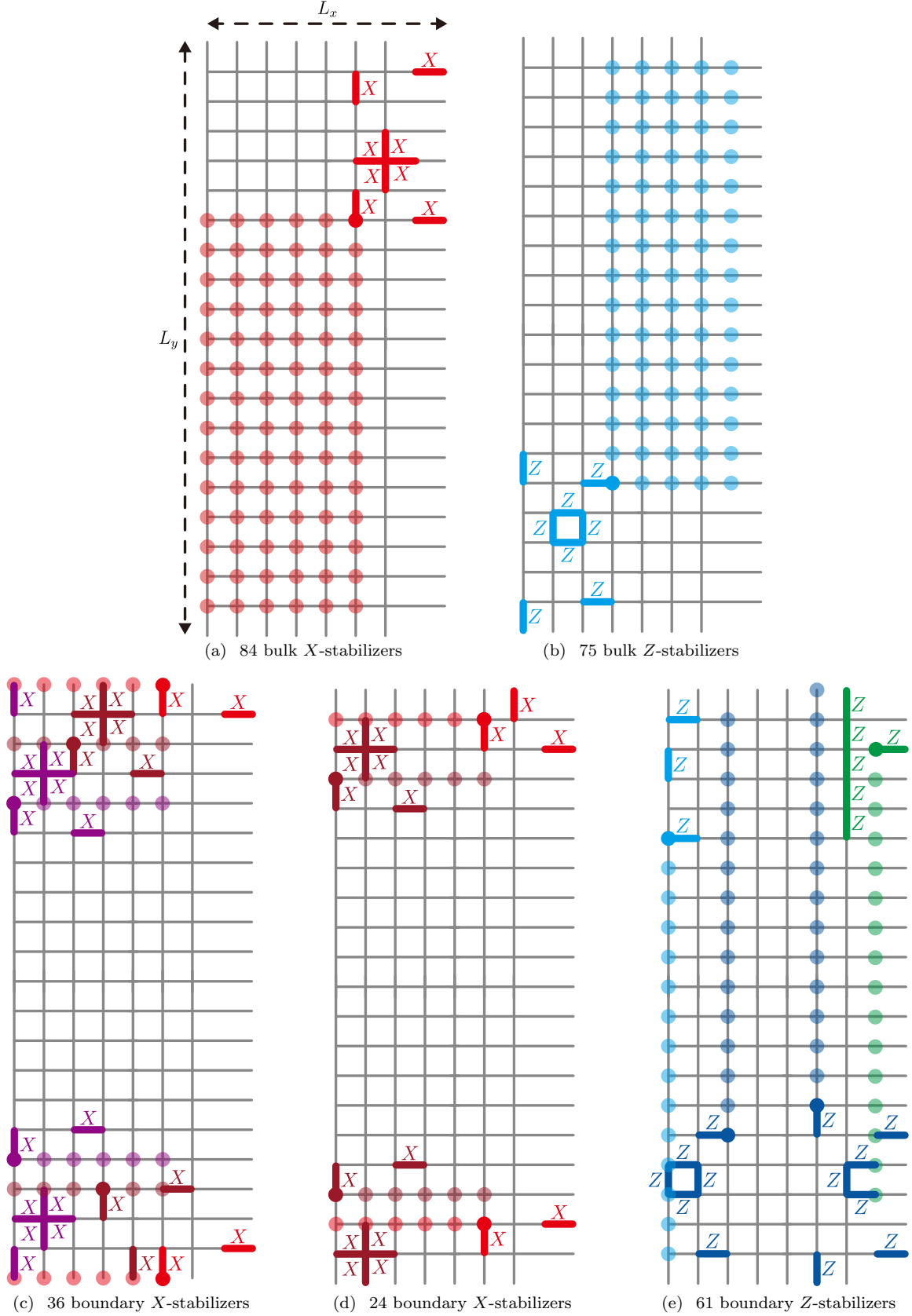

    \centering
    \vspace{-0.5cm}
    \hspace{0cm}
    \subfigure[~84 bulk $X$-stabilizers]{\raisebox{-0.05cm}{\includegraphics[scale=0.05]{X_bulk_292.pdf}}}
    \hspace{1cm}
    \subfigure[~75 bulk $Z$-stabilizers]{\includegraphics[scale=0.05]{Z_bulk_292.pdf}}\\
    \vspace{-0.1cm}
    \hspace{0cm}
    \subfigure[~36 boundary $X$-stabilizers]{\includegraphics[scale=0.05]{X_boundary_292_1.pdf}}
    \hspace{1cm}
    \subfigure[~24 boundary $X$-stabilizers]{\includegraphics[scale=0.05]{X_boundary_292_2.pdf}}
    \hspace{1cm}
    \subfigure[~61 boundary $Z$-stabilizers]{\includegraphics[scale=0.05]{Z_boundary_292.pdf}\label{fig:(e)secondary_boundary}}
    \vspace{-0.3cm}
    \caption{The $[[292, 12, 14]]$ code on an open square lattice.
    The bulk stabilizers correspond to $f(x,y) = y^2(1+x+x^2 y^3+x^2y^{-2}) $ and $g(x,y) = xy(1+y+x^{-1}y^{-1}+x^{-1}y^3)$.
    (a) 84 bulk X-stabilizers. (b) 75 bulk $Z$-stabilizers. (c) 36 boundary $X$-stabilizers near the top and bottom boundaries. (d) 24 boundary $X$-stabilizers near the top and bottom boundaries.
    (e) 61 boundary $Z$-stabilizers appear near the left and right boundaries; notably, the green terms cannot be obtained by truncating any bulk stabilizer.
    The logical dimension is $ k = 292-84-75-36-24-61=12$.
    This generates a code family of $k=12$: $[[2 L_x L_y - L_y, ~12, ~d(L_x, L_y)]]$, with code distances $d(L_x, L_y)$ listed in Table~\ref{tab: d(Lx, Ly) for [[292,12,14]]} in Appendix~\ref{app: Families of planar qLDPC codes}.
    }
    \label{fig: [[292, 12, 14]] code}
\end{figure*}

\subsection{$[[88,6,6]]$ planar code}\label{sec: [[88, 6, 6]] planar code}

Fig.~\ref{fig: [[88, 6, 6]] code} illustrates the $[[88,6,6]]$ planar code implemented on a $6 \times 8$ square lattice. Since the qubits along the right boundary are decoupled, the number of physical qubits is given by
\begin{equation}
    n = 2L_xL_y - L_y.
\end{equation}
Within this family, the logical dimension remains fixed at $k=6$.
Table~\ref{tab: d(Lx, Ly) for [[88,6,6]]} in Appendix~\ref{app: Families of planar qLDPC codes} details the distances $d(L_x, L_y)$, with several representative cases listed below:
\begin{enumerate}
    \item $L_x = 6$, $L_y = 8$: $[[88, 6, 6]]$, ~${kd^2}/{n} = 2.45$.
    \item $L_x = 8$, $L_y = 10$: $[[150, 6, 8]]$, ~${kd^2}/{n} = 2.56$.
    \item $L_x = 8$, $L_y = 11$: $[[165, 6, 9]]$, ~${kd^2}/{n} = 2.95$.
    \item $L_x = 10$, $L_y = 12$: $[[228, 6, 11]]$, ~${kd^2}/{n} = 3.18$.
    \item $L_x = 10$, $L_y = 13$: $[[247, 6, 12]]$, ~${kd^2}/{n} = 3.50$.
\end{enumerate}
Compared to the $[[188, 8, 9]]$ code family, for a fixed distance $d$, the number of physical qubits $n$ is slightly lower in this family; however, the logical dimension decreases from 8 to 6. Consequently, except for the $[[88,6,6]]$ code itself, the remaining members of this family exhibit suboptimal performance metrics $kd^2/n$.

\subsection{$[[292,12,14]]$ planar code (weight-8 stabilizers)}
\label{sec: [[292, 12, 14]] planar code}

Previously, we focused on weight-6 stabilizers and characterized broad families of planar qLDPC codes.
The same methodology extends to weight-8 stabilizers and can yield planar qLDPC codes with improved performance.
However, the search space grows exponentially in stabilizer weight, rendering an exhaustive examination for weight-8 codes impractical.  To manage complexity, we therefore constrain our search to codes with fixed logical dimension $k=12$, which leads to the discovery of a notable code family (see Fig.~\ref{fig: [[292, 12, 14]] code}).
Importantly, the boundary stabilizers in this family are not mere truncations of bulk operators: the green terms in Fig.~\ref{fig:(e)secondary_boundary} are secondary boundary gauge operators~\cite{liang2024operator}.

A representative member of this family is the $[[292,12,14]]$ code on an $8\times20$ square lattice. In the lattice configuration, the rightmost column of qubits decouples, so the total number of physical qubits is
\begin{equation}
    n = 2L_xL_y - L_y.
\end{equation}
Across this family, the logical dimension remains fixed at $k=12$.

Table~\ref{tab: d(Lx, Ly) for [[292,12,14]]} in Appendix~\ref{app: Families of planar qLDPC codes} details the distances $d(L_x, L_y)$, with several representative cases listed below:
\begin{enumerate}
    \item $L_x = 7$, $L_y = 18$: $[[227, 12, 11]]$, ~${kd^2}/{n} = 6.40$.
    \item $L_x = 7$, $L_y = 19$: $[[240, 12, 12]]$, ~${kd^2}/{n} = 7.20$.
    \item $L_x = 8$, $L_y = 20$: $[[292, 12, 14]]$, ~${kd^2}/{n} = 8.05$.
    \item $L_x = 9$, $L_y = 22$: $[[365, 12, 16]]$, ~${kd^2}/{n} = 8.42$.
    \item $L_x = 9$, $L_y = 23$: $[[382, 12, 17]]$, ~${kd^2}/{n} = 9.08$.
    \item $L_x = 9$, $L_y = 24$: $[[399, 12, 18]]$, ~${kd^2}/{n} = 9.74$.
\end{enumerate}

\section{Discussion}\label{sec: discussion}

In this section, we explore several intriguing properties and aspects of our construction that are worth further investigation.

\subsection{Bulk stabilizers of optimal code families} \label{sec:observationsonbulstabilizers}

We have presented the optimal families with logical dimensions ranging from $k=6$ to $k=13$. Each family is determined independently since the logical dimension is defined solely by the bulk stabilizers, and families with different $k$ values are not directly compared in Table~\ref{tab: GTC large weight and stabilizer}. Consequently, we did not expect any relation among the bulk stabilizers of different code families. However, our numerical results reveal that the optimal families for different $k$ share similar stabilizers.
For instance, the $k=6$ family corresponds to the bulk stabilizers
\begin{equation*}
    S^{k=6}_1 =\vcenter{\hbox{\includegraphics[scale=.125]{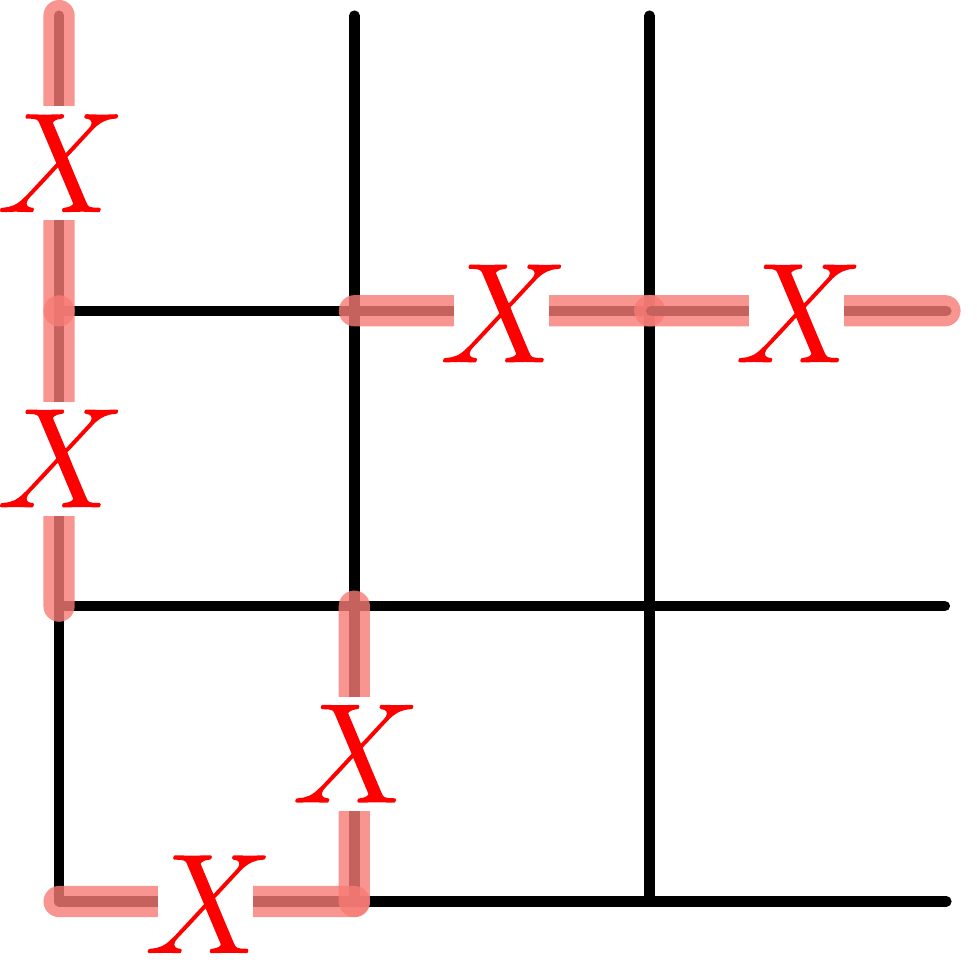}}},~
    S^{k=6}_2= \vcenter{\hbox{\includegraphics[scale=.125]{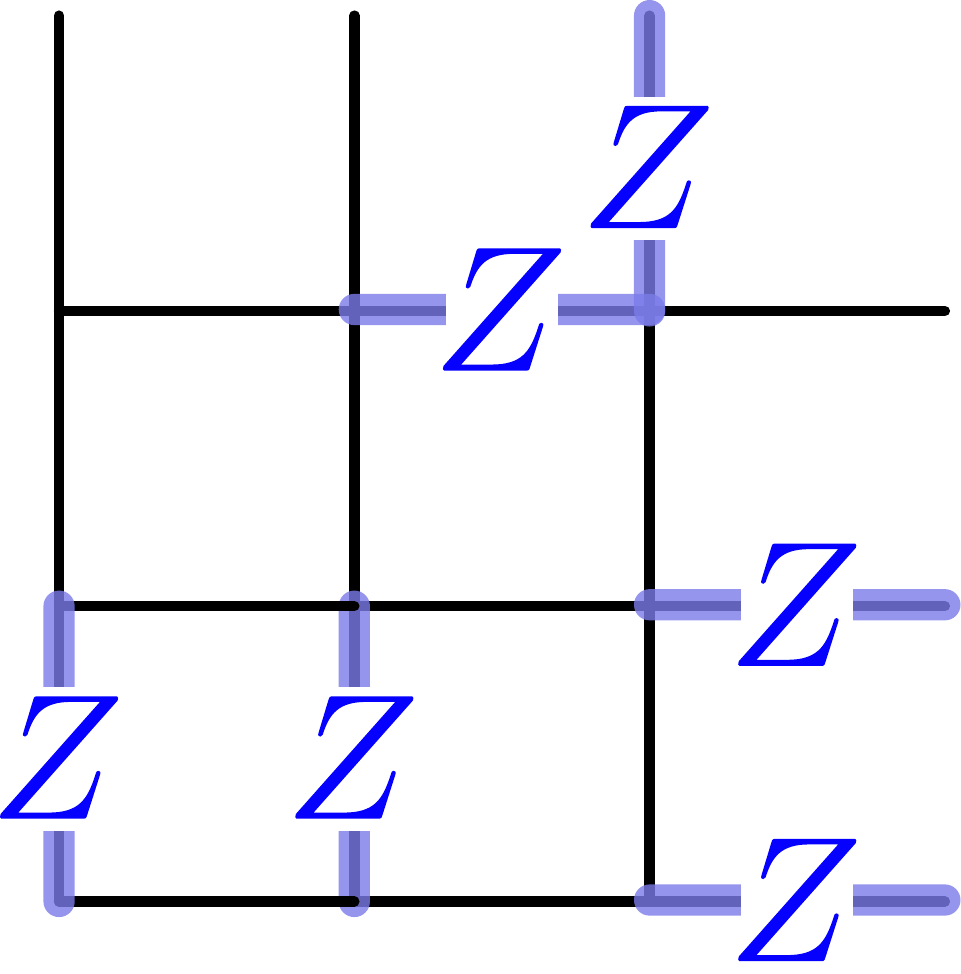}}},
\end{equation*}
while the $k=11$ code family is described by
\begin{equation*}
    S^{k=11}_1 =\vcenter{\hbox{\includegraphics[scale=.125]{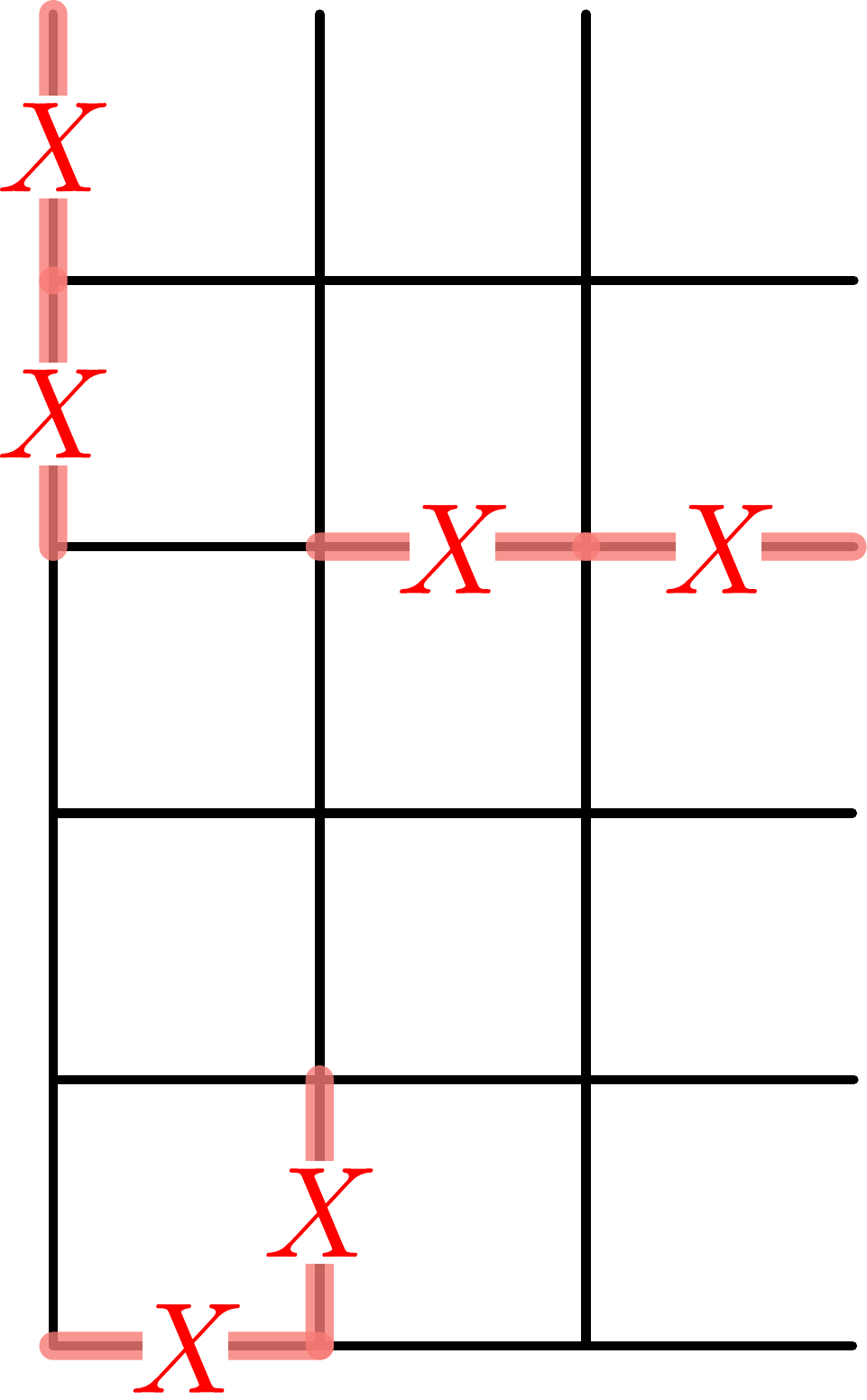}}},~
    S^{k=11}_2 =\vcenter{\hbox{\includegraphics[scale=.125]{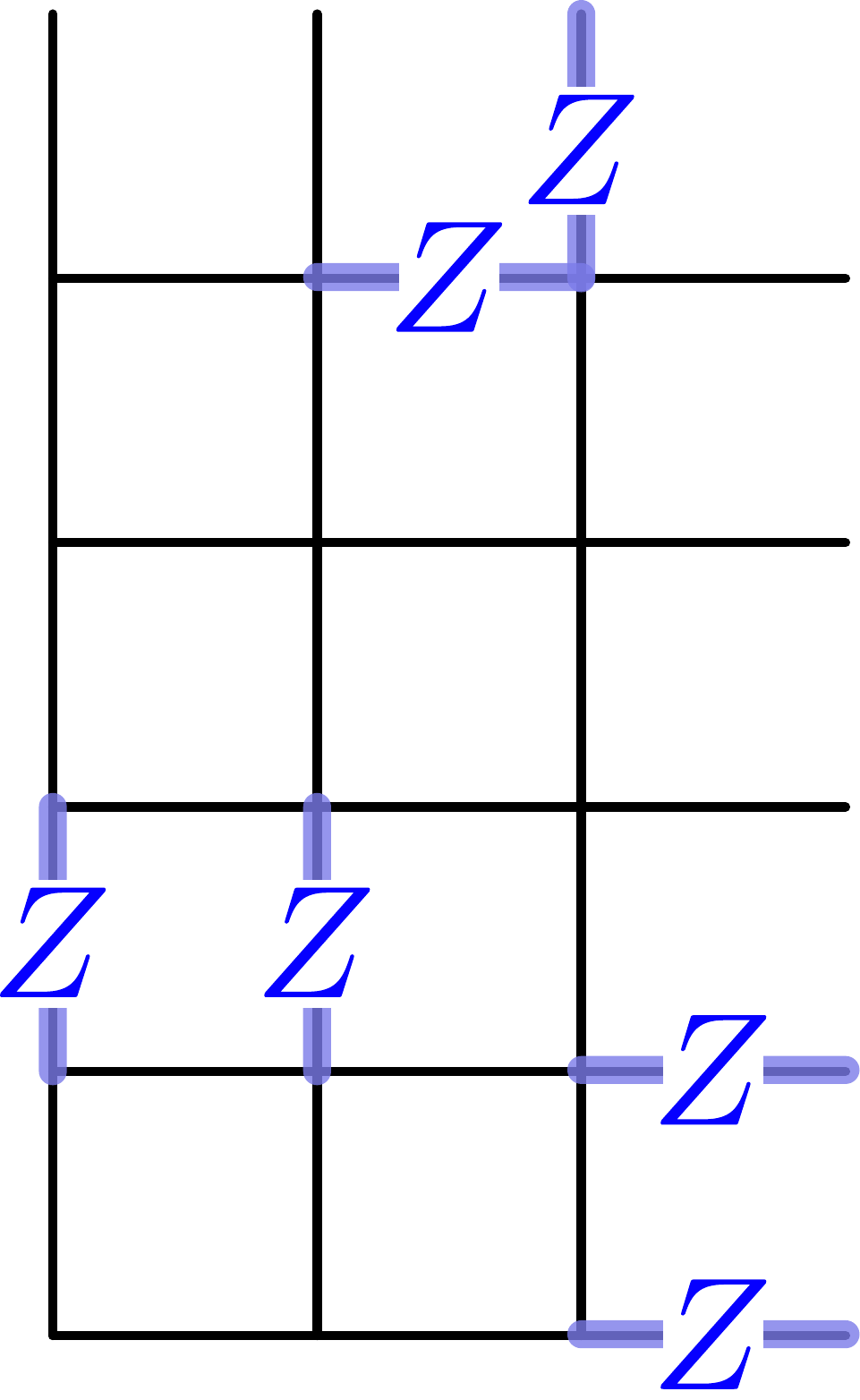}}}.
\end{equation*}
Remarkably, $S^{k=11}_1$ and $S^{k=11}_2$ can be interpreted as a vertical extension of $S^{k=6}_1$ and $S^{k=6}_2$, where each part of the stabilizer is shifted farther apart.

Furthermore, we compare the bulk stabilizers for the $k=7$, $k=8$ ($[[188, 8, 9]]$), $k=9$ ($[[441, 9, 15]]$), $k=10$ ($[[381, 10, 13]]$), $k=12$ ( $[[432, 12, 12]]$), and $k=13$ ($[[392, 13, 11]]$) families:
\begin{equation*}
S^{k=7}_1 =\vcenter{\hbox{\includegraphics[scale=.125]{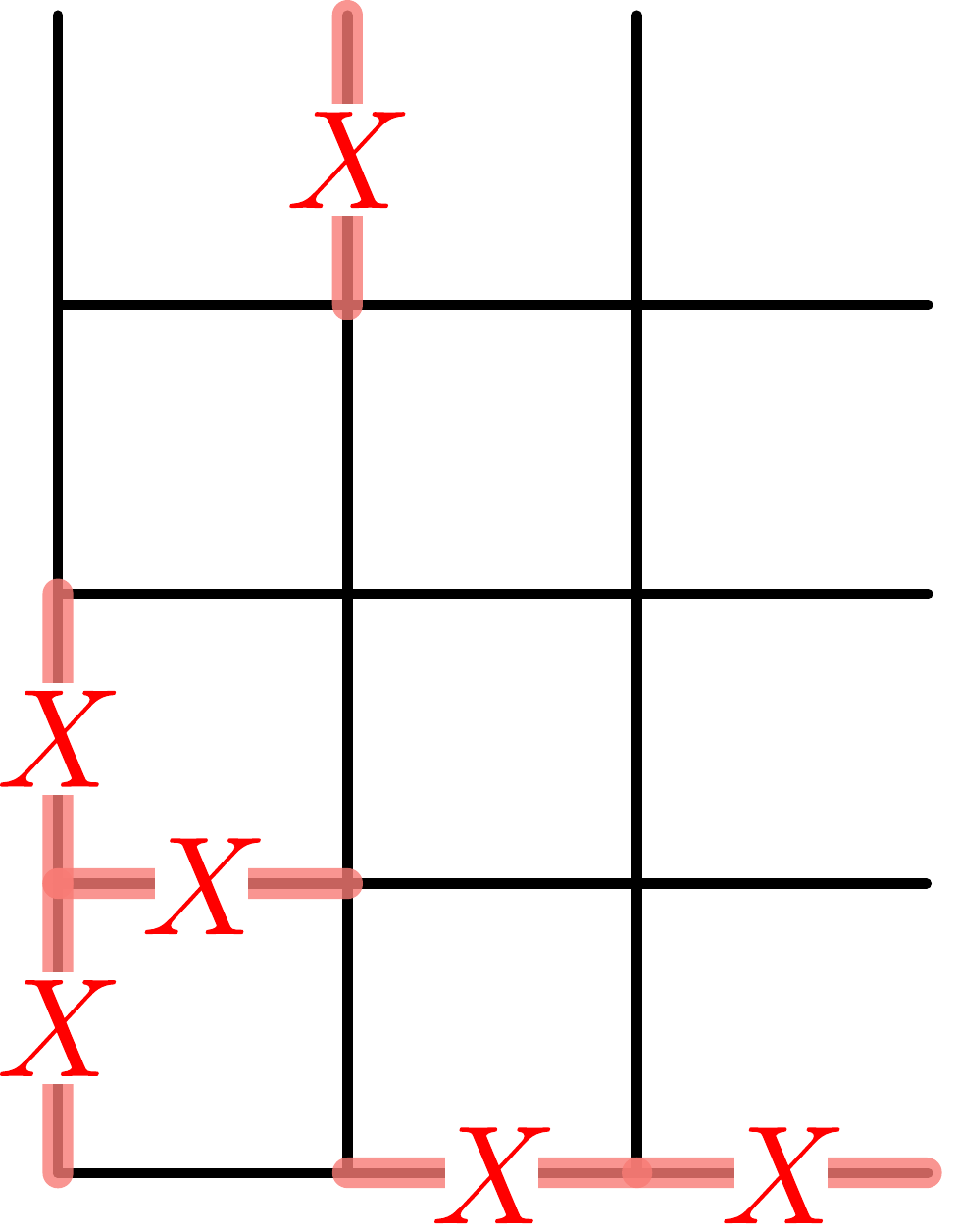}}},~
S^{k=7}_2 =\vcenter{\hbox{\includegraphics[scale=.125]{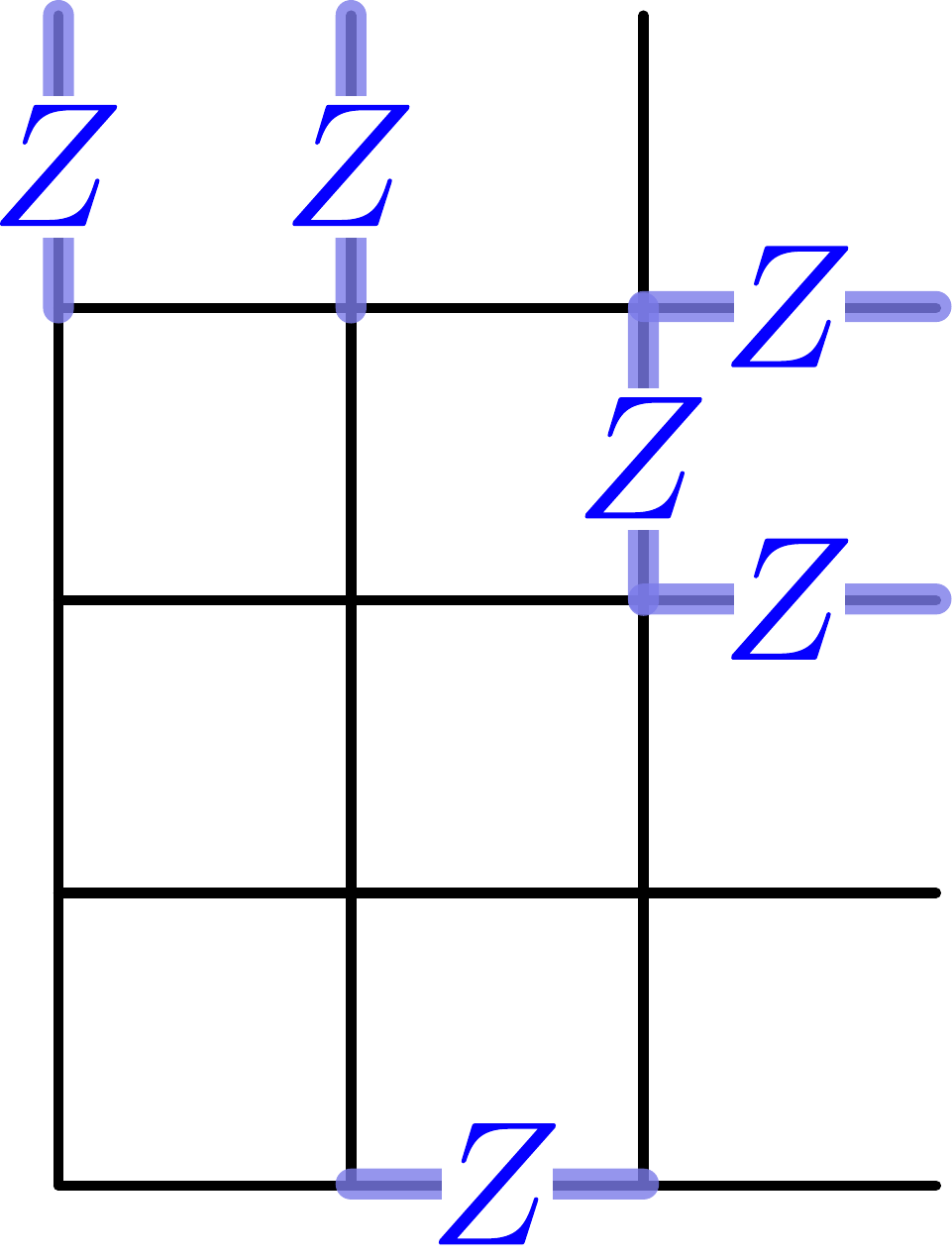}}},
\end{equation*}
\begin{equation*}
S^{k=8}_1 =\vcenter{\hbox{\includegraphics[scale=.125]{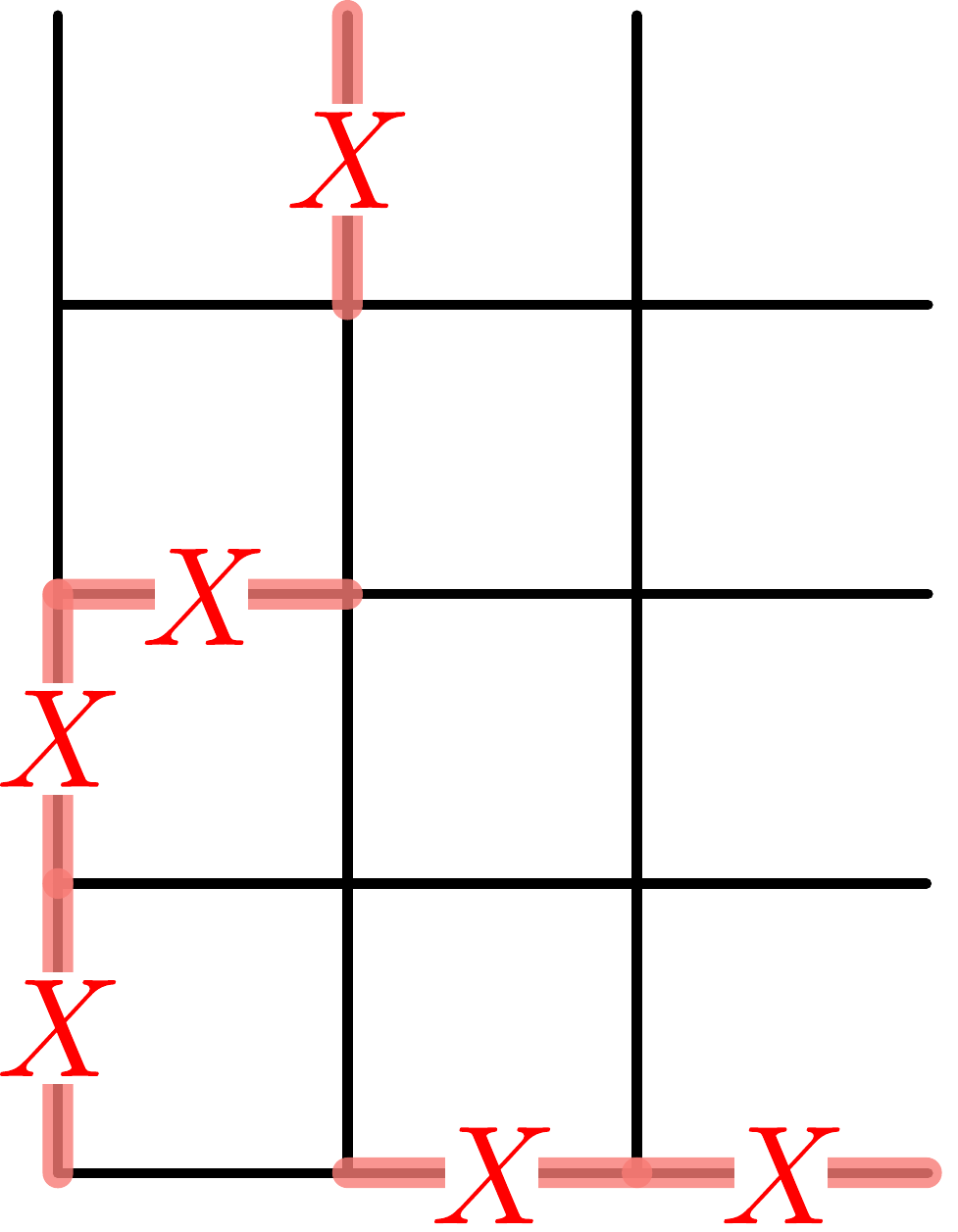}}},~
S^{k=8}_2 =\vcenter{\hbox{\includegraphics[scale=.125]{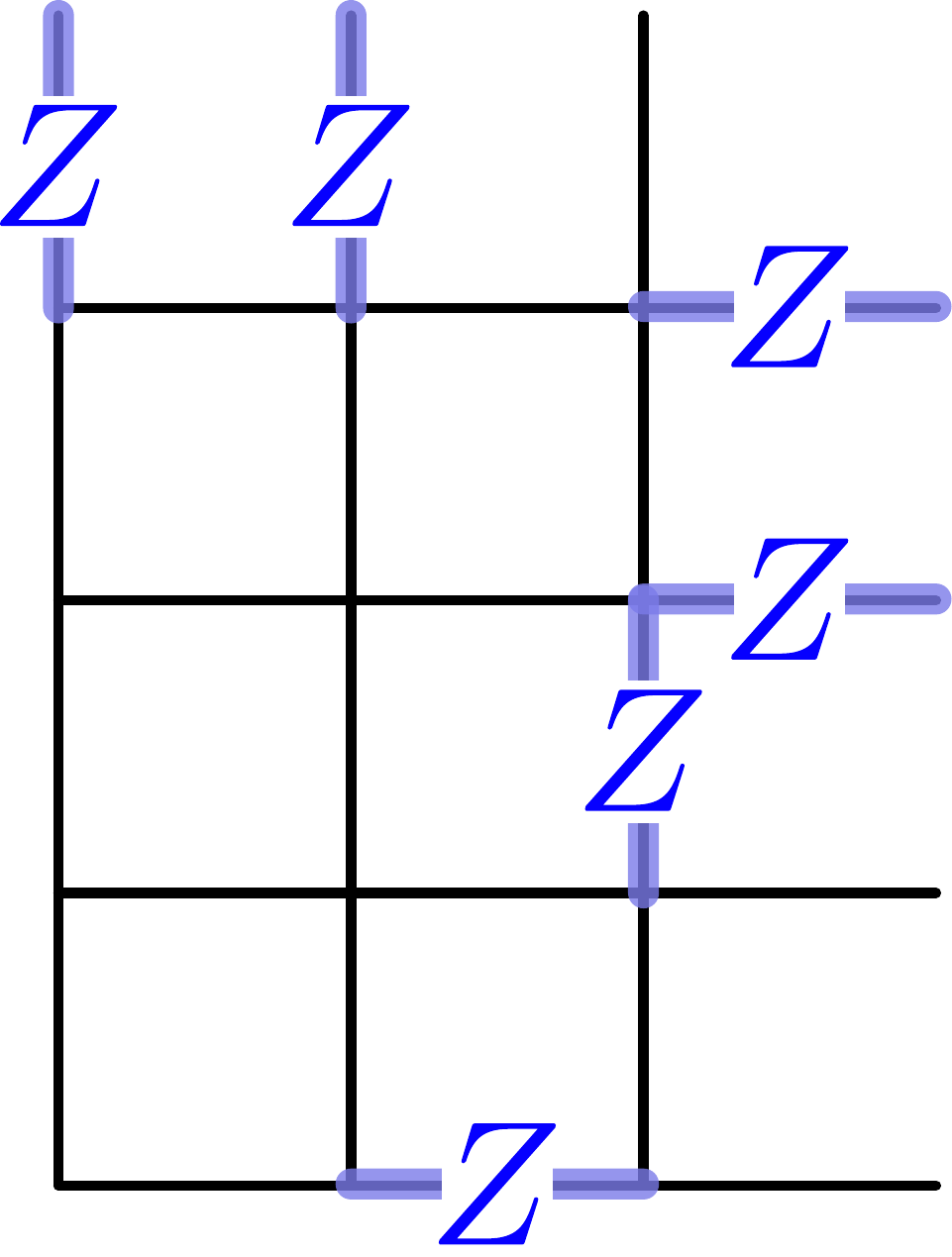}}},
\end{equation*}
\begin{equation*}
S^{k=9}_1 =\vcenter{\hbox{\includegraphics[scale=.125]{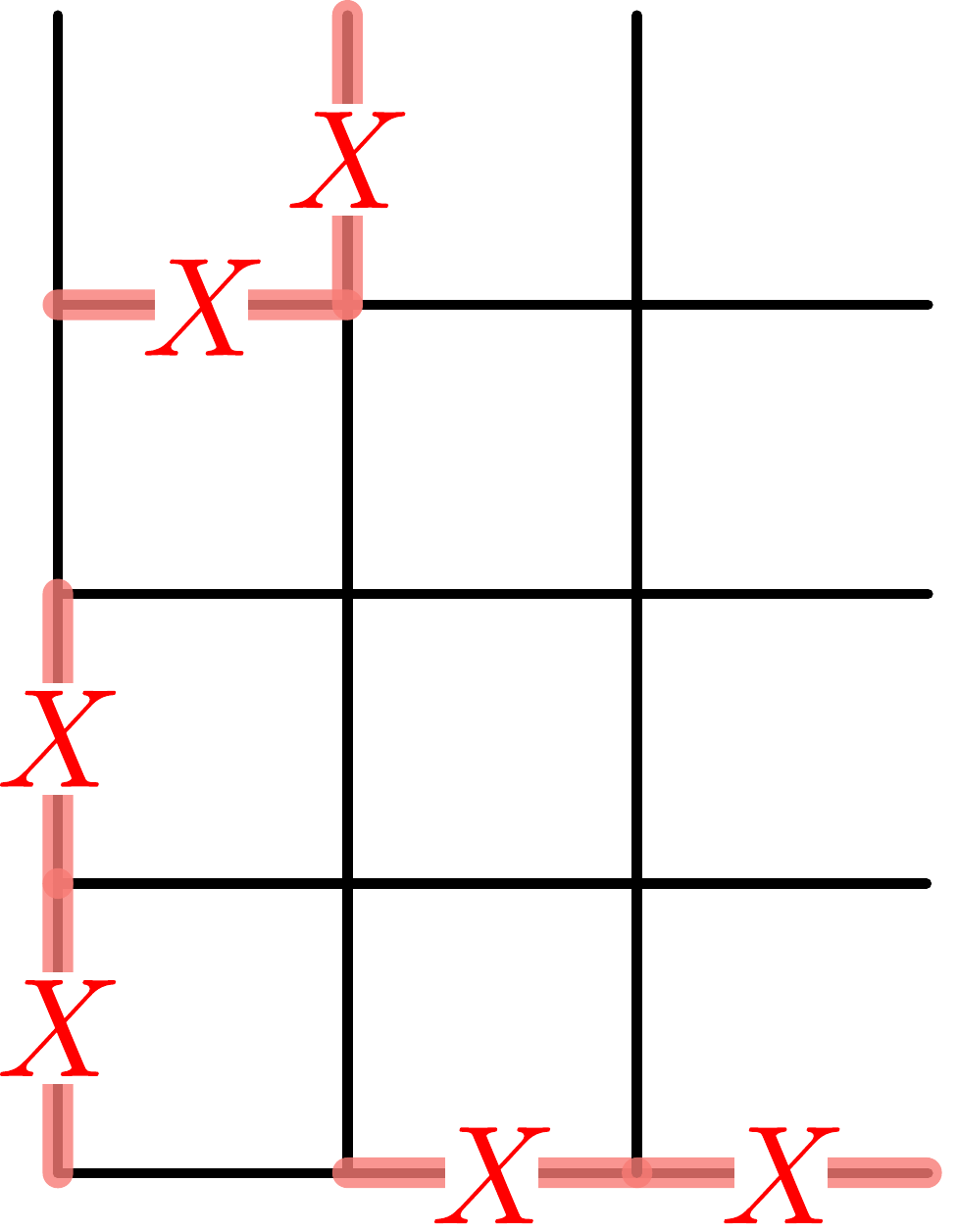}}},~
S^{k=9}_2 =\vcenter{\hbox{\includegraphics[scale=.125]{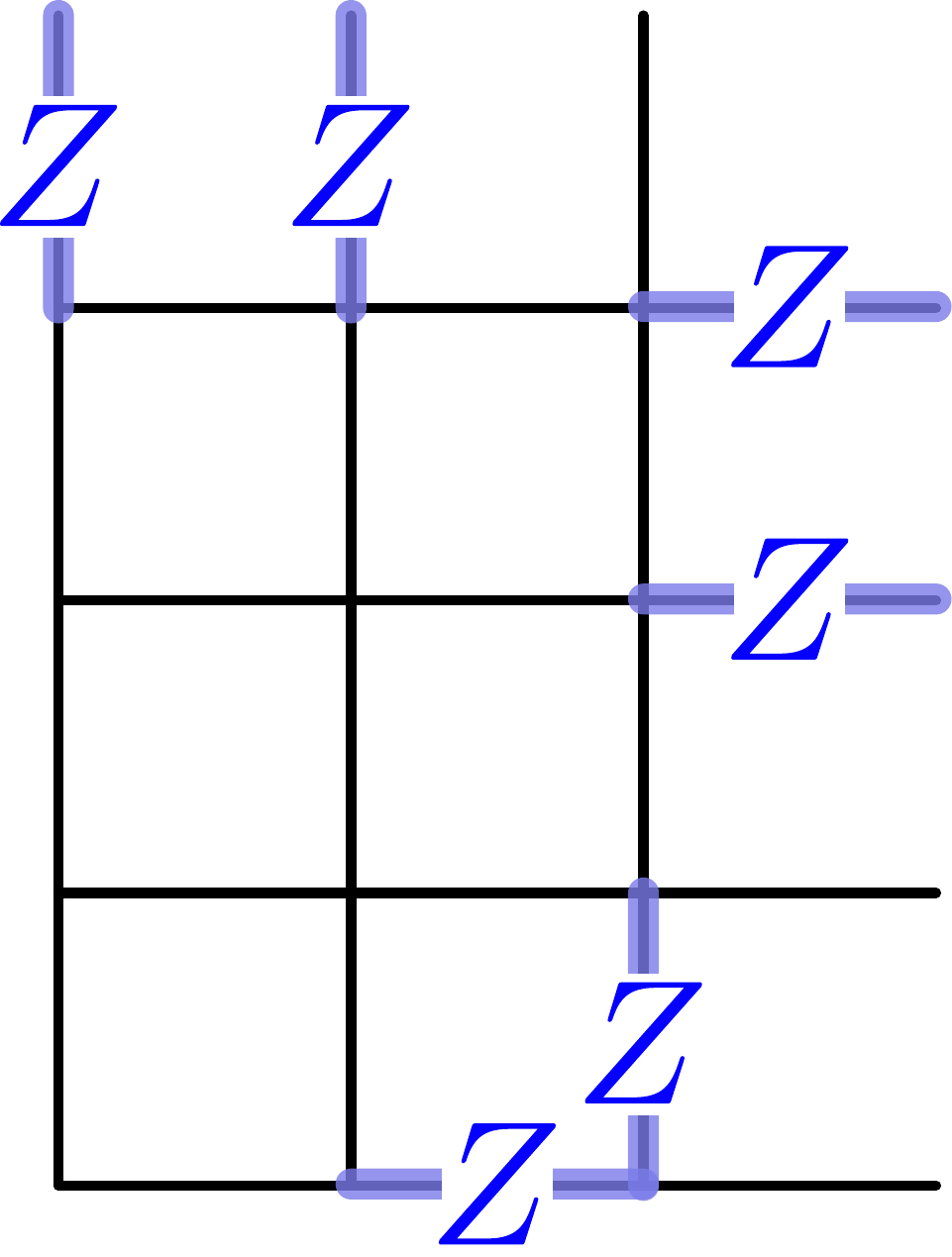}}},
\end{equation*}
\begin{equation*}
    S^{k=10}_1 =
    \vcenter{\hbox{\includegraphics[scale=.125]{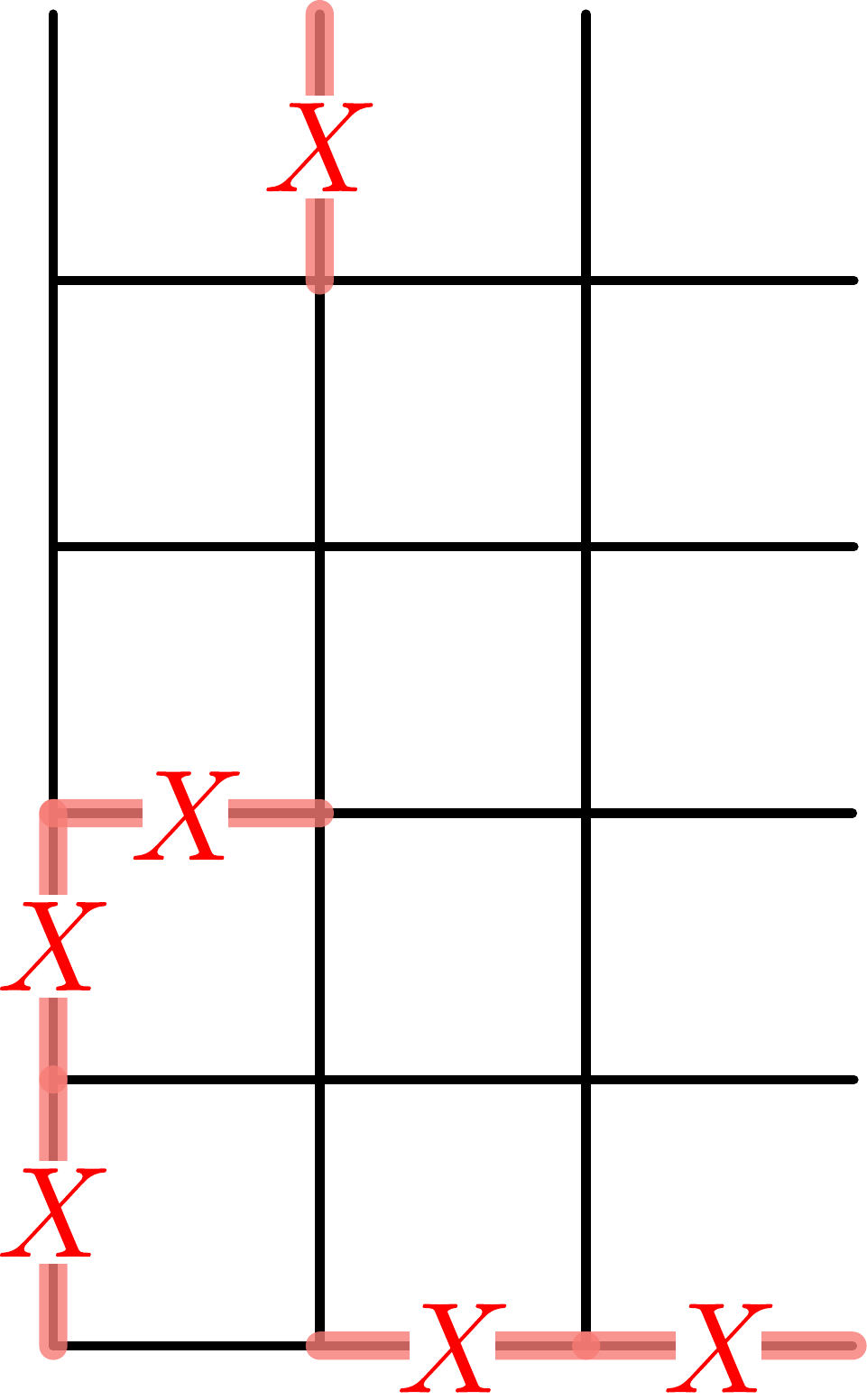}}},~
    S^{k=10}_2 =\vcenter{\hbox{\includegraphics[scale=.125]{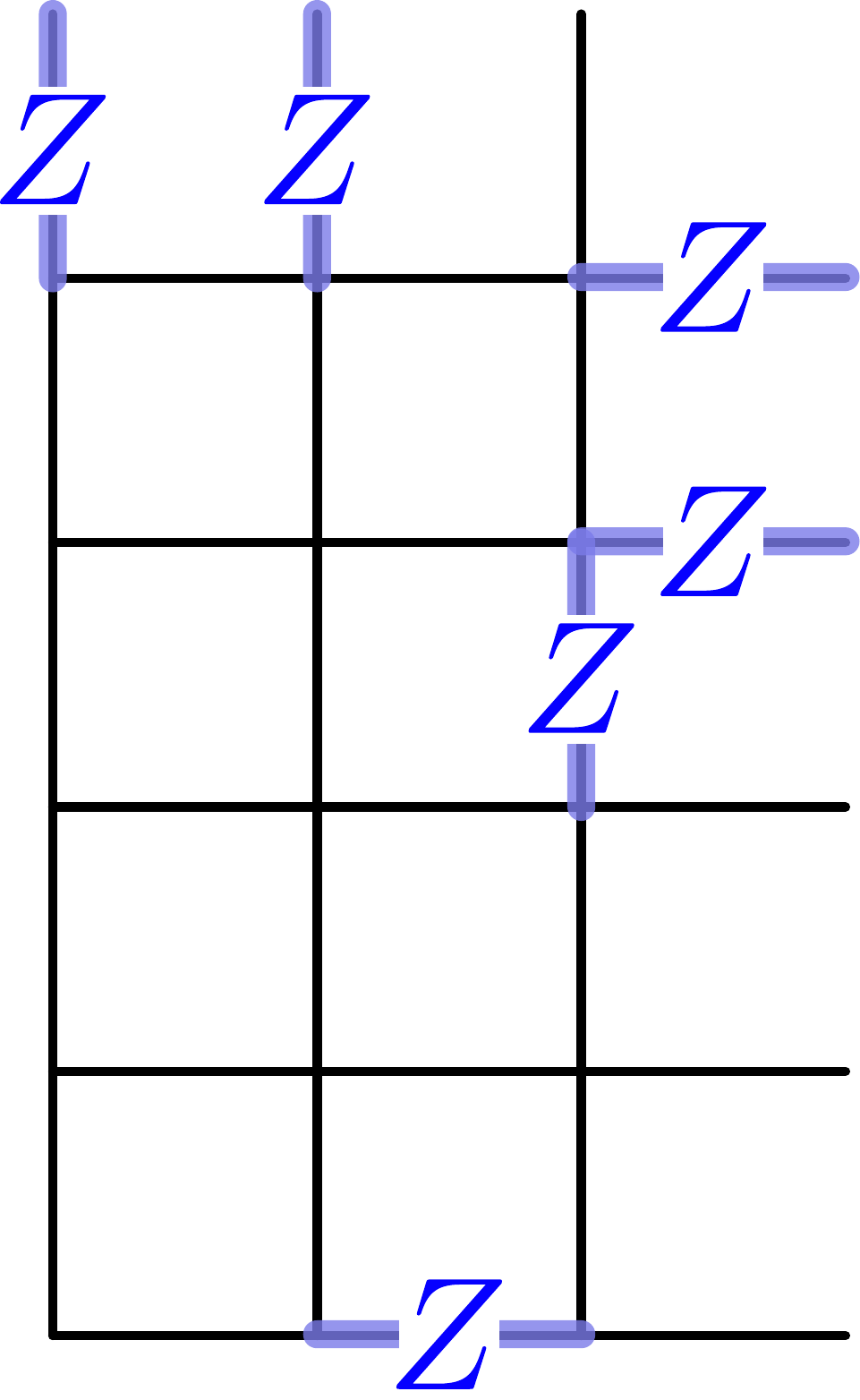}}},
\end{equation*}
\begin{equation*}
S^{k=12}_1 =\vcenter{\hbox{\includegraphics[scale=.125]{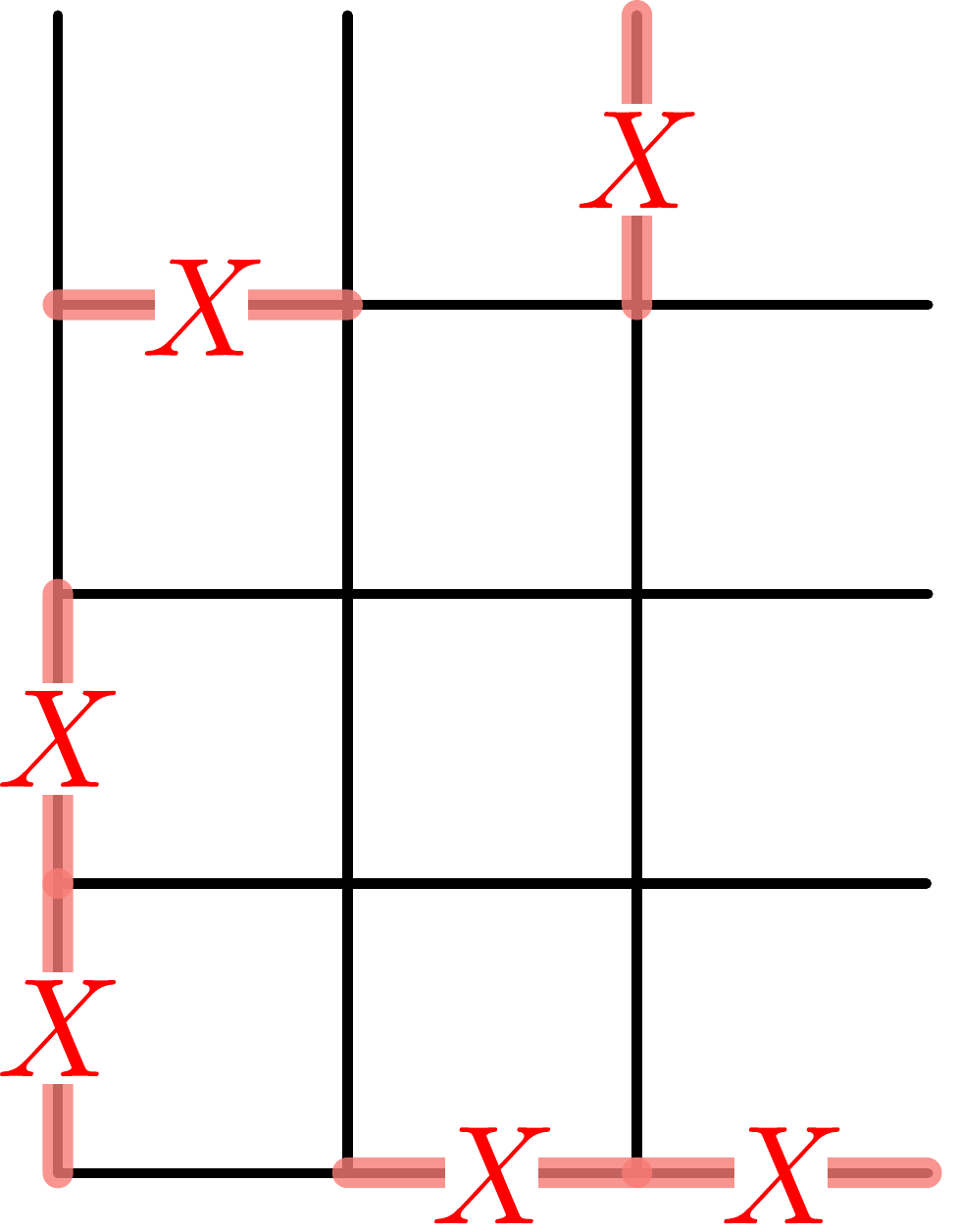}}},~
S^{k=12}_2 =\vcenter{\hbox{\includegraphics[scale=.125]{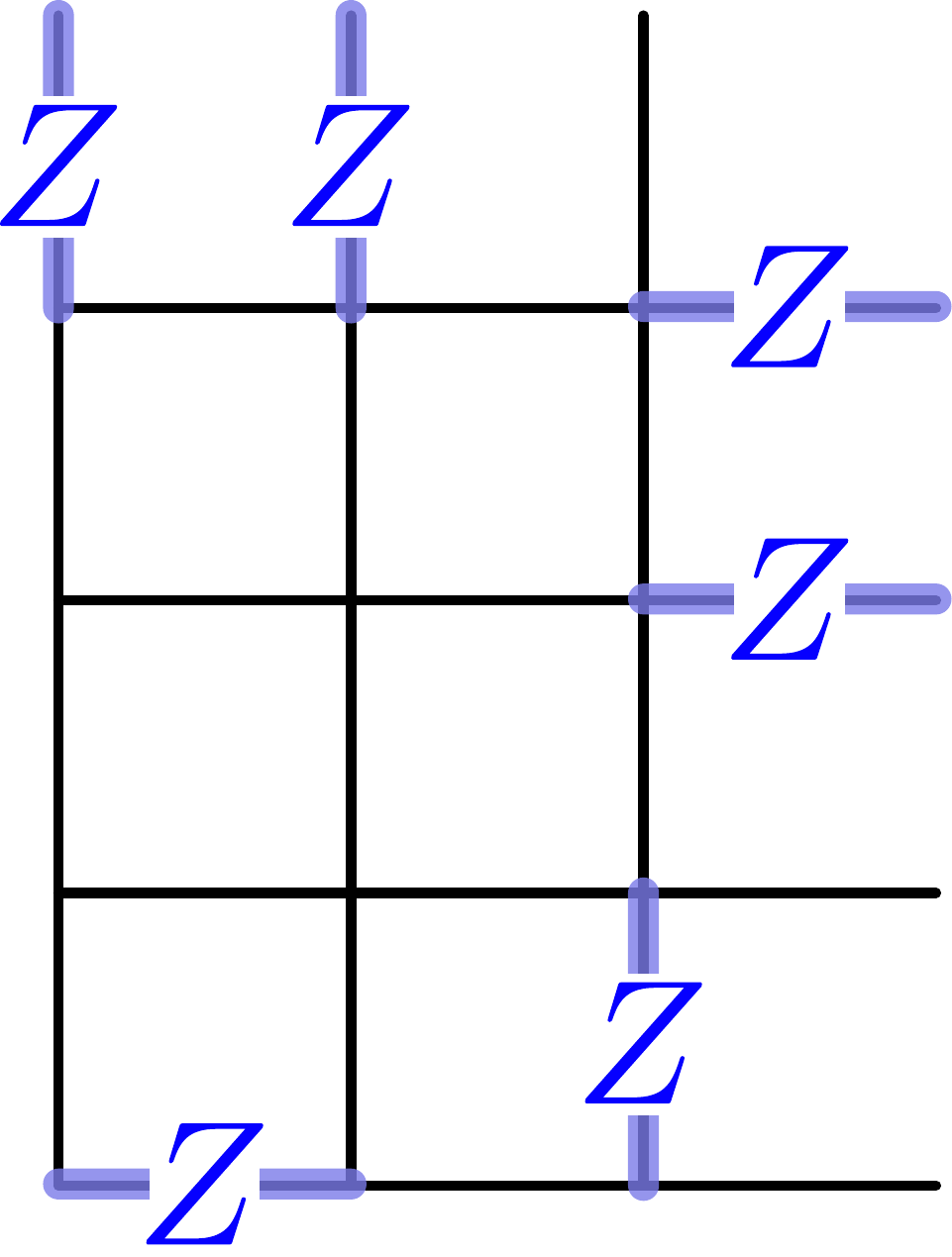}}},
\end{equation*}
\begin{equation*}
S^{k=13}_1 =\vcenter{\hbox{\includegraphics[scale=.125]{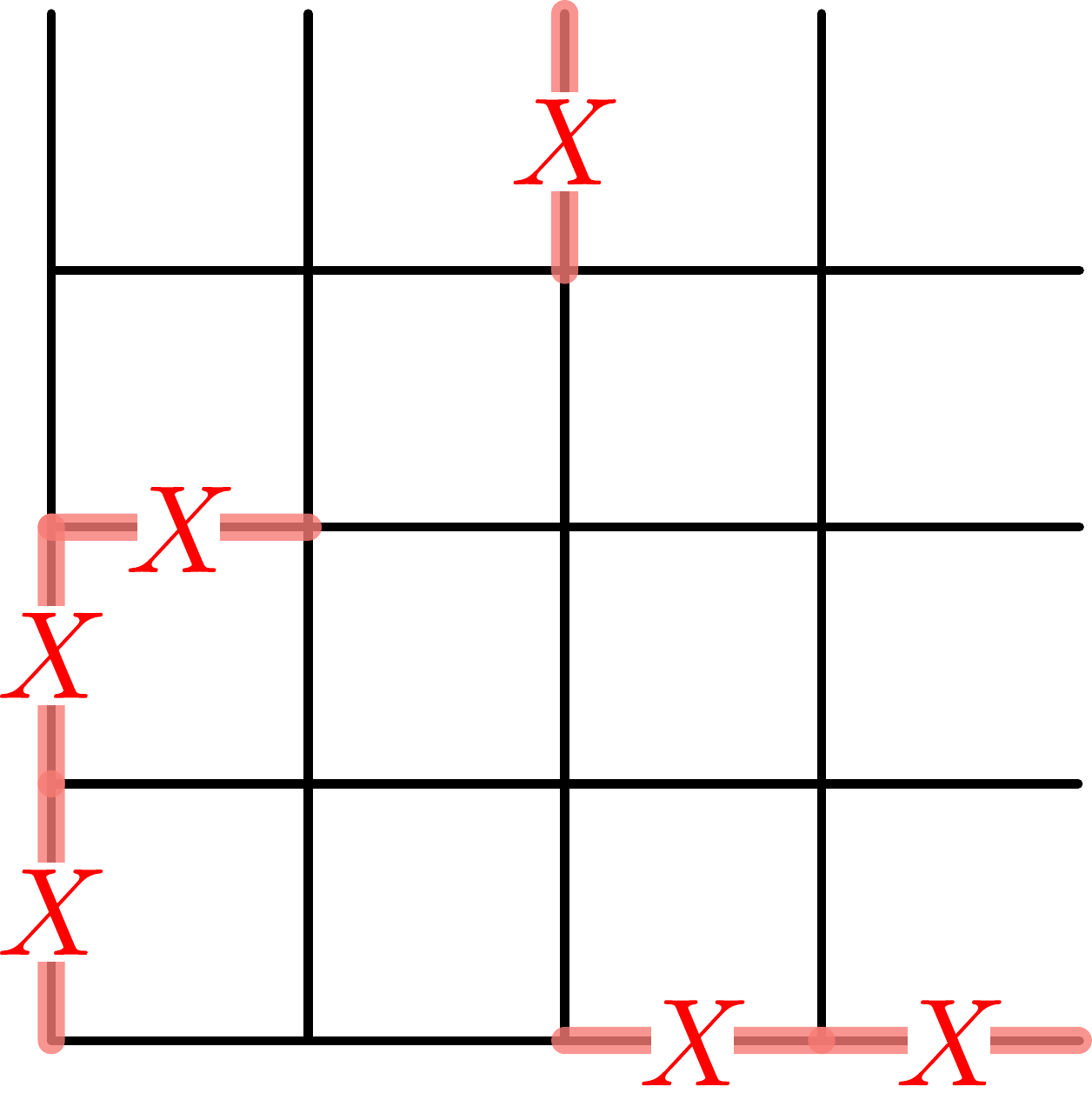}}},~
S^{k=13}_2 =\vcenter{\hbox{\includegraphics[scale=.125]{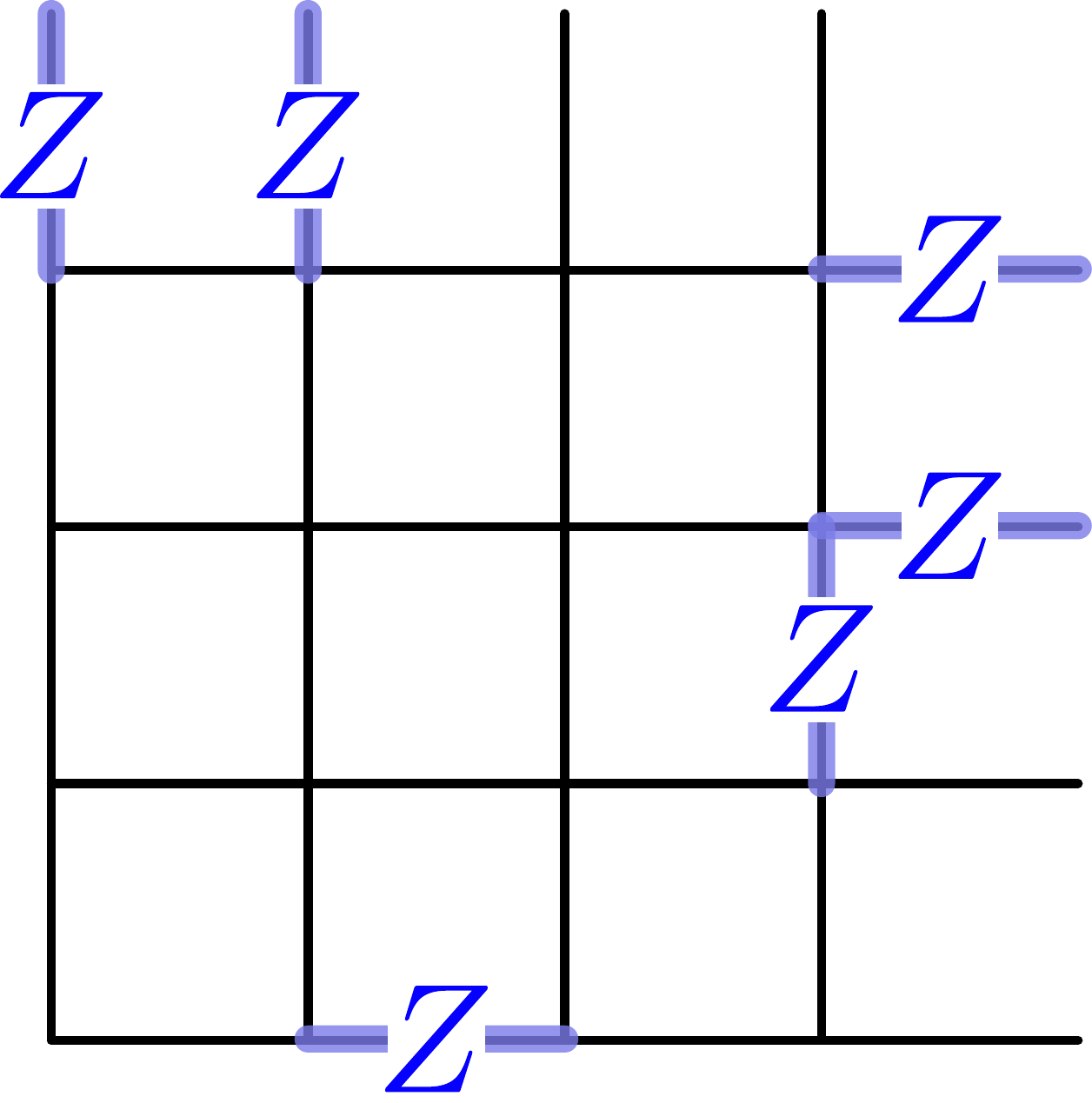}}}.
\end{equation*}
Notably, the $k=10$ family arises from a vertical extension of the $k=8$ stabilizers, while the $k=13$ family results from a horizontal extension of the $k=8$ stabilizers.
Moreover, the stabilizers of the $k=8$ family are derived from the $k=7$ family by shifting one edge; similarly, the $k=9$ stabilizers are formed by shifting one edge of the $k=8$ stabilizers, and the $k=12$ stabilizers are generated by an additional one-edge shift of the $k=9$ stabilizers.

Overall, these observations suggest that the bulk stabilizers are closely related. It would be of great interest to develop an analytical method to obtain these families directly rather than relying solely on numerical search, as this could provide valuable guidelines for designing high-performance families of planar qLDPC codes with larger logical dimensions $k$. We conjecture that the code families described above achieve an optimal balance between $k$ and the code distance $d$: the logical dimension $k$ scales with the range of the stabilizers (determined by the independent monomials after quotienting out $f(x,y)$ and $g(x,y)$), while the fractal operators favor smaller stabilizers in order to accommodate more small triangles. In other words, the bulk stabilizers presented above may be arranged to maximize the area of the Newton polytope associated with $f(x,y)$ and $g(x,y)$~\cite{chen2025anyon}, while remaining compact enough to allow the embedding of numerous small Sierpinski triangles within finite lattices, thereby achieving larger code distances. It would be worthwhile to formalize this trade-off more rigorously.

\subsection{Comparison with surface code and implementation}

Our planar qLDPC codes offer advantage over the surface code in theoretical resource efficiency, with a $kd^2/n$ ratio an order of magnitude higher. However, their implementation presents different challenges. While the surface code has a planar Tanner graph allowing single-layer 2D hardware implementation with only nearest-neighbor interactions, our codes, despite being 2D local, possess non-planar Tanner graphs. Still, it seems feasible to implement our codes on a two-layer architecture (e.g., via flip-chip bonding) while maintaining geometric locality within and between the layers.
This is more achievable than BB (toric) codes, which require long-range connections due to their periodic boundary conditions~\cite{Bravyi2024HighThreshold}. Therefore, our planar construction shifts the implementation challenge primarily to the development of practical two-layer quantum processor architectures. Evaluating the precise layout requirements and feasibility on specific multi-layer architectures is an important next step.

\subsection{Extending the numerical search for bulk stabilizers}

The numerical search for optimal bulk stabilizers presented in this work focused on a specific, relatively simple class of polynomials: $f(x,y) \propto 1+x+x^a y^b$ and $g(x,y) \propto 1+y+x^c y^d$, with exponents $(a,b,c,d)$ within a limited range. It is natural to ask if further improvements are possible by broadening the search space. Future numerical efforts could explore:
\begin{enumerate}
    \item Searching over larger values of $a,b,c,d$.
    \item Investigating polynomials with different structures, such as
    \begin{eqs}
        f(x,y) &= 1 + x^{a_1} y^{b_1} + x^{a_2} y^{b_2}, \\
        g(x,y) &= 1 + x^{c_1} y^{d_1} + x^{c_2} y^{d_2},
    \end{eqs}
    which corresponds to the same stabilizer weight (i.e., weight-6) but different geometries.
    \item Exploring stabilizers with weights greater than 6, such as weight-8 operators, while maintaining geometrical locality. This approach could examine the trade-off between stabilizer weight and code distance.
\end{enumerate}
Expanding the search in these directions could lead to the discovery of new code families with even better $[[n,k,d]]$ parameters or improved asymptotic scaling within the $\mathcal{O}(1)$ bound for $kd^2/n$. Moreover, the relationships between the bulk stabilizers observed in Sec.~\ref{sec:observationsonbulstabilizers} may offer valuable insights for guiding such extended searches.

\subsection{Exploring alternative boundary geometries}

Our construction methodology, based on anyon condensation at gapped boundaries, provides a systematic way to define boundaries for planar codes derived from translation-invariant systems. The examples presented here use standard $e$- and $m$-condensed boundaries on a rectangular grid. However, the flexibility of the TQFT framework invites exploration of alternative boundary geometries.
For example, one could explore boundaries rotated relative to the lattice axes, implement intricate interleaving of distinct Lagrangian subgroups (e.g.~$\mathcal{L}_1$, $\mathcal{L}_2$, and $\mathcal{L}_3$) along the boundary, or even introduce internal boundaries by puncturing holes within the lattice.
Systematically studying these variations could reveal new code families, potentially offer advantages in connectivity or robustness, or enable different types of logical operations. In the future, one could investigate how to tailor code properties through boundary engineering.

\subsection{Performance simulation and threshold estimation}

A crucial next step is to evaluate the performance of our codes using numerical simulations, such as Belief Propagation combined with Ordered Statistic Decoding (BP+OSD). The primary goal of these simulations would be to estimate the pseudo-threshold of our planar code families against relevant noise models (e.g., circuit-level noise).
Since our construction preserves both geometric locality and the LDPC structure, we conjecture that the (pseudo)-threshold will be comparable to that of the surface code and BB codes, whose thresholds around the range of $0.7\%$ to $1.1\%$~\cite{Bravyi2024HighThreshold}.


\section*{Acknowledgment}

We would like to thank Dave Aasen, Shin Ho Choe, Arpit Dua, Tyler D. Ellison, Yingfei Gu, Francisco Revson F. Pereira, Hao Song, Vincent Steffan, Nathanan Tantivasadakarn, Yi-Fei Wang, Dominic J. Williamson, and Bowen Yang for their valuable discussions.

This work is supported by the National Natural Science Foundation of China (Grant No.~12474491).

\bibliography{bibliography.bib}

\appendix

\begin{widetext}
\pagebreak 
{\change
\section{Review of the Laurent polynomial formalism}
\label{Appendix: Review of the Laurent polynomial formalism}

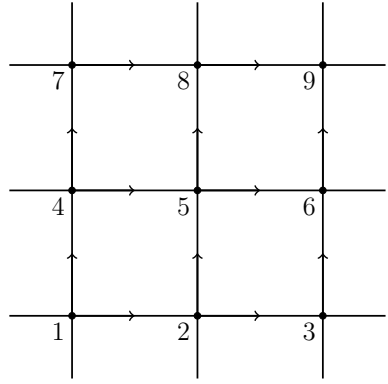
\begin{figure}[htb]
\centering
\resizebox{5cm}{!}{%
\begin{tikzpicture}
\draw[thick] (-3,0) -- (3,0);\draw[thick] (-3,-2) -- (3,-2);\draw[thick] (-3,2) -- (3,2);
\draw[thick] (0,-3) -- (0,3);\draw[thick] (-2,-3) -- (-2,3);\draw[thick] (2,-3) -- (2,3);
\draw[->] [thick](0,0) -- (1,0);\draw[->][thick] (0,2) -- (1,2);\draw[->][thick] (0,-2) -- (1,-2);
\draw[->][thick] (0,0) -- (0,1);\draw[->][thick] (2,0) -- (2,1);\draw[->][thick](-2,0) -- (-2,1);
\draw[->][thick] (-2,0) -- (-1,0);\draw[->][thick] (-2,2) -- (-1,2);\draw[->][thick](-2,-2) -- (-1,-2);
\draw[->][thick] (-2,-2) -- (-2,-1);\draw[->][thick] (0,-2) -- (0,-1);\draw[->] [thick](2,-2) -- (2,-1);
\filldraw [black] (-2,-2) circle (1.5pt) node[anchor=north east] {\large 1};
\filldraw [black] (0,-2) circle (1.5pt) node[anchor=north east] {\large 2};
\filldraw [black] (2,-2) circle (1.5pt) node[anchor=north east] {\large 3};
\filldraw [black] (-2,-0) circle (1.5pt) node[anchor=north east] {\large 4};
\filldraw [black] (0,0) circle (1.5pt) node[anchor=north east] {\large 5};
\filldraw [black] (2,0) circle (1.5pt) node[anchor=north east] {\large 6};
\filldraw [black] (-2,2) circle (1.5pt) node[anchor=north east] {\large 7};
\filldraw [black] (0,2) circle (1.5pt) node[anchor=north east] {\large 8};
\filldraw [black] (2,2) circle (1.5pt) node[anchor=north east] {\large 9};
\end{tikzpicture}}
\caption{A qudit is placed on each edge, with generalized Pauli operators $X_e$ and $Z_e$ acting on it.}
\label{fig:square}
\end{figure}

This appendix reviews the Laurent polynomial representation and its application to translation-invariant stabilizer codes. The formalism was first introduced in Refs.~\cite{haah_module_13, haah2016algebraic}. In this work, we adopt the conventions and notations of Refs.~\cite{liang2023extracting, liang2024operator, liang2025generalized}.  
We begin with a general $\ZZ_d$ qudit system, where the $d \times d$ generalized Pauli matrices are defined as
\begin{eqs}
    X = 
    \begin{bmatrix}
    0 & 0 & \cdots & 0 & 1 \\
    1 & 0 & \cdots & 0 & 0 \\
    0 & 1 & \cdots & 0 & 0 \\
    \vdots & \vdots & \ddots & \vdots & \vdots \\
    0 & 0 & \cdots & 1 & 0
    \end{bmatrix}, \quad
    Z = 
    \begin{bmatrix}
    1 & 0 & 0 & \cdots & 0 \\
    0 & \omega & 0 & \cdots & 0 \\
    0 & 0 & \omega^2 & \cdots & 0 \\
    \vdots & \vdots & \vdots & \ddots & \vdots \\
    0 & 0 & 0 & \cdots & \omega^{d-1}
    \end{bmatrix},
\end{eqs}
with $\omega := \exp(2 \pi i / d)$. These matrices satisfy the commutation relation
\begin{eqs}
    Z X = \omega X Z.
\end{eqs}
The qubit case is recovered by setting $d=2$.

We consider the case of two $\mathbb{Z}_d$ qudits per unit cell (e.g., one qudit on each edge of a square lattice), generalizable to $k$ qudits per cell. Any Pauli operator---a finite tensor product of Pauli matrices---can be represented, up to an overall phase, as a column vector over the polynomial ring
\begin{equation}
    R = \ZZ_d [x, y, x^{-1}, y^{-1}],
\end{equation}
which includes all polynomials in $x^{\pm1}$, $y^{\pm1}$ with coefficients in $\mathbb{Z}_d$~\cite{haah_module_13}.
We assign column vectors over $\ZZ_d$ to the (generalized) Pauli matrices $X_{12}$, $Z_{12}$, $X_{14}$, and $Z_{14}$, depicted in Fig.~\ref{fig:square}:
\begin{eqs}
    \mX_{12}=
    \left[\begin{array}{c}
        1 \\
        0 \\
        \hline
        0 \\
        0
    \end{array}\right],~
    \mZ_{12}=
    \left[\begin{array}{c}
        0 \\
        0 \\
        \hline
        1 \\
        0
    \end{array}\right],~
    \mX_{14}=
    \left[\begin{array}{c}
        0 \\
        1 \\
        \hline
        0 \\
        0
    \end{array}\right],~
    \mZ_{14}=
    \left[\begin{array}{c}
        0 \\
        0 \\
        \hline
        0 \\
        1
    \end{array}\right],
\end{eqs}
where column vector representations of operators are indicated using curly letters.
The coefficients in these vectors correspond to their powers:
\begin{eqs}
    \mathcal{P} =
    \left[\begin{array}{c}
        i \\
        j \\
        \hline
        k \\
        l
    \end{array}\right]
    ~\Rightarrow~
    \mathcal{P}^m =
    \left[\begin{array}{c}
        m i \\
        m j \\
        \hline
        m k \\
        m l
    \end{array}\right]
    \forall~m \in \ZZ_d.
\end{eqs}
The translation of operators is achieved using polynomials of $x$ and $y$ to denote translations in the $x$ and $y$ directions, respectively. To illustrate, translating the operator on edge $e_{12}$ to edge $e_{78}$ or to edge $e_{58}$ involves multiplying the column vector of the operator by $y^2$ or $xy$, respectively:
\begin{eqs}
    \mZ_{78}= y^2
    \mZ_{12}
    =
    \left[\begin{array}{c}
        0 \\
        0 \\
        \hline
        y^2 \\
        0
    \end{array}\right],~
    \mX_{58}= x y
    \mX_{14}
    =
    \left[\begin{array}{c}
        0 \\
        xy \\
        \hline
        0 \\
        0
    \end{array}\right].
\end{eqs}
A general Pauli operator can be expressed as
\begin{equation}
    P = \eta X^{a_1}_{e_1} X^{a_2}_{e_2} \cdots X^{a_n}_{e_n} Z^{b_1}_{e'_1} Z^{b_2}_{e'_2} \cdots Z^{b_m}_{e'_m},
\end{equation}
where $\eta$ represents a root of unity of order $2d$. After dropping the overall phase $\eta$, the corresponding column vector for this operator is a linear combination of individual Pauli matrices, expressed as
\begin{eqs}
    \mathcal{P} =  a_1 \mX_{e_1} + a_2 \mX_{e_2} + \cdots + a_n \mX_{e_n}  + b_1 \mZ_{e'_1} + b_2 \mZ_{e'_2} + \cdots + b_m \mZ_{e'_m}.
\end{eqs}
More examples are included in Fig.~\ref{fig:example_poly}.

\begin{figure}[thb]
    \centering
    \includegraphics[width=0.45\textwidth]{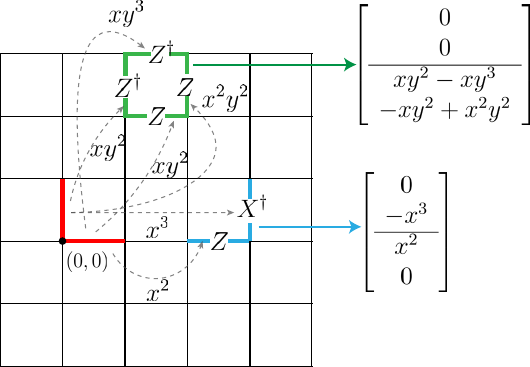}
    \caption{Examples of polynomial expressions for Pauli strings. The flux term on a plaquette and the $XZ$ term on edges are shown. The factors such as $x^2 y^2$ and $x^2$ represent the locations of the operators relative to the origin.}
\label{fig:example_poly}
\end{figure}

Next, we introduce the {\bf antipode map} that is a $\ZZ_d$-linear map from $R$ to $R$ defined by
\begin{eqs}
    x^a y^b \rightarrow \overline{x^a y^b}:=x^{-a} y^{-b}.
\end{eqs}
To determine whether two Pauli operators represented by vectors $v_1$ and $v_2$ commute or anti-commute, we define the dot product as
\begin{eqs}
    v_1 \cdot v_2 = \overline{v}_1^{T} \Lambda v_2,
\end{eqs}
where $T$ is the transpose operation on a matrix and
\begin{eqs}
    \Lambda=
    \left[\begin{array}{cc | cc}
        0 & 0 & 1 & 0 \\
        0 & 0 & 0 & 1 \\
        \hline
        -1 & 0 & 0 & 0 \\
        0 & -1 & 0 & 0 \\
    \end{array}\right]
\end{eqs}
is the matrix representation of the standard {\bf symplectic bilinear form}. For simplicity, we denote $\overline{(\cdots)}^T$ as $(\cdots)^\dagger$.

Two operators commute iff the constant term of $v_1 \cdot v_2$ is zero.  
For example:
\begin{eqs}
    \mX_{12} \cdot \mZ_{12} = 1 \quad (\text{anti-commute by the factor of $\omega$}), \quad
\mX_{58} \cdot \mZ_{14} = x^{-1}y^{-1} \quad (\text{commute}).
\end{eqs}
Furthermore, the physical interpretation of $\mX_{58} \cdot \mZ_{14} = x^{-1} y^{-1}$ is that shifting of $X_{58}$ in $-x$ and $-y$ directions by 1 step will anti-commute (by the factor of $\omega$) with $Z_{14}$. 

A translation-invariant stabilizer code corresponds to an $R$-submodule $\sigma$ such that
\begin{eqs}
    v_1 \cdot v_2 = v_1^\dagger \Lambda v_2 = 0, \quad \forall\, v_1, v_2 \in \sigma,
\end{eqs}
called the \textbf{stabilizer module}. 
The Hamiltonian could have two (or more) terms per square to have a unique ground state on a simply-connected manifold, denoted as $H = -\sum_{\text{cells}} (S_1 + S_2)$ with corresponding column vectors
\begin{eqs}
    \mS_1 = \left[\begin{array}{c}
        f_1(x,y) \\
        f_2 (x,y) \\
        \hline
        g_1(x,y) \\
        g_2 (x,y)
    \end{array}\right], \quad
    \mS_2 = \left[\begin{array}{c}
        h_1(x,y) \\
        h_2 (x,y) \\
        \hline
        k_1(x,y) \\
        k_2 (x,y)
    \end{array}\right].
\label{eq: generic S1 and S2}
\end{eqs}
For example, the trivial phase $H_0 = - \sum_e X_e$ is
\begin{eqs}
    \mS_1 = \left[\begin{array}{c}
        1 \\
        0 \\
        \hline
        0 \\
        0
    \end{array}\right], \quad
    \mS_2 = \left[\begin{array}{c}
        0 \\
        1 \\
        \hline
        0 \\
        0
    \end{array}\right],
\label{eq: trivial H0 SA SB}
\end{eqs}
and the standard $\ZZ_d$ toric code Hamiltonian 
\begin{eqs}
    H_{\text{TC}} = - \sum_v \vcenter{\hbox{\includegraphics[scale=.25]{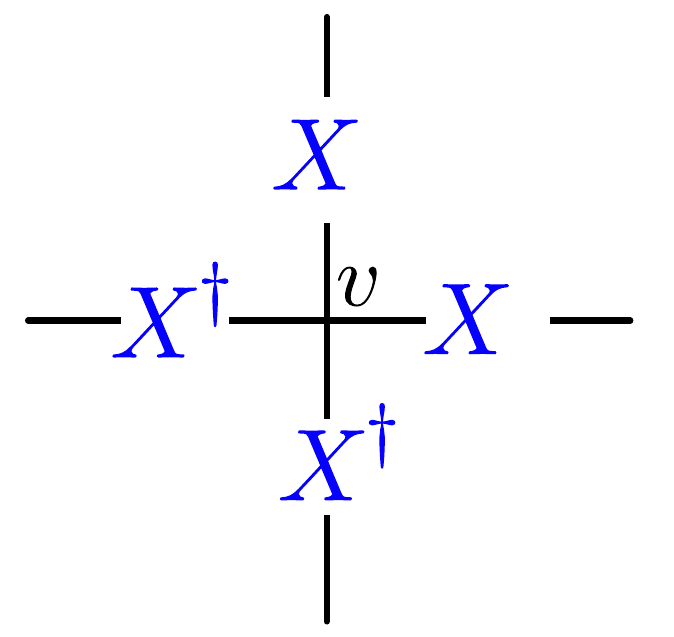}}} - \sum_p \vcenter{\hbox{\includegraphics[scale=.25]{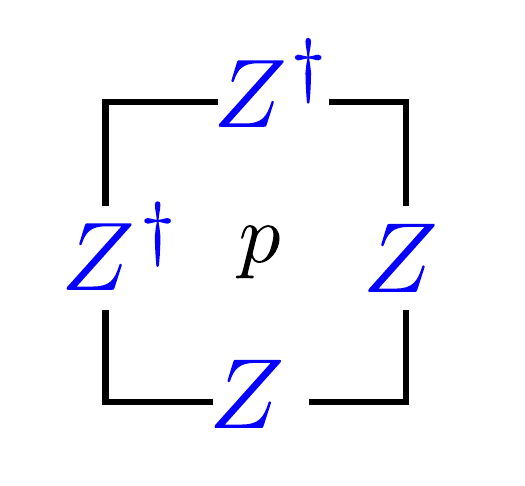}}},
\label{eq: toric code Hamiltonian}
\end{eqs}
corresponds to
\begin{eqs}
    \mS_1 = \left[\begin{array}{c}
        1-\bx \\
        1- \by \\
        \hline
        0 \\
        0
    \end{array}\right], \quad
    \mS_2 = \left[\begin{array}{c}
        0 \\
        0 \\
        \hline
        1-y \\
        -1+x
    \end{array}\right].
\label{eq: standard toric code SA SB}
\end{eqs}
In the main text, we consider a translation-invariant $\mathbb{Z}_2$ CSS code defined on the square lattice, with bulk stabilizer generators:
\begin{equation}
    A_v = 
    \begin{bmatrix}
        f(x,y) \\
        \rule{0pt}{1.1em}g(x,y) \\
        \hline
        0 \\
        0
    \end{bmatrix}, 
    \quad
    B_p = 
    \begin{bmatrix}
        0 \\
        0 \\
        \hline
        \rule{0pt}{1.1em}\overline{g(x,y)} \\
        \overline{f(x,y)}
    \end{bmatrix},
    \label{eq: stabilizer}
\end{equation}
with 
\begin{eqs}
    f(x,y) &= 1 + x + x^a y^b, \\
    g(x,y) &= 1 + y + x^c y^d.
\label{eq: a b c d generalized toric code}
\end{eqs}

We now introduce the {\bf excitation map} for a Pauli operator $\mP$ (represented as a column vector over $R$):
\begin{eqs}
    \eps(\mP):=\big[\, \mS_1 \cdot \mP , \; \mS_2 \cdot \mP \,\big],
\end{eqs}
which encodes how $\mP$ violates the stabilizers $S_1$ and $S_2$.  
For a single Pauli operator in the generic Hamiltonian \eqref{eq: generic S1 and S2}, the corresponding error syndromes are
\begin{eqs}
    \eps(\mX_1) &= [ \mS_1 \cdot \mX_{12}, \; \mS_2 \cdot \mX_{12}] = [ -\overline{g}_1, \; -\overline{k}_1 ], \\
    \eps(\mX_2) &= [ \mS_1 \cdot \mX_{14}, \; \mS_2 \cdot \mX_{14}] = [ -\overline{g}_2, \; -\overline{k}_2 ], \\
    \eps(\mZ_1) &= [ \mS_1 \cdot \mZ_{12}, \; \mS_2 \cdot \mZ_{12}] = [ \overline{f}_1, \; \overline{h}_1 ], \\
    \eps(\mZ_2) &= [ \mS_1 \cdot \mZ_{14}, \; \mS_2 \cdot \mZ_{14}] = [ \overline{f}_2, \; \overline{h}_2 ].
    \label{eq: get error syndromes}
\end{eqs}
Each syndrome is thus a two-component row vector over $\ZZ_d[x, y, x^{-1}, y^{-1}]$.  
In the more general case with $t$ stabilizer generators $S_1,\dots,S_t$, the error syndrome becomes a $t$-component row vector over the same ring.

With the excitation map $\eps$ in hand, the {\bf topological order (TO) condition} can be expressed as
\begin{eqs}
    \ker \eps = \sigma,
\label{eq: topological condition}
\end{eqs}
where $\sigma$ denotes the stabilizer module, and $\ker \eps$ represents all local Pauli operators that commute with the stabilizers.  
This requirement means that any such commuting operator must itself be a product of stabilizers (i.e., an element of $\sigma$).  
If the TO condition holds, no local observable can distinguish different ground states.  
In particular, the ground state degeneracy depends on the topology of the manifold: on a torus, it equals the number of anyon types, with each ground state associated with an anyon string operator.

}

{\change
\section{$[[54,6,4]]$ code and its grafted $[[44,6,4]]$ code}

In this appendix, we present a specific construction of the $[[54,6,4]]$ code, which achieves the smallest code parameters listed in Table~\ref{tab: GTC large weight and stabilizer}. For this construction, we use slightly different stabilizers $f(x,y)$ and $g(x,y)$ from those shown in Table~\ref{tab: GTC large weight and stabilizer}. The code and its grafted version, the $[[44,6,4]]$ code, are illustrated in Fig.~\ref{fig: [[54, 6, 4]] code} and Fig.~\ref{fig: [[44, 6, 4]] code}, respectively.
}

\begin{figure}[htb]
    \centering
    \subfigure[~The $\lbrack\lbrack54, 6, 4\rbrack\rbrack$ code]{{\includegraphics[scale=0.065]{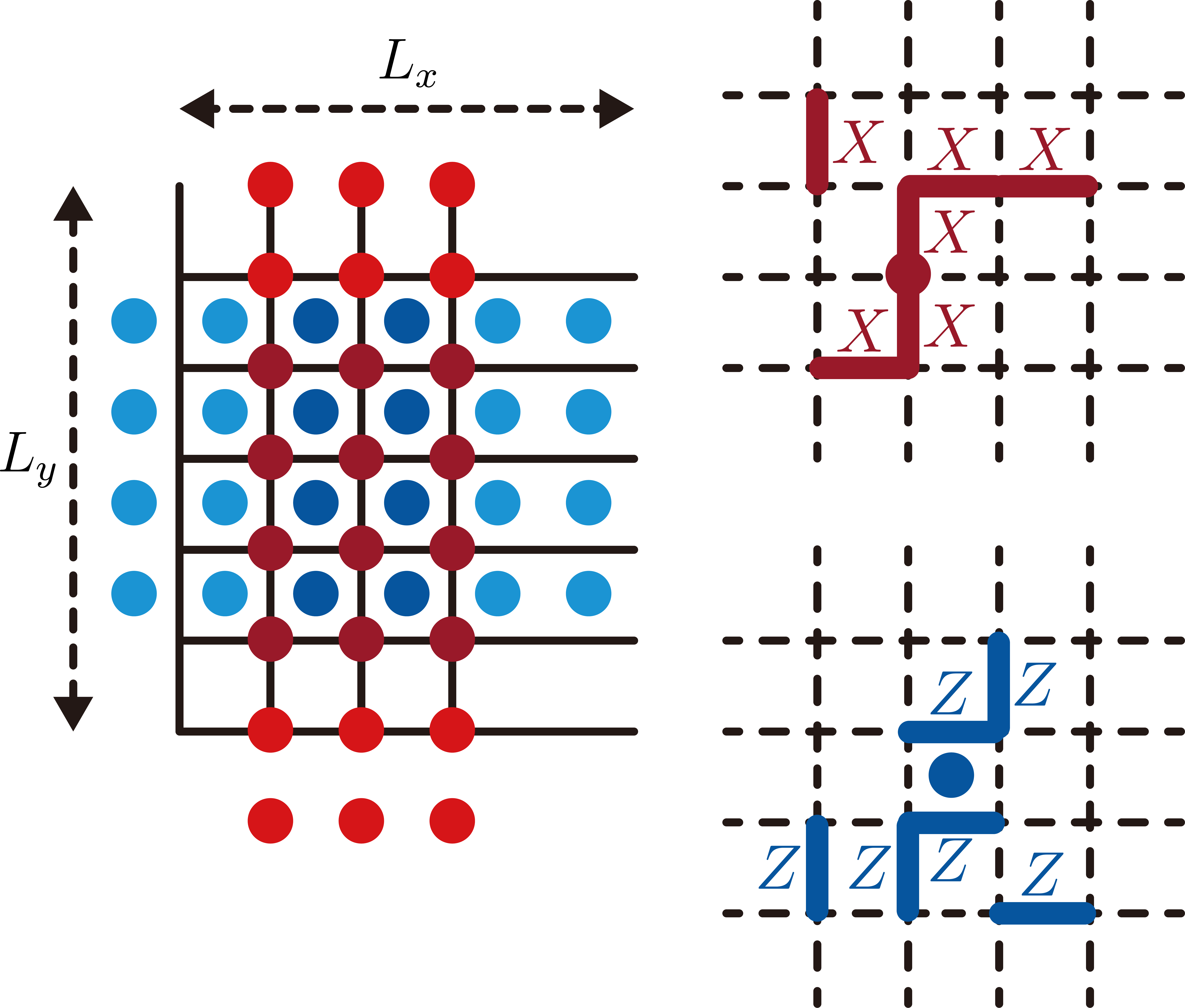}\label{fig: [[54, 6, 4]] code}}}
    \hspace{1cm}
    \subfigure[~The grafted $\lbrack\lbrack44, 6, 4\rbrack\rbrack$ code]{\includegraphics[scale=0.065]{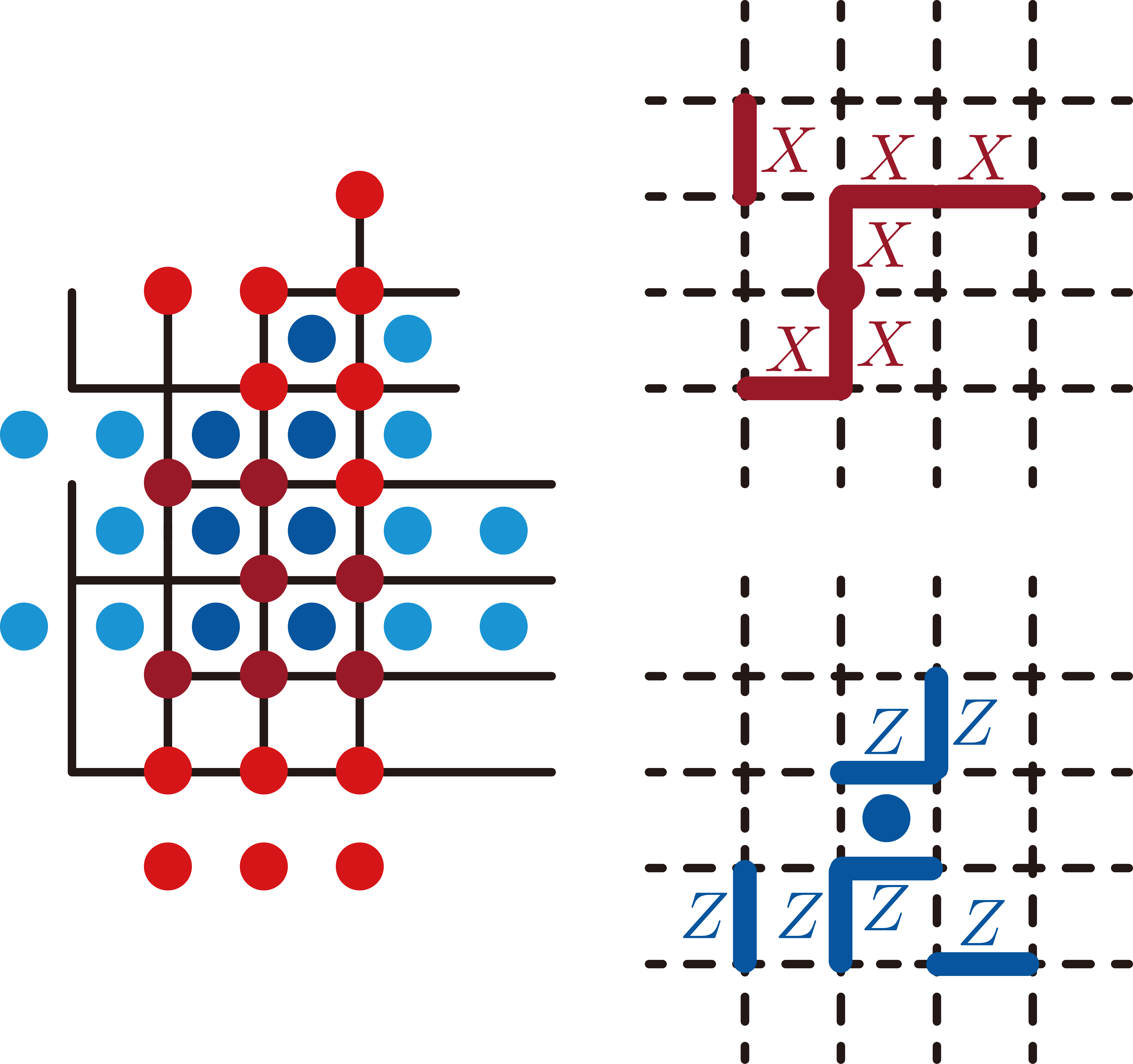}\label{fig: [[44, 6, 4]] code}}
    \caption{ {\change 
    (a) The $[[54,6,4]]$ code on an open square lattice with $L_x=5$ and $L_y=6$. The bulk stabilizers correspond to $f(x,y)=xy^2(1+x+x^{-1}y^{-2})$ and $g(x,y)=x(1+y+x^{-1}y^2)$. Each red (blue) dot represents an $X$ ($Z$) stabilizer: deep red/blue indicate bulk stabilizers fully supported on the pruned lattice, while lighter dots near the boundary denote stabilizers with truncated edges. There are 12 bulk $X$ (deep red), 8 bulk $Z$ (deep blue), 12 boundary $X$ (red), and 16 boundary $Z$ (blue) stabilizers, giving a logical dimension $k=54-12-8-12-16=6$.  
    (b) The grafted $[[44,6,4]]$ code, obtained by removing 10 qubits. The remaining stabilizers include 7 bulk $X$ (deep red), 7 bulk $Z$ (deep blue), 13 boundary $X$ (red), and 11 boundary $Z$ (blue), yielding the same logical dimension $k=44-7-7-13-11=6$.
    }}
\end{figure}



{\change
\section{Qubit numbering}

In this appendix, we describe our convention for labeling qubits on the open square lattice, as illustrated in Fig.~\ref{fig: number_convention}. This labeling allows all of our planar code constructions to be expressed as parity-check matrices. The supplementary material provides the complete parity-check matrices for all codes presented in this manuscript, where $H_X$ and $H_Z$ denote the parity-check matrices corresponding to $X$-type and $Z$-type stabilizers, respectively, since we focus on CSS codes. Interested readers may use these matrices to verify the code properties with standard tools for qLDPC codes.

}

\begin{figure}[htb]
    \centering
    \includegraphics[width=.25\textwidth]{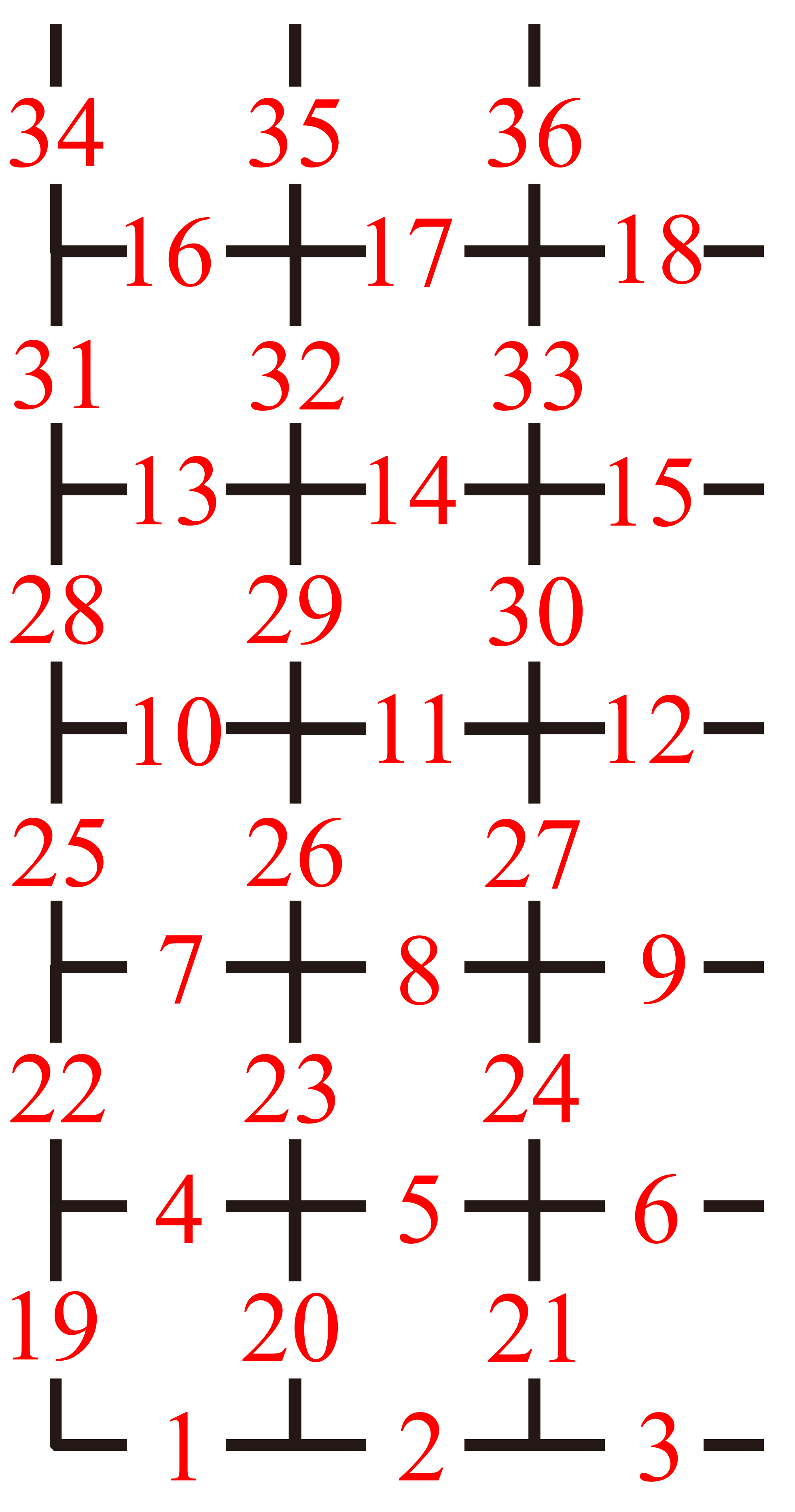}
    \caption{{\change Sequential qubit numbering from $1$ to $36$ on a $3\times6$ lattice. Indices are first assigned to qubits on horizontal edges, followed by those on vertical edges, proceeding from the bottom-left to the top-right. After lattice grafting, the indexing is obtained by removing the corresponding qubits while preserving the same ordering convention.}
    }
    \label{fig: number_convention}
\end{figure}

\section{Families of planar qLDPC codes}\label{app: Families of planar qLDPC codes}

In this section, we first present tables of the code distances $d(L_x, L_y)$ as functions of the lattice sizes $L_x$ and $L_y$ for the families of planar qLDPC codes derived from the examples in Sec.~\ref{sec: examples of planar qLDPC codes}:
\begin{itemize}
    \item The $k=6$ code family, derived from the $[[88, 6, 6]]$ code (shown in Fig.~\ref{fig: [[88, 6, 6]] code}) in Sec.~\ref{sec: [[88, 6, 6]] planar code}, has its code distances listed in Table~\ref{tab: d(Lx, Ly) for [[88,6,6]]}.
    \item The $k=7$ code family, derived from the $[[131, 7, 7]]$ code (shown in Fig.~\ref{fig: [[131, 7, 7]] code}) in Sec.~\ref{sec: [[131, 7, 7]] planar code}, has its code distances listed in Table~\ref{tab: d(Lx, Ly) for [[131,7,7]]}.
    \item The $k=8$ code family, derived from the $[[188, 8, 9]]$ code (shown in Fig.~\ref{fig: [[188, 8, 9]] code}) in Sec.~\ref{sec: [[188, 8, 9]] planar code}, has its code distances listed in Table~\ref{tab: d(Lx, Ly) for [[188,8,9]]}.
    \item The $k=8$ code family, derived from the $[[288, 8, 12]]$ code (shown in Fig.~\ref{fig: [[288, 8, 12]] code}) in Sec.~\ref{sec: [[288, 8, 12]] planar code}, has its code distances listed in Table~\ref{tab: d(Lx, Ly) for [[288,8,12]]}.
    \item The $k=12$ code family (weight-8 stabilizers), derived from the $[[292, 12, 14]]$ code (shown in Fig.~\ref{fig: [[292, 12, 14]] code}) in Sec.~\ref{sec: [[292, 12, 14]] planar code}, has its code distances listed in Table~\ref{tab: d(Lx, Ly) for [[292,12,14]]}.
\end{itemize}

Furthermore, we introduce additional families of planar qLDPC codes with logical dimensions ranging from $k=9$ to $k=13$:
\begin{itemize}
    \item The $k=9$ code family, derived from the $[[441, 9, 15]]$ code (shown in Fig.~\ref{fig: [[441, 9, 15]]}), has its code distances listed in Table~\ref{tab: d(Lx, Ly) for [[441, 9, 15]]}.
    \item The $k=10$ code family, derived from the $[[381, 10, 13]]$ code (shown in Fig.~\ref{fig: [[381, 10, 13]]}), has its code distances listed in Table~\ref{tab: d(Lx, Ly) for [[381, 10, 13]]}.
    \item The $k=11$ code family, derived from the $[[494, 11, 15]]$ code (shown in Fig.~\ref{fig: [[494, 11, 15]]}), has its code distances listed in Table~\ref{tab: d(Lx, Ly) for [[494, 11, 15]]}.
    \item The $k=12$ code family, derived from the $[[432, 12, 12]]$ code (shown in Fig.~\ref{fig: [[432, 12, 12]]}), has its code distances listed in Table~\ref{tab: d(Lx, Ly) for [[432, 12, 12]]}.
    \item The $k=13$ code family, derived from the $[[392, 13, 11]]$ code (shown in Fig.~\ref{fig: [[392, 13, 11]]}), has its code distances listed in Table~\ref{tab: d(Lx, Ly) for [[392, 13, 11]]}.
\end{itemize}
We have enumerated these codes for various values of $L_x$ and $L_y$ so that the code distances fall within the range $4 \leq d \leq 16$. {\change This regime allows us to compute the code distances precisely, using the integer programming approach~\cite{landahl2011fault}.} In contrast, for codes with larger distances, the exact method is computationally prohibitive, so the distances are estimated using a probabilistic algorithm~\cite{Pryadko2022GAP}, which provides only an upper bound and may not be tight.

\section{Grafted planar qLDPC codes}\label{app: Grafted planar qLDPC codes}

In this section, we apply the ``lattice grafting'' optimization to the examples from the previous code families. The grafted codes are presented below. It is important to note that lattice grafting involves randomized choices (since enumerating exponentially many iterative cases is impractical); therefore, we run the algorithm several hundred times for each parent code in Appendix~\ref{app: Families of planar qLDPC codes} and select the instance with the smallest number of physical qubits $n$.
\begin{itemize}
    \item The grafted $[[78, 6, 6]]$ code, with ${kd^2}/{n} = 2.77$, is illustrated in Fig.~\ref{fig: [[78, 6, 6]] code} and is obtained from the $[[88, 6, 6]]$ code (shown in Fig.~\ref{fig: [[88, 6, 6]] code}) by removing 10 physical qubits.
    \item The grafted $[[107, 7, 7]]$ code, with ${kd^2}/{n} = 3.21$, is illustrated in Fig.~\ref{fig: [[107, 7, 7]] code} and is obtained from the $[[131, 7, 7]]$ code (shown in Fig.~\ref{fig: [[131, 7, 7]] code}) by removing 24 physical qubits.
    \item The grafted $[[173, 8, 9]]$ code, with ${kd^2}/{n} = 3.75$, is illustrated in Fig.~\ref{fig: [[173, 8, 9]] code} and is obtained from the $[[188, 8, 9]]$ code (shown in Fig.~\ref{fig: [[188, 8, 9]] code}) by removing 15 physical qubits.
    \item The grafted $[[268, 8, 12]]$ code, with ${kd^2}/{n} = 4.30$, is illustrated in Fig.~\ref{fig: [[268, 8, 12]] code} and is obtained from the $[[288, 12, 12]]$ code (shown in Fig.~\ref{fig: [[288, 8, 12]] code}) by removing 20 physical qubits.
    \item The grafted $[[405, 9, 15]]$ code, with ${kd^2}/{n} = 5.00$, is illustrated in Fig.~\ref{fig: [[405, 9, 15]] code} and is obtained from the $[[441, 9, 15]]$ code (shown in Fig.~\ref{fig: [[441, 9, 15]]}) by removing 36 physical qubits.
    \item The grafted $[[348, 10, 13]]$ code, with ${kd^2}/{n} = 4.86$, is illustrated in Fig.~\ref{fig: [[348, 10, 13]] code} and is obtained from the $[[381, 10, 13]]$ code (shown in Fig.~\ref{fig: [[381, 10, 13]]}) by removing 33 physical qubits.
    \item The grafted $[[450, 11, 15]]$ code, with ${kd^2}/{n} = 5.50$, is illustrated in Fig.~\ref{fig: [[450, 11, 15]] code} and is obtained from the $[[494, 11, 15]]$ code (shown in Fig.~\ref{fig: [[494, 11, 15]]}) by removing 44 physical qubits.
    \item The grafted $[[386, 12, 12]]$ code, with ${kd^2}/{n} = 4.48$, is illustrated in Fig.~\ref{fig: [[386, 12, 12]] code} and is obtained from the $[[432, 12, 12]]$ code (shown in Fig.~\ref{fig: [[432, 12, 12]]}) by removing 46 physical qubits.
    \item The grafted $[[362, 13, 11]]$ code, with ${kd^2}/{n} = 4.35$, is illustrated in Fig.~\ref{fig: [[362, 13, 11]] code} and is obtained from the $[[392, 13, 11]]$ code (shown in Fig.~\ref{fig: [[392, 13, 11]]}) by removing 30 physical qubits.
    \item The grafted $[[282, 12, 14]]$ code (weight-8 stabilizers), with ${kd^2}/{n} = 8.34$, is illustrated in Fig.~\ref{fig: [[282, 12, 14]] code} and is obtained from the $[[292, 12, 14]]$ code (shown in Fig.~\ref{fig: [[292, 12, 14]] code}) by removing 10 physical qubits.
\end{itemize}
This lattice grafting process removes a subset of the physical qubits while preserving the code distances, geometric locality, and maximum stabilizer weights. Although the process typically produces highly irregular lattice boundaries, the qubits can be rearranged to regularize the boundaries for experimental implementations.


\begin{table*}[b]
\setlength{\tabcolsep}{0pt} 
\begin{tabular}{|c|c|c|c|c|c|c|c|c|c|c|c|c|}
    \hline
    \diagbox[height=0.7cm, innerwidth=0.7cm]{$L_x$}{\raisebox{-0.1em}{$L_y$}}  & 6       & 7       & 8       & 9       & 10      & 11      & 12       & 13       & 14       & 15       & 16       & 17       \\ \hline
    5  & \cellcolor{green!30}(54,4)& (63,4)  & (72,4)  & (81,4)  & (90,4)  & (99,4)  & (108,4)  & (117,4)  & (126,4)  & (135,4)  & (144,4)  & (153,4)  \\ \hline
    6  & (66,4)  & \cellcolor{green!30}(77,5)& \cellcolor{green!30}(88,6)& (99,6)  & (110,6) & (121,6) & (132,6)  & (143,6)  & (154,6)  & (165,6)  & (176,6)  & (187,6)  \\ \hline
    7  & (78,4)  & (91,5)  & (104,6) & \cellcolor{green!30}(117,7)& (130,7) & (143,7) & (156,7)  & (169,7)  & (182,7)  & (195,7)  & (208,7)  & (221,7)  \\ \hline
    8  & (90,4)  & (105,5) & (120,6) & (135,7) & \cellcolor{green!30}(150,8)& \cellcolor{green!30}(165,9)& (180,9)  & (195,9)  & (210,9)  & (225,9)  & (240,9)  & (255,9)  \\ \hline
    9  & (102,4) & (119,5) & (136,6) & (153,7) & (170,8) & (187,9) & \cellcolor{green!30}(204,10)& (221,10) & (238,10) & (255,10) & (272,10) & (289,10) \\ \hline
    10 & (114,4) & (133,5) & (152,6) & (171,7) & (190,8) & (209,9) & \cellcolor{green!30}(228,11)& \cellcolor{green!30}(247,12)& (266,12) & (285,12) & (304,12) & (323,12) \\ \hline
    11 & (126,4) & (147,5) & (168,6) & (189,7) & (210,8) & (231,9) & (252,11) & (273,12) & \cellcolor{green!30}(294,13)& (315,13) & (336,13) & (357,13) \\ \hline
    12 & (138,4) & (161,5) & (184,6) & (207,7) & (230,8) & (253,9) & (276,11) & (299,12) & (322,13) & \cellcolor{green!30}(345,14)& (368,14) & (391,14) \\ \hline
    13 & (150,4) & (175,5) & (200,6) & (225,7) & (250,8) & (275,9) & (300,11) & (325,12) & (350,13) & (375,14) & \cellcolor{green!30}(400,15)& \cellcolor{green!30}(425,16)\\ \hline
    14 & ~(162,4)~ & ~(189,5)~ & ~(216,6)~ & ~(243,7)~ & ~(270,8)~ & ~(297,9)~ & ~(324,11)~ & ~(351,12)~ & ~(378,13)~ & ~(405,14)~ & ~(432,15)~ & ~(459,16)~ \\ \hline
\end{tabular}
\caption{$(n,d)$ for the $k=6$ family from the $[[88,6,6]]$ code (Fig.~\ref{fig: [[88, 6, 6]] code}) for different $L_x$ and $L_y$.
}
\label{tab: d(Lx, Ly) for [[88,6,6]]}

\vspace{2.5em} 

\begin{tabular}{|c|c|c|c|c|c|c|c|c|c|c|c|c|c|}
\hline
\diagbox[height=0.7cm, innerwidth=0.7cm]{$L_x$}{\raisebox{-0.1em}{$L_y$}}  & 7     & 8     & 9     & 10     & 11     & 12     & 13     & 14     & 15     & 16     & 17     & 18     & 19     \\ \hline
5  & \cellcolor{green!30}(55,4)& (64,4)  & (73,4)  & (82,4)  & (91,4)  & (100,4) & (109,4)  & (118,4)  & (127,4)  & (136,4)  & (145,4)  & (154,4)  & (163,4)  \\ \hline
6  & (67,4)  & (78,4)  & \cellcolor{green!30}(89,5)& (100,5) & (111,5) & (122,5) & (133,5)  & (144,5)  & (155,5)  & (166,5)  & (177,5)  & (188,5)  & (199,5)  \\ \hline
7  & (79,4)  & (92,4)  & (105,5) & \cellcolor{green!30}(118,6)& \cellcolor{green!30}(131,7)& (144,7) & (157,7)  & (170,7)  & (183,7)  & (196,7)  & (209,7)  & (222,7)  & (235,7)  \\ \hline
8  & (91,4)  & (106,4) & (121,5) & (136,6) & (151,7) & \cellcolor{green!30}(166,8)& (181,8)  & (196,8)  & (211,8)  & (226,8)  & (241,8)  & (256,8)  & (271,8)  \\ \hline
9  & (103,4) & (120,4) & (137,5) & (154,6) & (171,7) & (188,8) & \cellcolor{green!30}(205,10)& (222,10) & (239,10) & (256,10) & (273,10) & (290,10) & (307,10) \\ \hline
10 & (115,4) & (134,4) & (153,5) & (172,6) & (191,7) & (210,8) & (229,10) & \cellcolor{green!30}(248,11)& \cellcolor{green!30}(267,12)& (286,12) & (305,12) & (324,12) & (343,12) \\ \hline
11 & (127,4) & (148,4) & (169,5) & (190,6) & (211,7) & (232,8) & (253,10) & (274,11) & (295,12) & \cellcolor{green!30}(316,13)& \cellcolor{green!30}(337,14)& (358,14) & (379,14) \\ \hline
12 & (139,4) & (162,4) & (185,5) & (208,6) & (231,7) & (254,8) & (277,10) & (300,11) & (323,12) & (346,13) & (369,14) & \cellcolor{green!30}(392,15)& (415,15) \\ \hline
13 & ~(151,4)~ & ~(176,4)~ & ~(201,5)~ & ~(226,6)~ & ~(251,7)~ & ~(276,8)~ & ~(301,10)~ & ~(326,11)~ & ~(351,12)~ & ~(376,13)~ & ~(401,14)~ & ~(426,15)~ & \cellcolor{green!30}~(451,16)~ \\ \hline
\end{tabular}
\caption{$(n,d)$ for the $k=7$ family from the $[[131,7,7]]$ code (Fig.~\ref{fig: [[131, 7, 7]] code}) for different $L_x$ and $L_y$.
}
\label{tab: d(Lx, Ly) for [[131,7,7]]}
\end{table*}

\begin{table*}[]
\setlength{\tabcolsep}{0pt} 
\begin{tabular}{|c|c|c|c|c|c|c|c|c|c|c|c|c|c|}
\hline
\diagbox[height=0.7cm, innerwidth=0.7cm]{$L_x$}{\raisebox{-0.1em}{$L_y$}} & 7     & 8     & 9     & 10     & 11     & 12     & 13    & 14     & 15     & 16     & 17     & 18     & 19     \\ \hline
4  & (46,3)  & (53,3)  & (60,3)  & (67,3)  & (74,3)  & (81,3)  & (88,3)  & (95,3)  & (102,3)  & (109,3)  & (116,3)  & (123,3)  & (130,3)  \\ \hline
5  & \cellcolor{green!30}(59,4)& (68,4)  & (77,4)  & (86,4)  & (95,4)  & (104,4) & (113,4) & (122,4)  & (131,4)  & (140,4)  & (149,4)  & (158,4)  & (167,4)  \\ \hline
6  & (72,4)  & (83,4)  & \cellcolor{green!30}(94,5)& \cellcolor{green!30}(105,6)& (116,6) & (127,6) & (138,6) & (149,6)  & (160,6)  & (171,6)  & (182,6)  & (193,6)  & (204,6)  \\ \hline
7  & (85,4)  & (98,4)  & (111,5) & (124,6) & \cellcolor{green!30}(137,7)& (150,7) & (163,7) & (176,7)  & (189,7)  & (202,7)  & (215,7)  & (228,7)  & (241,7)  \\ \hline
8  & (98,4)  & (113,4) & (128,5) & (143,6) & (158,7) & \cellcolor{green!30}(173,8)& \cellcolor{green!30}(188,9)& (203,9)  & (218,9)  & (233,9)  & (248,9)  & (263,9)  & (278,9)  \\ \hline
9  & (111,4) & (128,4) & (145,5) & (162,6) & (179,7) & (196,8) & (213,9) & \cellcolor{green!30}(230,10)& (247,10) & (264,10) & (281,10) & (298,10) & (315,10) \\ \hline
10 & (124,4) & (143,4) & (162,5) & (181,6) & (200,7) & (219,8) & (238,9) & (257,10) & \cellcolor{green!30}(276,11)& \cellcolor{green!30}(295,12)& (314,12) & (333,12) & (352,12) \\ \hline
11 & (137,4) & (158,4) & (179,5) & (200,6) & (221,7) & (242,8) & (263,9) & (284,10) & (305,11) & \cellcolor{green!30}(326,13)& \cellcolor{green!30}(347,14)& \cellcolor{green!30}(368,15)& (389,15) \\ \hline
12 & ~(150,4)~ & ~(173,4)~ & ~(196,5)~ & ~(219,6)~ & ~(242,7)~ & ~(265,8)~ & ~(288,9)~ & ~(311,10)~ & ~(334,11)~ & ~(357,13)~ & ~(380,14)~ & ~(403,15)~ & \cellcolor{green!30}~(426,16)~ \\ \hline
\end{tabular}
\caption{$(n,d)$ for the $k=8$ family from the $[[188,8,9]]$ code (Fig.~\ref{fig: [[188, 8, 9]] code}) for different $L_x$ and $L_y$.
}
\label{tab: d(Lx, Ly) for [[188,8,9]]}
\vspace{2.5em} 
\begin{tabular}{|c|c|c|c|c|c|c|c|c|c|c|c|}
\hline
\diagbox[height=0.7cm, innerwidth=0.7cm]{$L_x$}{\raisebox{-0.1em}{$L_y$}} & 5     & 6     & 7     & 8     & 9     & 10     & 11      & 12    & 13     & 14     & 15     \\ \hline
5  & \cellcolor{green!30}(50,3)& (60,3)  & (70,3)  & (80,3)  & (90,3)  & (100,3) & (110,3)  & (120,3)  & (130,3)  & (140,3)  & (150,3)  \\ \hline
6  & (60,3)  & \cellcolor{green!30}(72,4)& (84,4)  & (96,4)  & (108,4) & (120,4) & (132,4)  & (144,4)  & (156,4)  & (168,4)  & (180,4)  \\ \hline
7  & (70,3)  & (84,4)  & \cellcolor{green!30}(98,5)& (112,5) & (126,5) & (140,5) & (154,5)  & (168,5)  & (182,5)  & (196,5)  & (210,5)  \\ \hline
8  & (80,3)  & (96,4)  & (112,5) & \cellcolor{green!30}(128,6)& (144,6) & (160,6) & (176,6)  & (192,6)  & (208,6)  & (224,6)  & (240,6)  \\ \hline
9  & (90,3)  & (108,4) & (126,5) & (144,6) & \cellcolor{green!30}(162,7)& (180,7) & (198,7)  & (216,7)  & (234,7)  & (252,7)  & (270,7)  \\ \hline
10 & (100,3) & (120,4) & (140,5) & (160,6) & (180,7) & \cellcolor{green!30}(200,9)& (220,9)  & (240,9)  & (260,9)  & (280,9)  & (300,9)  \\ \hline
11 & (110,3) & (132,4) & (154,5) & (176,6) & (198,7) & (220,9) & \cellcolor{green!30}(242,10)& (264,10) & (286,10) & (308,10) & (330,10) \\ \hline
12 & (120,3) & (144,4) & (168,5) & (192,6) & (216,7) & (240,9) & (264,10) & \cellcolor{green!30}(288,12)& (312,12) & (336,12) & (360,12) \\ \hline
13 & (130,3) & (156,4) & (182,5) & (208,6) & (234,7) & (260,9) & (286,10) & (312,12) & \cellcolor{green!30}(338,13)& (364,13) & (390,13) \\ \hline
14 & (140,3) & (168,4) & (196,5) & (224,6) & (252,7) & (280,9) & (308,10) & (336,12) & (364,13) & \cellcolor{green!30}(392,15)& (420,15) \\ \hline
15 & ~(150,3)~ & ~(180,4)~ & ~(210,5)~ & ~(240,6)~ & ~(270,7)~ & ~(300,9)~ & ~(330,10)~ & ~(360,12)~ & ~(390,13)~ & ~(420,15)~ & \cellcolor{green!30}~(450,16)~ \\ \hline
\end{tabular}
\caption{$(n,d)$ for the $k=8$ family from the $[[288,8,12]]$ code (Fig.~\ref{fig: [[288, 8, 12]] code}) for different $L_x$ and $L_y$.
}
\label{tab: d(Lx, Ly) for [[288,8,12]]}
\vspace{2.5em} 
\begin{tabular}{|c|c|c|c|c|c|c|c|c|c|c|c|}
\hline
\diagbox[height=0.7cm, innerwidth=0.7cm]{$L_x$}{\raisebox{-0.1em}{$L_y$}} 
   & 14       & 15       & 16       & 17        & 18        & 19       & 20       & 21       & 22       & 23       & 24       \\ \hline
4  & (94,5)  & (101,5) & (108,5) & (115,5)  & (122,5)  & (129,5)  & (136,5)  & (143,5)  & (150,5)  & (157,5)  & (164,5)  \\ \hline
5  & (121,6) & (130,6) & (139,6) & (148,6)  & (157,6)  & (166,6)  & (175,6)  & (184,6)  & (193,6)  & (202,6)  & (211,6)  \\ \hline
6  & (148,6) & (159,8) & \cellcolor{green!30}(170,9) & \cellcolor{green!30}(181,10) & (192,10) & (203,10) & (214,10) & (225,10) & (236,10) & (247,10) & (258,10) \\ \hline
7  & (175,6) & (188,7) & (201,9) & (214,9)  & \cellcolor{green!30}(227,11) & \cellcolor{green!30}(240,12) & (253,12) & (266,12) & (279,12) & (292,12) & (305,12) \\ \hline
8  & (202,6) & (217,7) & (232,9) & (247,9)  & (262,11) & (277,12) & \cellcolor{green!30}(292,14) & (307,14) & (322,14) & (337,14) & (352,14) \\ \hline
9  & (229,6) & (246,7) & (263,9) & (280,9)  & (297,11) & (314,12) & (331,14) & \cellcolor{green!30}(348,15) & \cellcolor{green!30}(365,16) & \cellcolor{green!30}(382,17) & \cellcolor{green!30}(399,18) \\ \hline
10 & (256,6) & (275,7) & (294,9) & (313,9)  & (332,11) & (351,12) & (370,14) & (389,15) & (408,16) & (427,17) & (446,18) \\ \hline
11 & (283,6) & (304,7) & (325,9) & (346,9)  & (367,11) & (388,12) & (409,14) & (430,15) & (451,16) & (472,17) & (493,18) \\ \hline
12 & (310,6) & (333,7) & (356,9) & (379,9)  & (402,11) & (425,12) & (448,14) & (471,15) & (494,16) & (517,17) & (540,18) \\\hline
\end{tabular}
\caption{$(n,d)$ for the $k=12$ family (weight-8 stabilizers) from the $[[292,12,14]]$ code (Fig.~\ref{fig: [[292, 12, 14]] code}) for different $L_x$ and $L_y$.
}
\label{tab: d(Lx, Ly) for [[292,12,14]]}
\end{table*}


\begin{figure}[!htbp]
    \centering
    \includegraphics[scale=0.065]{code_441_9_15.pdf}
    \vspace{-1em} 
    \caption{The $[[441, 9, 15]]$ code on an open square lattice. Bulk stabilizers are specified by two polynomials $f(x,y) = x(1+x+x^{-1}y^3)$ and $g(x,y) = 1+y+x y^3$~\cite{liang2025generalized}. Vertical edges of the rightmost column have been removed.
    Each red (blue) dot represents an $X$ ($Z$) stabilizer. Deep red/blue dots correspond to bulk stabilizers fully supported on the lattice. In contrast, dots near the boundary represent stabilizers with truncated edges (i.e., the bulk stabilizer pattern positioned at the boundary dot, retaining only the Pauli operators on edges within the lattice).
    The design ensures that all truncated $X$ and $Z$ operators still commute, forming valid boundary stabilizers. In total, there are 162 bulk $X$-stabilizers (deep red), 144 bulk $Z$-stabilizers (deep blue), 54 boundary $X$-stabilizers (red), and 72 boundary $Z$-stabilizers (blue), yielding a logical dimension of $k=441-162-144-54-72=9$.
    This generates a code family of $k=9$: $[[2L_x L_y - L_y, ~9, ~d(L_x, L_y)]]$, with code distances $d(L_x, L_y)$ listed in Table~\ref{tab: d(Lx, Ly) for [[441, 9, 15]]} below.}
    \label{fig: [[441, 9, 15]]}
    
    \vspace{1em} 
    \setlength{\tabcolsep}{0pt} 
    \begin{tabular}{|c|c|c|c|c|c|c|c|c|c|c|c|c|c|c|}
    \hline
    \diagbox[height=0.7cm, innerwidth=0.7cm]{$L_x$}{\raisebox{-0.1em}{$L_y$}}& 9     & 10     & 11     & 12     & 13    & 14    & 15    & 16     & 17     & 18     & 19     & 20     & 21     & 22     \\ \hline
    5  & \cellcolor{green!30}~(81,4)~& ~(90,4)~  & ~(99,4)~  & ~(108,4)~ & ~(117,4)~ & ~(126,4)~ & ~(135,4)~ & ~(144,4)~  & ~(153,4)~  & ~(162,4)~  & ~(171,4)~  & ~(180,4)~  & ~(189,4)~  & ~(198,4)~  \\ \hline
    6  & (99,4)  & \cellcolor{green!30}(110,5)& (121,5) & \cellcolor{green!30}(132,6)& (143,6) & (154,6) & (165,6) & (176,6)  & (187,6)  & (198,6)  & (209,6)  & (220,6)  & (231,6)  & (242,6)  \\ \hline
    7  & (117,4) & (130,5) & (143,5) & (156,6) & \cellcolor{green!30}(169,7)& (182,7) & (195,7) & (208,7)  & (221,7)  & (234,7)  & (247,7)  & (260,7)  & (273,7)  & (286,7)  \\ \hline
    8  & (135,4) & (150,5) & (165,5) & (180,6) & (195,7) & \cellcolor{green!30}(210,8)& \cellcolor{green!30}(225,9)& (240,9)  & (255,9)  & (270,9)  & (285,9)  & (300,9)  & (315,9)  & (330,9)  \\ \hline
    9  & (153,4) & (170,5) & (187,5) & (204,6) & (221,7) & (238,8) & (255,9) & \cellcolor{green!30}(272,10)& (289,10) & (306,10) & (323,10) & (340,10) & (357,10) & (374,10) \\ \hline
    10 & (171,4) & (190,5) & (209,5) & (228,6) & (247,7) & (266,8) & (285,9) & (304,10) & (323,10) & \cellcolor{green!30}(342,11)& \cellcolor{green!30}(361,12)& (380,12) & (399,12) & (418,12) \\ \hline
    11 & (189,4) & (210,5) & (231,5) & (252,6) & (273,7) & (294,8) & (315,9) & (336,10) & (357,10) & (378,11) & (399,12) & \cellcolor{green!30}(420,13)& \cellcolor{green!30}(441,15)& (462,15) \\ \hline
    12 & (207,4) & (230,5) & (253,5) & (276,6) & (299,7) & (322,8) & (345,9) & (368,10) & (391,10) & (414,11) & (437,12) & (460,13) & (483,15) & \cellcolor{green!30}(506,16) \\ \hline
    \end{tabular}
    \captionof{table}{$(n,d)$ for the $k=9$ family from the $[[441, 9, 15]]$ code for different $L_x$ and $L_y$.}
    \label{tab: d(Lx, Ly) for [[441, 9, 15]]}
\end{figure}

\begin{figure}[!htbp]

    \centering
    \includegraphics[scale=0.06]{code_432_9_14.pdf}
    \caption{The $[[432, 9, 14]]$ code on an open square lattice. Bulk stabilizers are specified by two polynomials $f(x,y) = x^3y(1+x+x^{-3}y^{-1})$ and $g(x,y) = y(1+y+xy^{-1})$~\cite{liang2025generalized}. Vertical edges of the rightmost column have been removed.
    In total, there are 144 bulk $X$-stabilizers (deep red), 135 bulk $Z$-stabilizers (deep blue), 72 boundary $X$-stabilizers (red), and 72 boundary $Z$-stabilizers (blue), yielding a logical dimension of $k=432-144-135-72\times2=9$.
    This generates a code family of $k=9$: $[[2L_x L_y - L_x- 3L_y +3, ~9, ~d(L_x, L_y)]]$, with code distances $d(L_x, L_y)$ listed in Table~\ref{tab: d(Lx, Ly) for [[432, 9, 14]]} below.}
    \label{fig: [[432, 9, 14]]}
    
    \vspace{1em} 
    \setlength{\tabcolsep}{0pt} 
    \begin{tabular}{|c|c|c|c|c|c|c|c|c|c|c|}
    \hline
    \diagbox[height=0.7cm, innerwidth=0.7cm]{$L_x$}{\raisebox{-0.1em}{$L_y$}}& 5       & 6       & 7       & 8       & 9        & 10        & 11        & 12       & 13       & 14       \\ \hline
    10  & \cellcolor{green!30}(78,4)& (95,4)  & (112,4) & (129,4) & (146,4)  & (163,4)  & (180,4)  & (197,4)  & (214,4)  & (231,4)  \\ \hline
    11  & (87,4)  & (106,4) & (125,4) & (144,4) & (163,4)  & (182,4)  & (201,4)  & (220,4)  & (239,4)  & (258,4)  \\ \hline
    12  & (96,4)  & \cellcolor{green!30}(117,5)& (138,5) & (159,5) & (180,5)  & (201,5)  & (222,5)  & (243,5)  & (264,5)  & (285,5)  \\ \hline
    13  & (105,4) & (128,5) & \cellcolor{green!30}(151,6)& (174,6) & (197,6)  & (220,6)  & (243,6)  & (266,6)  & (289,6)  & (312,6)  \\ \hline
    14 & (114,4) & (139,5) & \cellcolor{green!30}(164,7)& (189,7) & (214,7)  & (239,7)  & (264,7)  & (289,7)  & (314,7)  & (339,7)  \\ \hline
    15 & (123,4) & (150,5) & (177,7) & \cellcolor{green!30}(204,8)& (231,8)  & (258,8)  & (285,8)  & (312,8)  & (339,8)  & (366,8)  \\ \hline
    16 & (132,4) & (161,5) & (190,7) & (219,8) & \cellcolor{green!30}(248,9)& (277,9)  & (306,9)  & (335,9)  & (364,9)  & (393,9)  \\ \hline
    17 & (141,4) & (172,5) & (203,7) & (234,8) & \cellcolor{green!30}(265,10)& (296,10) & (327,10) & (358,10) & (389,10) & (420,10) \\ \hline
    18 & (150,4) & (183,5) & (216,7) & (249,8) & (282,10) & (315,10) & (348,10) & (381,10) & (414,10) & (447,10) \\ \hline
    19 & (159,4) & (194,5) & (229,7) & (264,8) & (299,10) & \cellcolor{green!30}(334,11)& (369,11) & (404,11) & (439,11) & (474,11) \\ \hline
    20 & (168,4) & (205,5) & (242,7) & (279,8) & (316,10) & \cellcolor{green!30}(353,12)& (390,12) & (427,12) & (464,12) & (501,12) \\ \hline
    21 & (177,4) & (216,5) & (255,7) & (294,8) & (333,10) & (372,12) & \cellcolor{green!30}(411,13)& (450,13) & (489,13) & (528,13) \\ \hline
    22 & (186,4) & (227,5) & (268,7) & (309,8) & (350,10) & (391,12) & \cellcolor{green!30}(432,14)& (473,14) & (514,14) & (555,14) \\ \hline
    23& (195,4) & (238,5) & (281,7) & (324,8) & (367,10) & (410,12) & (453,14) & \cellcolor{green!30}(496,15)& (539,15) & (582,15) \\ \hline
    24 & (204,4) & (249,5) & (294,7) & (339,8) & (384,10) & (429,12) & (474,14) & (519,15) & (564,15) & (609,15) \\ \hline
    25 & ~(213,4)~ & ~(260,5)~ & ~(307,7)~ & ~(354,8) ~& ~(401,10)~ & ~(448,12)~ & ~(495,14)~ & ~\cellcolor{green!30}(542,16)~ & ~(589,16)~ & ~(636,16)~ \\ \hline
    \end{tabular}
    \captionof{table}{$(n,d)$ for the $k=9$ family from the $[[432, 9, 14]]$ code for different $L_x$ and $L_y$.}
    \label{tab: d(Lx, Ly) for [[432, 9, 14]]}
\end{figure}

\begin{figure*}[!htbp]
    \centering
    \includegraphics[scale=0.065]{code_324_10_11.pdf} 
    \caption{The $[[324, 10, 11]]$ code on an open square lattice. Bulk stabilizers are specified by two polynomials $f(x,y) = x^2y^2(1+x+x^{-2}y^{-2})$ and $g(x,y) = x^2(1+y+x^{-2}y^2)$~\cite{liang2025generalized}. Vertical edges of the rightmost column have been removed.
    In total, there are 110 bulk $X$-stabilizers (deep red), 100 bulk $Z$-stabilizers (deep blue), 44 boundary $X$-stabilizers (red), and 60 boundary $Z$-stabilizers (blue), yielding a logical dimension of $k=324-110-100-44-60=10$.
    This generates a code family of $k=10$: $[[2L_x L_y - L_y, ~10, ~d(L_x, L_y)]]$, with code distances $d(L_x, L_y)$ listed in Table~\ref{tab: d(Lx, Ly) for [[324, 10, 11]]} below.}
    \label{fig: [[324, 10, 11]]}

    \vspace{1em} 

    \setlength{\tabcolsep}{0pt} 
    \begin{tabular}{|c|c|c|c|c|c|c|c|c|c|c|}
    \hline
    \diagbox[height=0.7cm, innerwidth=0.7cm]{$L_x$}{\raisebox{-0.1em}{$L_y$}}   & 6     & 7     & 8     & 9     & 10     & 11      & 12     & 13     & 14     & 15     \\ \hline
    7  & \cellcolor{green!30}(78,4)& (91,4)  & (104,4) & (117,4) & (130,4) & (143,4)  & (156,4)  & (169,4)  & (182,4)  & (195,4)  \\ \hline
    8  & (90,4)  & (105,4) & (120,4) & (135,4) & (150,4) & (165,4)  & (180,4)  & (195,4)  & (210,4)  & (225,4)  \\ \hline
    9  & (102,4) & \cellcolor{green!30}(119,5)& (136,5) & (153,5) & (170,5) & (187,5)  & (204,5)  & (221,5)  & (238,5)  & (255,5)  \\ \hline
    10 & (114,4) & (133,5) & \cellcolor{green!30}(152,6)& (171,6) & (190,6) & (209,6)  & (228,6)  & (247,6)  & (266,6)  & (285,6)  \\ \hline
    11 & (126,4) & (147,5) & (168,6) & \cellcolor{green!30}(189,7)& (210,7) & (231,7)  & (252,7)  & (273,7)  & (294,7)  & (315,7)  \\ \hline
    12 & (138,4) & (161,5) & (184,6) & (207,7) & \cellcolor{green!30}(230,9)& (253,9)  & (276,9)  & (299,9)  & (322,9)  & (345,9)  \\ \hline
    13 & (150,4) & (175,5) & (200,6) & (225,7) & (250,9) & \cellcolor{green!30}(275,10)& (300,10) & (325,10) & (350,10) & (375,10) \\ \hline
    14 & (162,4) & (189,5) & (216,6) & (243,7) & (270,9) & (297,10) & \cellcolor{green!30}(324,11)& (351,11) & (378,11) & (405,11) \\ \hline
    15 & (174,4) & (203,5) & (232,6) & (261,7) & (290,9) & (319,10) & (348,11) & (377,11) & (406,11) & (435,11) \\ \hline
    16 & (186,4) & (217,5) & (248,6) & (279,7) & (310,9) & (341,10) & \cellcolor{green!30}(372,12)& \cellcolor{green!30}(403,13)& (434,13) & (465,13) \\ \hline
    17 & (198,4) & (231,5) & (264,6) & (297,7) & (330,9) & (363,10) & (396,12) & (429,13) & \cellcolor{green!30}(462,14)& (495,14) \\ \hline
    18 & (210,4) & (245,5) & (280,6) & (315,7) & (350,9) & (385,10) & (420,12) & (455,13) & \cellcolor{green!30}(490,15)& (525,15) \\ \hline
    19 & ~(222,4)~ & ~(259,5)~ & ~(296,6)~ & ~(333,7)~ & ~(370,9)~ & ~(407,10)~ & ~(444,12)~ & ~(481,13)~ & ~(518,15)~ & \cellcolor{green!30}~(555,16)~ \\ \hline
    \end{tabular}
    \captionof{table}{$(n,d)$ for the $k=10$ family from the $[[324, 10, 11]]$ code for different $L_x$ and $L_y$.}
    \label{tab: d(Lx, Ly) for [[324, 10, 11]]}
\end{figure*}

\begin{figure*}[!htbp]
    \centering
    \includegraphics[scale=0.065]{code_381_10_13.pdf} 
    \caption{The $[[381, 10, 13]]$ code on an open square lattice. Bulk stabilizers are specified by two polynomials $f(x,y) = x(1+x+x^{-1}y^{2})$ and $g(x,y) = 1+y+xy^4$~\cite{liang2025generalized}. Vertical edges of the rightmost column have been removed.
    In total, there are 120 bulk $X$-stabilizers (deep red), 119 bulk $Z$-stabilizers (deep blue), 64 boundary $X$-stabilizers (red), and 68 boundary $Z$-stabilizers (blue), yielding a logical dimension of $k=381-120-119-64-68=10$.
    This generates a code family of $k=10$: $[[2L_x L_y - 2L_x- L_y+2, ~10, ~d(L_x, L_y)]]$, with code distances $d(L_x, L_y)$ listed in Table~\ref{tab: d(Lx, Ly) for [[381, 10, 13]]} below.}
    \label{fig: [[381, 10, 13]]}

    \vspace{1em} 

    \setlength{\tabcolsep}{0pt} 
    \begin{tabular}{|c|c|c|c|c|c|c|c|c|c|c|c|c|c|c|c|}
        \hline
          \diagbox[height=0.7cm, innerwidth=0.7cm]{$L_x$}{\raisebox{-0.1em}{$L_y$}} & 10& 11& 12& 13& 14& 15& 16& 17& 18& 19& 20& 21& 22& 23& 24\\ \hline
        5  & \cellcolor{green!30}(82,4)  & (91,4)  & (100,4) & (109,4) & (118,4) & (127,4) & (136,4) & (145,4)  & (154,4)  & (163,4)  & (172,4)  & (181,4)  & (190,4)  & (199,4)  & (208,4)  \\ \hline
        6  & (100,4) & (111,4) & \cellcolor{green!30}(122,5) & \cellcolor{green!30}(133,6)& (144,6) & (155,6) & (166,6) & (177,6)  & (188,6)  & (199,6)  & (210,6)  & (221,6)  & (232,6)  & (243,6)  & (254,6)  \\ \hline
        7  & (118,4) & (131,4) & (144,5) & (157,6)& \cellcolor{green!30}(170,7)& (183,7) & (196,7) & (209,7)  & (222,7)  & (235,7)  & (248,7)  & (261,7)  & (274,7)  & (287,7)  & (300,7)  \\ \hline
        8  & (136,4) & (151,4) & (166,5) & (181,6) & (196,7) & \cellcolor{green!30}(211,8)& \cellcolor{green!30}(226,9)& (241,9)  & (256,9)  & (271,9)  & (286,9)  & (301,9)  & (316,9)  & (331,9)  & (346,9)  \\ \hline
        9  & (154,4) & (171,4) & (188,5) & (205,6) & (222,7) & (239,8) & (256,9) & \cellcolor{green!30}(273,10)& (290,10) & (307,10) & (324,10) & (341,10) & (358,10) & (375,10) & (392,10) \\ \hline
        10  & (172,4) & (191,4) & (210,5) & (229,6) & (248,7) & (267,8) & (286,9) & (305,10) & (324,10) & \cellcolor{green!30}(343,11)& \cellcolor{green!30}(362,12)& \cellcolor{green!30}(381,13)& (400,13) & (419,13) & (438,13) \\ \hline
        11  & (190,4) & (211,4) & (232,5) & (253,6) & (274,7) & (295,8) & (316,9) & (337,10) & (358,10) & (379,11) & (400,12) & (421,13) & \cellcolor{green!30}(442,14)& (463,14) & (484,14) \\ \hline
        12 & (208,4) & (231,4) & (254,5) & (277,6) & (300,7) & (323,8) & (346,9) & (369,10) & (392,10) & (415,11) & (438,12) & (461,13) & (484,14) & \cellcolor{green!30}(507,15)& \cellcolor{green!30}(530,16)\\ \hline
        13 & (226,4) & (251,4) & (276,5) & (301,6) & (326,7) & (351,8) & (376,9) & (401,10) & (426,10) & (451,11) & (476,12) & (501,13) & (526,14) & (551,15) & (576,16) \\ \hline
        14 & (244,4) & (271,4) & (298,5) & (325,6) & (352,7) & (379,8) & (406,9) & (433,10) & (460,10) & (487,11) & (514,12) & (541,13) & (568,14) & (595,15) & (622,16) \\ \hline
        15 & (262,4) & (291,4) & (320,5) & (349,6) & (378,7) & (407,8) & (436,9) & (465,10) & (494,10) & (523,11) & (552,12) & (581,13) & (610,14) & (639,15) & (668,16) \\ \hline
        \end{tabular}
    \captionof{table}{$(n,d)$ for the $k=10$ family from the $[[381, 10, 13]]$ code for different $L_x$ and $L_y$.}
    \label{tab: d(Lx, Ly) for [[381, 10, 13]]}
\end{figure*}

\begin{figure}[!htbp]
    \centering
    \includegraphics[scale=0.06]{code_494_11_15.pdf}
    \caption{The $[[494, 11, 15]]$ code on an open square lattice. Bulk stabilizers are specified by two polynomials $f(x,y) = xy^3(1+x+x^{-1}y^{-3})$ and $g(x,y) = y^3(1+y+x y^{-3})$~\cite{liang2025generalized}. Vertical edges of the two rightmost columns and the bottommost row have been removed.
    There are 171 bulk $X$-stabilizers (deep red), 160 bulk $Z$-stabilizers (deep blue), 72 boundary $X$-stabilizers (red), and 80 boundary $Z$-stabilizers (blue), yielding a logical dimension of $k=494-171-160-72-80=11$.
    This generates a code family of $k=11$: $[[2L_x L_y - L_x - L_y +1, ~11, ~d(L_x, L_y)]]$, with code distances $d(L_x, L_y)$ listed in Table~\ref{tab: d(Lx, Ly) for [[494, 11, 15]]} below.}
    \label{fig: [[494, 11, 15]]}

    \vspace{1em} 

    \setlength{\tabcolsep}{0pt} 
\begin{tabular}{|c|c|c|c|c|c|c|c|c|c|c|c|c|c|c|c|c|}
\hline
  \diagbox[height=0.7cm, innerwidth=0.7cm]{$L_x$}{\raisebox{-0.1em}{$L_y$}} & 10& 11& 12& 13& 14& 15& 16& 17& 18& 19& 20& 21& 22& 23& 24& 25\\ \hline
5  & \cellcolor{green!30}(86,4)& (95,4)  & (104,4) & (113,4) & (122,4) & (131,4) & (140,4) & (149,4) & (158,4)  & (167,4)  & (176,4)  & (185,4)  & (194,4)  & (203,4)  & (212,4)  & (221,4)  \\ \hline
6  & (105,4) & (116,4) & \cellcolor{green!30}(127,5)& \cellcolor{green!30}(138,6)& (149,6) & (160,6) & (171,6) & (182,6) & (193,6)  & (204,6)  & (215,6)  & (226,6)  & (237,6)  & (248,6)  & (259,6)  & (270,6)  \\ \hline
7  & (124,4) & (137,4) & (150,5) & (163,6) & \cellcolor{green!30}(176,7)& (189,7) & (202,7) & (215,7) & (228,7)  & (241,7)  & (254,7)  & (267,7)  & (280,7)  & (293,7)  & (306,7)  & (319,7)  \\ \hline
8  & (143,4) & (158,4) & (173,5) & (188,6) & (203,7) & (218,7) & \cellcolor{green!30}(233,8)& \cellcolor{green!30}(248,9)& (263,9)  & (278,9)  & (293,9)  & (308,9)  & (323,9)  & (338,9)  & (353,9)  & (368,9)  \\ \hline
9  & (162,4) & (179,4) & (196,5) & (213,6) & (230,7) & (247,7) & (264,8) & (281,9) & \cellcolor{green!30}(298,10)& (315,10) & (332,10) & (349,10) & (366,10) & (383,10) & (400,10) & (417,10) \\ \hline
10  & (181,4) & (200,4) & (219,5) & (238,6) & (257,7) & (276,7) & (295,8) & (314,9) & (333,10) & \cellcolor{green!30}(352,11)& \cellcolor{green!30}(371,12)& (390,12) & (409,12) & (428,12) & (447,12) & (466,12) \\ \hline
11  & (200,4) & (221,4) & (242,5) & (263,6) & (284,7) & (305,7) & (326,8) & (347,9) & (368,10) & (389,11) & (410,12) & \cellcolor{green!30}(431,13)& (452,13) & \cellcolor{green!30}(473,14)& \cellcolor{green!30}(494,15)& (515,15) \\ \hline
12 & (219,4) & (242,4) & (265,5) & (288,6) & (311,7) & (334,7) & (357,8) & (380,9) & (403,10) & (426,11) & (449,12) & (472,13) & (495,13) & (518,14) & (541,15) & \cellcolor{green!30}(564,16)\\ \hline
13 & (238,4) & (263,4) & (288,5) & (313,6) & (338,7) & (363,7) & (388,8) & (413,9) & (438,10) & (463,11) & (488,12) & (513,13) & (538,13) & (563,14) & (588,15) & (613,16) \\ \hline
14 & (257,4) & (284,4) & (311,5) & (338,6) & (365,7) & (392,7) & (419,8) & (446,9) & (473,10) & (500,11) & (527,12) & (554,13) & (581,13) & (608,14) & (635,15) & (662,16) \\ \hline
\end{tabular}
    \captionof{table}{$(n,d)$ for the $k=11$ family from the $[[494,11,15]]$ code for different $L_x$ and $L_y$.}
    \label{tab: d(Lx, Ly) for [[494, 11, 15]]}
\end{figure}

\begin{figure}[!htbp]
    \centering
    \includegraphics[scale=0.065]{code_432_12_12.pdf}
    \caption{The $[[432, 12, 12]]$ code on an open square lattice. Bulk stabilizers are specified by two polynomials $f(x,y) = x(1+x+x^{-1}y^{3})$ and $g(x,y) = 1+y+x^2 y^3$~\cite{liang2025generalized}.
    There are 150 bulk $X$-stabilizers (deep red), 150 bulk $Z$-stabilizers (deep blue), 60 boundary $X$-stabilizers (red), and 60 boundary $Z$-stabilizers (blue), yielding a logical dimension of $k=432-150\times2-60\times2=12$.
    This generates a code family of $k=12$: $[[2L_x L_y, ~12, ~d(L_x, L_y)]]$, with code distances $d(L_x, L_y)$ listed in Table~\ref{tab: d(Lx, Ly) for [[432, 12, 12]]} below.}
    \label{fig: [[432, 12, 12]]}

    \vspace{1em} 

    \setlength{\tabcolsep}{0pt} 
    \begin{tabular}{|c|c|c|c|c|c|c|c|c|c|c|c|c|c|c|}
    \hline
    \diagbox[height=0.7cm, innerwidth=0.7cm]{$L_x$}{\raisebox{-0.1em}{$L_y$}}   & 9     & 10    & 11    & 12     & 13    & 14    & 15     & 16     & 17     & 18     & 19     & 20     & 21      &22\\ \hline
    6  & \cellcolor{green!30}~(108,4)~& ~(120,4)~ & ~(132,4)~ & ~(144,4)~ & ~(156,4)~ & ~(168,4)~ & ~(180,4)~ & ~(192,4)~ & ~(204,4)~ & ~(216,4)~ & ~(228,4)~ & ~(240,4)~ & ~(252,4)~ & ~(264,4)~ \\ \hline
    7  & (126,4) & \cellcolor{green!30}(140,5)& (154,5) & (168,5) & (182,5) & (196,5) & (210,5) & (224,5) & (238,5) & (252,5) & (266,5) & (280,5) & (294,5) & (308,5) \\ \hline
    8  & (144,4) & (160,5) & (176,5) & \cellcolor{green!30}(192,6)& (208,6) & (224,6) & (240,6) & (256,6) & (272,6) & (288,6) & (304,6) & (320,6) & (336,6) & (352,6) \\ \hline
    9  & (162,4) & (180,5) & (198,5) & (216,6) & \cellcolor{green!30}(234,7)& (252,7) & (270,7) & (288,7) & (306,7) & (324,7) & (342,7) & (360,7) & (378,7) & (396,7) \\ \hline
    10 & (180,4) & (200,5) & (220,5) & (240,6) & (260,7) & \cellcolor{green!30}(280,8)& \cellcolor{green!30}(300,9)& (320,9) & (340,9) & (360,9) & (380,9) & (400,9) & (420,9) & (440,9) \\ \hline
    11 & (198,4) & (220,5) & (242,5) & (264,6) & (286,7) & (308,8) & \cellcolor{green!30}(330,10)& (352,10) & (374,10) & (396,10) & (418,10) & (440,10) & (462,10) & (484,10) \\ \hline
    12 & (216,4) & (240,5) & (264,5) & (288,6) & (312,7) & (336,8) & (360,10) & \cellcolor{green!30}(384,11)& (408,11) & \cellcolor{green!30}(432,12)& (456,12) & (480,12) & (504,12) & (528,12) \\ \hline
    13 & (234,4) & (260,5) & (286,5) & (312,6) & (338,7) & (364,8) & (390,10) & (416,11) & (442,11) & (468,12) & \cellcolor{green!30}(494,13)& (520,13) & (546,13) & (572,13) \\ \hline
    14 & (252,4) & (280,5) & (308,5) & (336,6) & (364,7) & (392,8) & (420,10) & (448,11) & (476,11) & (504,12) & (532,13) & \cellcolor{green!30}(560,14)& \cellcolor{green!30}(588,15)& (616,15) \\ \hline
    15 & ~(270,4)~ & ~(300,5)~ & ~(330,5)~ & ~(360,6)~ & ~(390,7)~ & ~(420,8)~ & (450,10) & (480,11) & (510,11) & (540,12) & (570,13) & (600,14) & (630,15) & \cellcolor{green!30}(660,16) \\ \hline
    \end{tabular}
    \captionof{table}{$(n,d)$ for the $k=12$ family from the $[[432,12,12]]$ code for different $L_x$ and $L_y$.}
    \label{tab: d(Lx, Ly) for [[432, 12, 12]]}
\end{figure}

\begin{figure}[!htbp]
    \centering
    \includegraphics[scale=0.065]{code_279_12_9.pdf}
    \caption{The $[[279, 12, 9]]$ code on an open square lattice. Bulk stabilizers are specified by two polynomials $f(x,y) = x(1+x+x^{-1}y^{2})$ and $g(x,y) = 1+y+xy^5$~\cite{liang2025generalized}.
    There are 72 bulk $X$-stabilizers (deep red), 75 bulk $Z$-stabilizers (deep blue), 60 boundary $X$-stabilizers (red), and 60 boundary $Z$-stabilizers (blue), yielding a logical dimension of $k=279-72-75-60\times2=12$.
    This generates a code family of $k=12$: $[[2L_x L_y-3L_x-L_y+3, ~12, ~d(L_x, L_y)]]$, with code distances $d(L_x, L_y)$ listed in Table~\ref{tab: d(Lx, Ly) for [[279, 12, 9]]} below.}
    \label{fig: [[279, 12, 9]]}

    \vspace{1em} 

    \setlength{\tabcolsep}{0pt} 
    \scalebox{0.85}{
    \begin{tabular}{|c|c|c|c|c|c|c|c|c|c|c|c|c|c|c|c|c|c|c|c|}
        \hline
        \diagbox[height=0.7cm, innerwidth=0.7cm]{$L_x$}{\raisebox{-0.1em}{$L_y$}}& 13& 14& 15& 16& 17& 18& 19& 20& 21& 22& 23& 24& 25& 26& 27& 28& 29& 30& 31\\
        \hline
        5& \cellcolor{green!30}(105,4)& (114,4) & (123,4) & (132,4) & (141,4) & (150,4) & (159,4) & (168,4) & (177,4)  & (186,4)  & (195,4)  & (204,4)  & (213,4)  & (222,4)  & (231,4)  & (240,4)  & (249,4)  & (258,4)  & (267,4)  \\ \hline
        6& (128,4) & (139,4) & \cellcolor{green!30}(150,5)& (161,5) & \cellcolor{green!30}(172,6)& (183,6) & (194,6) & (205,6) & (216,6)  & (227,6)  & (238,6)  & (249,6)  & (260,6)  & (271,6)  & (282,6)  & (293,6)  & (304,6)  & (315,6)  & (326,6)  \\ \hline
        7& (151,4) & (164,4) & (177,5) & (190,5) & (203,6) & \cellcolor{green!30}(216,7)& (229,7) & (242,7) & (255,7)  & (268,7)  & (281,7)  & (294,7)  & (307,7)  & (320,7)  & (333,7)  & (346,7)  & (359,7)  & (372,7)  & (385,7)  \\ \hline
        8  & (174,4) & (189,4) & (204,5) & (219,5) & (234,6) & (249,7) & \cellcolor{green!30}(264,8)& \cellcolor{green!30}(279,9)& (294,9)  & (309,9)  & (324,9)  & (339,9)  & (354,9)  & (369,9)  & (384,9)  & (399,9)  & (414,9)  & (429,9)  & (444,9)  \\ \hline
        9  & (197,4) & (214,4) & (231,5) & (248,5) & (265,6) & (282,7) & (299,8) & (316,9) & \cellcolor{green!30}(333,10)& (350,10) & (367,10) & (384,10) & (401,10) & (418,10) & (435,10) & (452,10) & (469,10) & (486,10) & (503,10) \\ \hline
        10  & (220,4) & (239,4) & (258,5) & (277,5) & (296,6) & (315,7) & (334,8) & (353,9) & (372,10) & (391,10) & (410,10) & (429,10) & \cellcolor{green!30}(448,11) & \cellcolor{green!30}(467,12) & \cellcolor{green!30}(486,13) & (505,13) & (524,13) & (543,13) & (562,13) \\ \hline
        11  & (243,4) & (264,4) & (285,5) & (306,5) & (327,6) & (348,7) & (369,8) & (390,9) & (411,10) & (432,10) & (453,10) & (474,10) & (495,11) & (516,12) & (537,13) & (558,13) & \cellcolor{green!30}(579,14) & (600,14) & (621,14) \\ \hline
        12 & (266,4) & (289,4) & (312,5) & (335,5) & (358,6) & (381,7) & (404,8) & (427,9) & (450,10) & (473,10) & (496,10) & (519,10) & (542,11) & (565,12) & (588,13) & (611,13) & (634,14) & \cellcolor{green!30}(657,15) & \cellcolor{green!30}(680,16) \\ \hline
    \end{tabular}
    }
    \captionof{table}{$(n,d)$ for the $k=12$ family from the $[[279,12,9]]$ code for different $L_x$ and $L_y$.}
    \label{tab: d(Lx, Ly) for [[279, 12, 9]]}
\end{figure}

\begin{figure}[!htbp]
    \centering
    \includegraphics[scale=0.065]{code_392_13_11.pdf}
    \caption{The $[[392, 13, 11]]$ code on an open square lattice. Bulk stabilizers are specified by two polynomials $f(x,y) = x^2(1+x+x^{-2}y^{2})$ and $g(x,y) = 1+y+x^2 y^3$~\cite{liang2025generalized}. Vertical edges of the rightmost column and the bottommost row have been removed.
    There are 120 bulk $X$-stabilizers (deep red), 121 bulk $Z$-stabilizers (deep blue), 72 boundary $X$-stabilizers (red), and 66 boundary $Z$-stabilizers (blue), yielding a logical dimension of $k=392-120-121-72-66=13$.
    This generates a code family of $k=13$: $[[2L_x L_y - L_x - L_y +1, ~13, ~d(L_x, L_y)]]$, with code distances $d(L_x, L_y)$ listed in Table~\ref{tab: d(Lx, Ly) for [[392, 13, 11]]} below.}
    \label{fig: [[392, 13, 11]]}

    \vspace{1em} 

    \setlength{\tabcolsep}{0pt} 
    \begin{tabular}{|c|c|c|c|c|c|c|c|c|c|c|c|c|}
    \hline
    \diagbox[height=0.7cm, innerwidth=0.7cm]{$L_x$}{\raisebox{-0.1em}{$L_y$}}   & 7     & 8     & 9     & 10     & 11     & 12     & 13     & 14     & 15     & 16     & 17     & 18     \\ \hline
    8  & \cellcolor{green!30}(98,4)& (113,4) & (128,4) & (143,4) & (158,4) & (173,4) & (188,4) & (203,4) & (218,4) & (233,4) & (248,4) & (263,4) \\ \hline
    9  & (111,4) & (128,4) & \cellcolor{green!30}(145,5)& (162,5) & (179,5) & (196,5) & (213,5) & (230,5) & (247,5) & (264,5) & (281,5) & (298,5) \\ \hline
    10 & (124,4) & (143,4) & (162,5) & (181,5) & (200,5) & (219,5) & (238,5) & (257,5) & (276,5) & (295,5) & (314,5) & (333,5) \\ \hline
    11 & (137,4) & (158,4) & (179,5) & \cellcolor{green!30}(200,6)& (221,6) & (242,6) & (263,6) & (284,6) & (305,6) & (326,6) & (347,6) & (368,6) \\ \hline
    12 & (150,4) & (173,4) & (196,5) & (219,6) & \cellcolor{green!30}(242,7)& (265,7) & (288,7) & (311,7) & (334,7) & (357,7) & (380,7) & (403,7) \\ \hline
    13 & (163,4) & (188,4) & (213,5) & (238,6) & (263,7) & \cellcolor{green!30}(288,9)& (313,9) & (338,9) & (363,9) & (388,9) & (413,9) & (438,9) \\ \hline
    14 & (176,4) & (203,4) & (230,5) & (257,6) & (284,7) & (311,9) & (338,9) & (365,9) & (392,9) & (419,9) & (446,9) & (473,9) \\ \hline
    15 & (189,4) & (218,4) & (247,5) & (276,6) & (305,7) & (334,9) & \cellcolor{green!30}(363,10)& \cellcolor{green!30}(392,11)& (421,11) & (450,11) & (479,11) & (508,11) \\ \hline
    16 & (202,4) & (233,4) & (264,5) & (295,6) & (326,7) & (357,9) & (388,10) & (419,11) & (450,11) & \cellcolor{green!30}(481,12)& (512,12) & (543,12) \\ \hline
    17 & (215,4) & (248,4) & (281,5) & (314,6) & (347,7) & (380,9) & (413,10) & (446,11) & (479,11) & \cellcolor{green!30}(512,13)& (545,13) & (578,13) \\ \hline
    18 & (228,4) & (263,4) & (298,5) & (333,6) & (368,7) & (403,9) & (438,10) & (473,11) & (508,11) & (543,13) & \cellcolor{green!30}(578,14)& (613,14) \\ \hline
    19 & ~(241,4)~ & ~(278,4)~ & ~(315,5)~ & ~(352,6)~ & ~(389,7)~ & ~(426,9)~ & ~(463,10)~ & ~(500,11)~ & ~(537,11)~ & ~(574,13)~ & ~(611,14)~ & \cellcolor{green!30}~(648,16)~ \\ \hline
    \end{tabular}
    \captionof{table}{$(n,d)$ for the $k=13$ family from the $[[392, 13,11]]$ code for different $L_x$ and $L_y$.}
    \label{tab: d(Lx, Ly) for [[392, 13, 11]]}
\end{figure}

\begin{figure}[!htbp]
    \vspace{1em}
    \hspace{-2em} 
    \includegraphics[scale=0.048]{code_495_13_13.pdf}
    \vspace{1em}
    \caption{ The $[[495, 13, 13]]$ code on an open square lattice. Bulk stabilizers are specified by two polynomials $f(x,y) = 1+x+x^{5}y$ and $g(x,y) = y(1+y+x^3 y^{-1})$~\cite{liang2025generalized}. Vertical edges of the rightmost column and the bottommost row have been removed.
    There are 154 bulk $X$-stabilizers (deep red), 160 bulk $Z$-stabilizers (deep blue), 88 boundary $X$-stabilizers (red), and 80 boundary $Z$-stabilizers (blue), yielding a logical dimension of $k=495-154-160-88-80=13$.
    This generates a code family of $k=13$: $[[2L_x L_y - L_x - 2L_y +2, ~13, ~d(L_x, L_y)]]$, with code distances $d(L_x, L_y)$ listed in Table~\ref{tab: d(Lx, Ly) for [[495, 13, 13]]} below.}
    \label{fig: [[495, 13, 13]]}

    \vspace{1em} 

    \setlength{\tabcolsep}{0pt} 
    \begin{tabular}{|c|c|c|c|c|c|c|c|c|c|}
\hline
   \diagbox[height=0.7cm, innerwidth=0.7cm]{$L_x$}{\raisebox{-0.1em}{$L_y$}}& 5& 6& 7& 8& 9& 10& 11& 12& 13\\ \hline
13& \cellcolor{green!30}(109,4)& (133,4) & (157,4) & (181,4) & (205,4)  & (229,4)  & (253,4)  & (277,4)  & (301,4)  \\ \hline
14& (118,4) & (144,4) & (170,4) & (196,4) & (222,4)  & (248,4)  & (274,4)  & (300,4)  & (326,4)  \\ \hline
15& (127,4) & \cellcolor{green!30}(155,5)& (183,5) & (211,5) & (239,5)  & (267,5)  & (295,5)  & (323,5)  & (351,5)  \\ \hline
16& (136,4) & (166,5) & (196,5) & (226,5) & (256,5)  & (286,5)  & (316,5)  & (346,5)  & (376,5)  \\ \hline
17& (145,4) & \cellcolor{green!30}(177,6)& (209,6) & (241,6) & (273,6)  & (305,6)  & (337,6)  & (369,6)  & (401,6)  \\ \hline
18& (154,4) & (188,6) & \cellcolor{green!30}(222,7)& (256,7) & (290,7)  & (324,7)  & (358,7)  & (392,7)  & (426,7)  \\ \hline
19& (163,4) & (199,6) & (235,7) & \cellcolor{green!30}(271,8)& (307,8)  & (343,8)  & (379,8)  & (415,8)  & (451,8)  \\ \hline
20& (172,4) & (210,6) & (248,7) & \cellcolor{green!30}(286,9)& (324,9)  & (362,9)  & (400,9)  & (438,9)  & (476,9)  \\ \hline
21& (181,4) & (221,6) & (261,7) & (301,9) & \cellcolor{green!30}(341,10)& (381,10) & (421,10) & (461,10) & (501,10) \\ \hline
22& (190,4) & (232,6) & (274,7) & (316,9) & (358,10) & (400,10) & (442,10) & (484,10) & (526,10) \\ \hline
23& (199,4) & (243,6) & (287,7) & (331,9) & (375,10) & (419,10) & (463,10) & (507,10) & (551,10) \\ \hline
24& (208,4) & (254,6) & (300,7) & (346,9) & (392,10) & (438,10) & (484,10) & (530,10) & (576,10) \\ \hline
25& (217,4) & (265,6) & (313,7) & (361,9) & (409,10) & \cellcolor{green!30}(457,11)& (505,11) & (553,11) & (601,11) \\ \hline
26& (226,4) & (276,6) & (326,7) & (376,9) & (426,10) & \cellcolor{green!30}(476,12)& (526,12) & (576,12) & (626,12) \\ \hline
27& (235,4) & (287,6) & (339,7) & (391,9) & (443,10) & \cellcolor{green!30}(495,13)& (547,13) & (599,13) & (651,13) \\ \hline
28& (244,4) & (298,6) & (352,7) & (406,9) & (460,10) & (514,13) & (568,13) & (622,13) & (676,13) \\ \hline
29& (253,4) & (309,6) & (365,7) & (421,9) & (477,10) & (533,13) & \cellcolor{green!30}(589,14)& (645,14) & (701,14)  \\ \hline
30& (262,4) & (320,6) & (378,7) & (436,9) & (494,10) & (552,13) & \cellcolor{green!30}(610,15)& (668,15) & (726,15)  \\ \hline
31& ~(271,4)~ & ~(331,6)~ & ~(391,7)~ & ~(451,9)~ & ~(511,10)~ & ~(571,13)~ & ~(631,15)~ & ~\cellcolor{green!30}(691,16)~ & ~(751,16)~  \\ \hline
\end{tabular}
    \captionof{table}{$(n,d)$ for the $k=13$ family from the $[[495, 13,13]]$ code for different $L_x$ and $L_y$.}
    \label{tab: d(Lx, Ly) for [[495, 13, 13]]}
\end{figure}

\begin{figure}
\centering
    \includegraphics[scale=0.07]{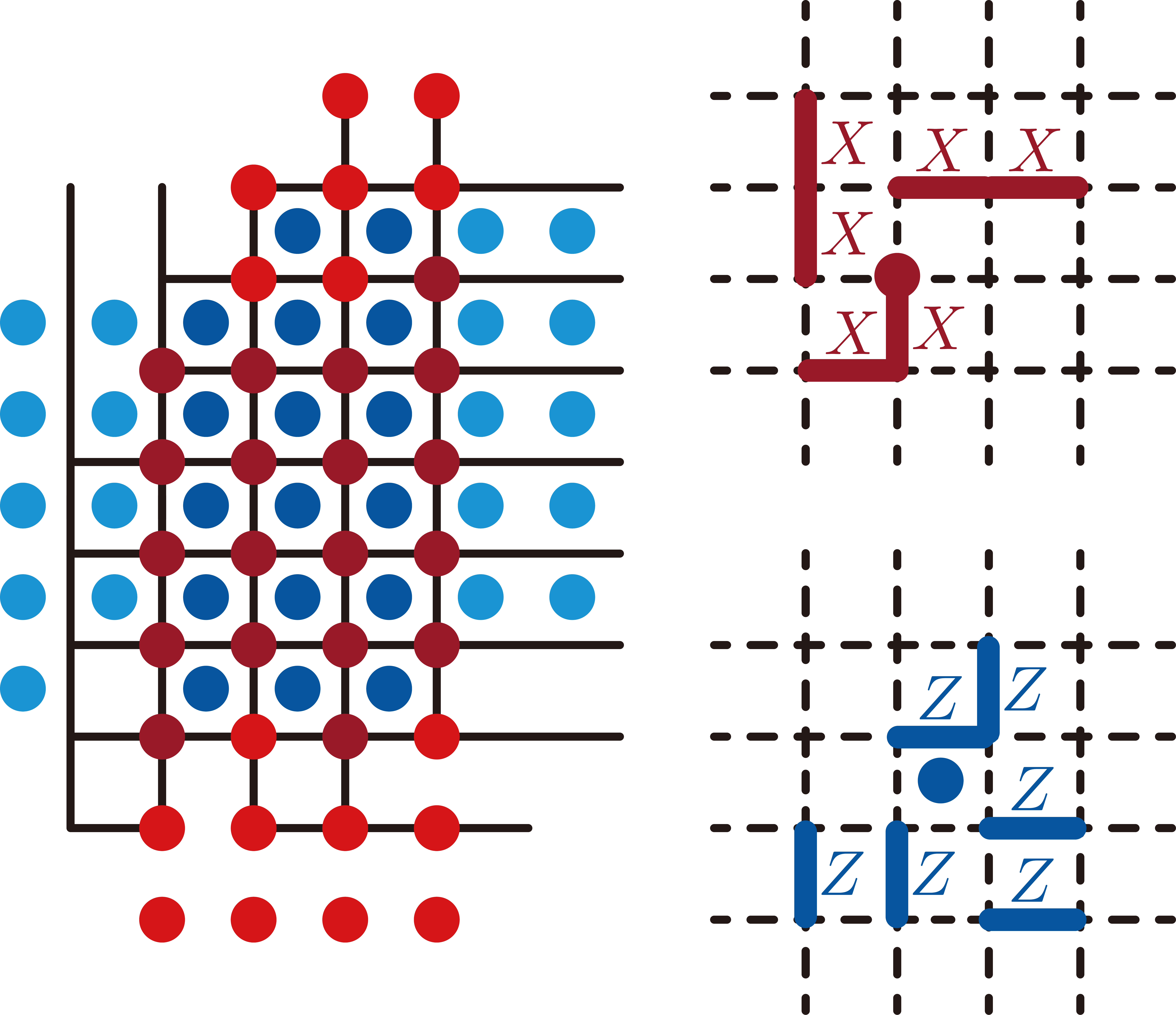}
    \caption{ The grafted $[[78,6,6]]$ code. Each red (blue) dot represents an $X$ ($Z$) stabilizer. Deep red/blue dots correspond to bulk stabilizers that are fully supported on the pruned lattice, while dots near the boundary represent stabilizers with truncated edges (i.e., the bulk stabilizer pattern positioned at the boundary dot, retaining only the Pauli operators on edges within the pruned lattice).
    The design ensures that all truncated $X$ and $Z$ operators still commute, forming valid boundary stabilizers. In total, there are 19 bulk $X$-stabilizers (deep red), 17 bulk $Z$-stabilizers (deep blue), 17 boundary $X$-stabilizers (red), and 19 boundary $Z$-stabilizers (blue), yielding a logical dimension of $k=78-19\times2-17\times 2=6$.}
    \label{fig: [[78, 6, 6]] code}
\end{figure}


\begin{figure}
\centering
    \includegraphics[scale=0.07]{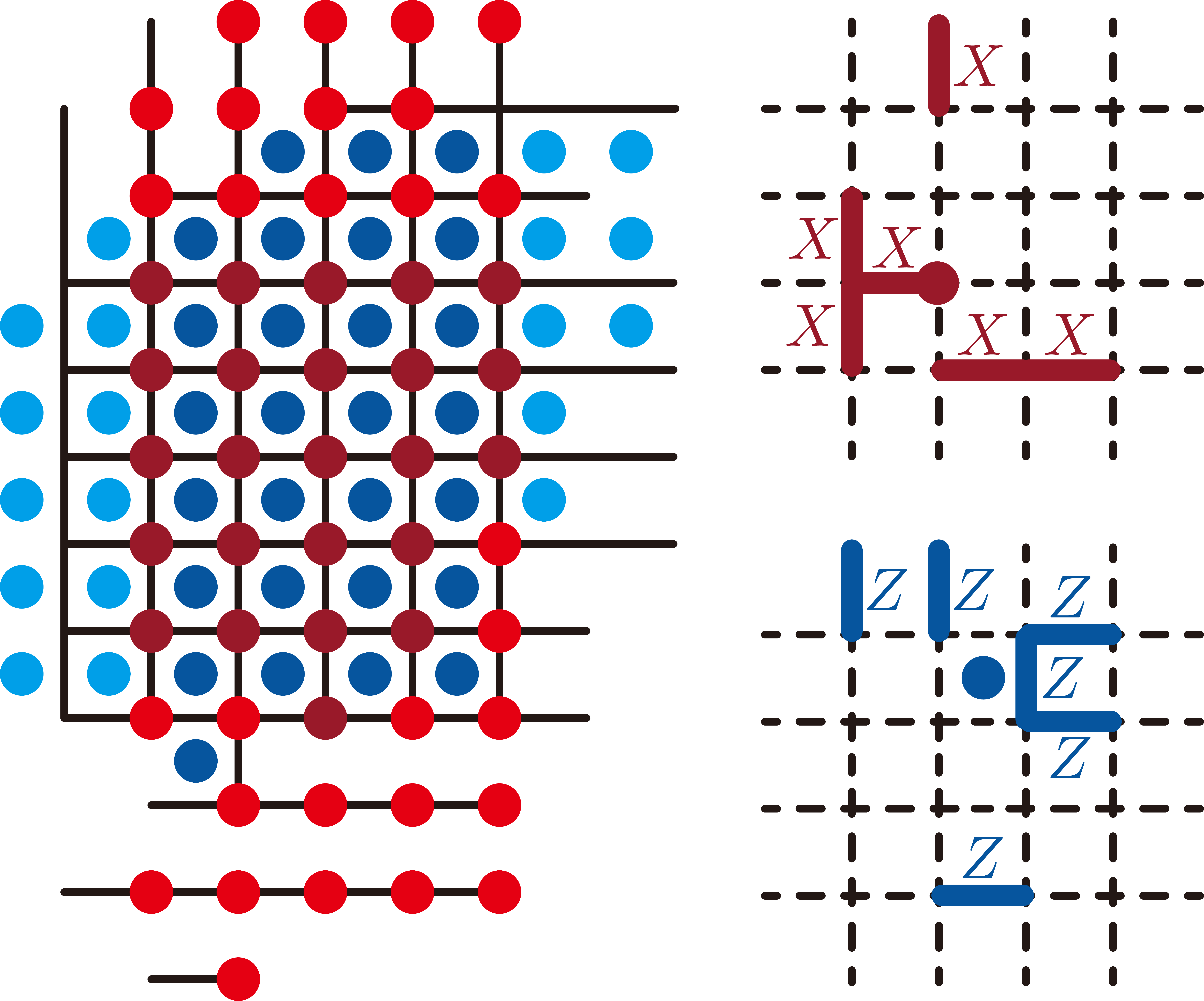}
    \caption{The grafted $[[107, 7, 7]]$ code. Each red (blue) dot represents an $X$ ($Z$) stabilizer. Deep red/blue dots correspond to bulk stabilizers that are fully supported on the pruned lattice, while dots near the boundary represent stabilizers with truncated edges. There are 24 bulk $X$-stabilizers (deep red), 28 bulk $Z$-stabilizers (deep blue), 29 boundary $X$-stabilizers (red), and 19 boundary $Z$-stabilizers (blue), yielding a logical dimension of $k=107-24-28-29-19=7$. }
    \label{fig: [[107, 7, 7]] code}
\end{figure}

\begin{figure}
\centering
    \includegraphics[scale=0.07]{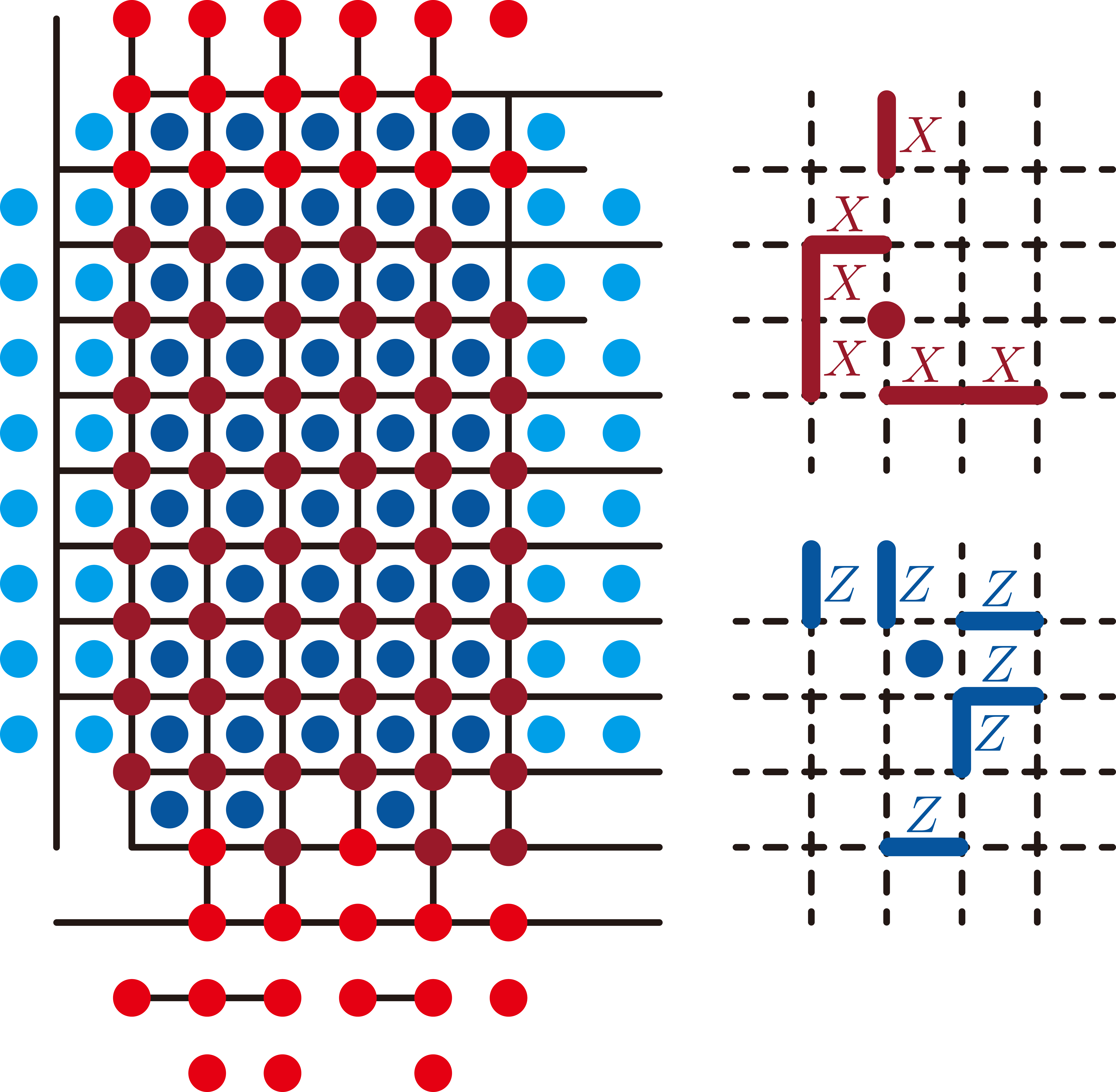}
    \vspace{-0.75em}
    \caption{The grafted $[[173, 8, 9]]$ code. Each red (blue) dot represents an $X$ ($Z$) stabilizer. Deep red/blue dots correspond to bulk stabilizers that are fully supported on the pruned lattice, while dots near the boundary represent stabilizers with truncated edges. There are 50 bulk $X$-stabilizers (deep red), 48 bulk $Z$-stabilizers (deep blue), 33 boundary $X$-stabilizers (red), and 34 boundary $Z$-stabilizers (blue), yielding a logical dimension of $k=173-50-48-33-34=8$.}
    \label{fig: [[173, 8, 9]] code}
    \vspace{1em}
    \centering
    \includegraphics[scale=0.07]{code_268_8_12.pdf}
    \vspace{-0.75em}
    \caption{The grafted $[[268, 8, 12]]$ code. Each red (blue) dot represents an $X$ ($Z$) stabilizer. Deep red/blue dots correspond to bulk stabilizers that are fully supported on the pruned lattice, while dots near the boundary represent stabilizers with truncated edges. There are 89 bulk $X$-stabilizers (deep red), 87 bulk $Z$-stabilizers (deep blue), 41 boundary $X$-stabilizers (red), and 43 boundary $Z$-stabilizers (blue), yielding a logical dimension of $k=268-89-87-41-43=8$.}
    \label{fig: [[268, 8, 12]] code}
\end{figure}

\begin{figure}
\centering
    \includegraphics[scale=0.07]{code_405_9_15.pdf}
    \caption{ The grafted $[[405, 9, 15]]$ code. Each red (blue) dot represents an $X$ ($Z$) stabilizer. Deep red/blue dots correspond to bulk stabilizers that are fully supported on the pruned lattice, while dots near the boundary represent stabilizers with truncated edges. There are 143 bulk $X$-stabilizers (deep red), 136 bulk $Z$-stabilizers (deep blue), 58 boundary $X$-stabilizers (red), and 59 boundary $Z$-stabilizers (blue), yielding a logical dimension of $k=405-143-136-58-59=9$. }
    \label{fig: [[405, 9, 15]] code}
\end{figure}

\begin{figure}
\centering
    \includegraphics[scale=0.07]{code_348_10_13.pdf}
    \caption{  The grafted $[[348, 10, 13]]$ code. Each red (blue) dot represents an $X$ ($Z$) stabilizer. Deep red/blue dots correspond to bulk stabilizers that are fully supported on the pruned lattice, while dots near the boundary represent stabilizers with truncated edges. There are 114 bulk $X$-stabilizers (deep red), 109 bulk $Z$-stabilizers (deep blue), 57 boundary $X$-stabilizers (red), and 58 boundary $Z$-stabilizers (blue), yielding a logical dimension of $k=348-114-109-57-58=10$. }
    \label{fig: [[348, 10, 13]] code}
\end{figure}

\begin{figure}
\centering
    \includegraphics[scale=0.07]{code_450_11_15.pdf}
    \caption{ The grafted $[[450, 11, 15]]$ code. Each red (blue) dot represents an $X$ ($Z$) stabilizer. Deep red/blue dots correspond to bulk stabilizers that are fully supported on the pruned lattice, while dots near the boundary represent stabilizers with truncated edges. There are 154 bulk $X$-stabilizers (deep red), 148 bulk $Z$-stabilizers (deep blue), 66 boundary $X$-stabilizers (red), and 71 boundary $Z$-stabilizers (blue), yielding a logical dimension of $k=450-154-148-66-71=11$. }
    \label{fig: [[450, 11, 15]] code}
\end{figure}

\begin{figure}
\centering
    \includegraphics[scale=0.07]{code_386_12_12.pdf}
    \caption{The grafted $[[386, 12, 12]]$ code. Each red (blue) dot represents an $X$ ($Z$) stabilizer. Deep red/blue dots correspond to bulk stabilizers that are fully supported on the pruned lattice, while dots near the boundary represent stabilizers with truncated edges. There are 121 bulk $X$-stabilizers (deep red), 128 bulk $Z$-stabilizers (deep blue), 71 boundary $X$-stabilizers (red), and 54 boundary $Z$-stabilizers (blue), yielding a logical dimension of $k=386-121-128-71-54=12$.}
    \label{fig: [[386, 12, 12]] code}
\end{figure}

\begin{figure}
\centering
    \includegraphics[scale=0.07]{code_362_13_11.pdf}
    \caption{The grafted $[[362, 13, 11]]$ code. Each red (blue) dot represents an $X$ ($Z$) stabilizer. Deep red/blue dots correspond to bulk stabilizers that are fully supported on the pruned lattice, while dots near the boundary represent stabilizers with truncated edges. There are 107 bulk $X$-stabilizers (deep red), 110 bulk $Z$-stabilizers (deep blue), 76 boundary $X$-stabilizers (red), and 56 boundary $Z$-stabilizers (blue), yielding a logical dimension of $k=362-107-110-76-56=13$.}
    \label{fig: [[362, 13, 11]] code}
\end{figure}

\begin{figure}
\centering
    \includegraphics[scale=0.07]{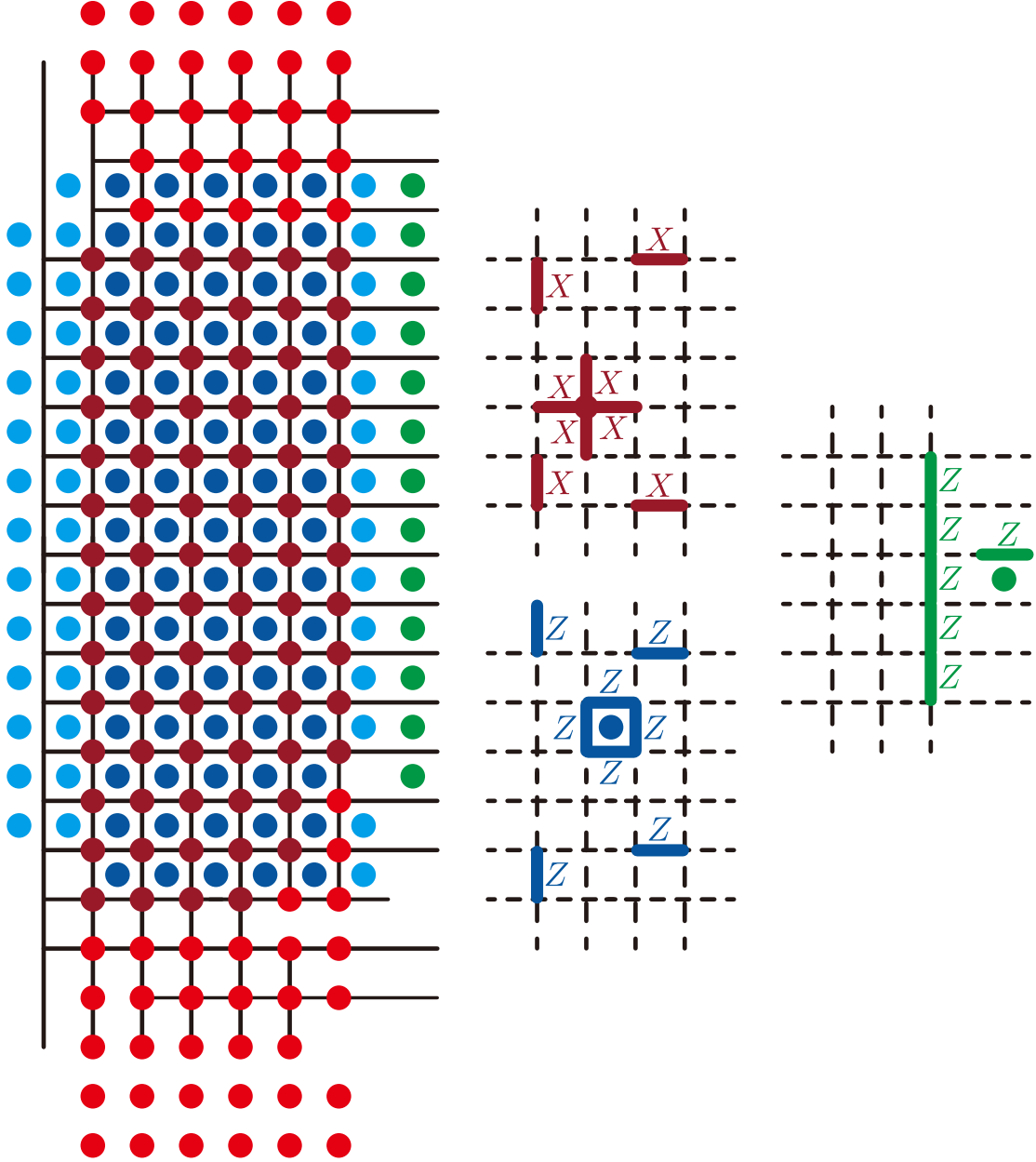}
    \caption{The grafted $[[282, 12, 14]]$ code. Each red (blue) dot represents an $X$ ($Z$) stabilizer. Deep red/blue dots correspond to bulk stabilizers that are fully supported on the pruned lattice, while dots near the boundary represent stabilizers with truncated edges. Furthermore, the green terms indicate the secondary boundary gauge operators that do not originate from truncating the bulk stabilizers.
    There are 80 bulk $X$-stabilizers (deep red), 75 bulk $Z$-stabilizers (deep blue), 61 boundary $X$-stabilizers (red), 41 boundary $Z$-stabilizers (blue) and 13 boundary $Z$-stabilizers (green), yielding a logical dimension of $k=282-80-75-61-41-13=12$.}
    \label{fig: [[282, 12, 14]] code}
\end{figure}

\end{widetext}

\end{document}